\documentclass[useAMS,usenatbib]{mn2e}

\usepackage{epsfig,color,pifont}

\usepackage{natbib}
\bibpunct{(}{)}{;}{a}{}{,}
\def\spose#1{\hbox to 0pt{#1\hss}}
\def\ltsimm{\mathrel{\spose{\lower 3pt\hbox{$\sim$}}
        \raise 2.0pt\hbox{$<$}}}
\def\gtsimm{\mathrel{\spose{\lower 3pt\hbox{$\sim$}}
        \raise 2.0pt\hbox{$>$}}}

\def\cm{{\rm\thinspace cm}}

\def\s{{\rm\thinspace s}}
\def\yr{{\rm\thinspace yr}}
\def\g{{\rm\thinspace g}}

\def\erg{{\rm\thinspace erg}}
\def\Hz{{\rm\thinspace Hz}}

\def\ster{{\rm\thinspace ster}}
\def\ergps{\hbox{${\rm\erg\s^{-1}\,}$}}

\def\Msol{\hbox{${\rm\thinspace M_{\odot}}$}}

\def\Msolpyr{\hbox{${\rm\Msol\yr^{-1}\,}$}}
\def\pcm{\hbox{${\rm\cm^{-1}\,}$}}
\def\pcm2{\hbox{${\rm\cm^{-2}\,}$}}
\def\pcm3{\hbox{${\rm\cm^{-3}\,}$}}
\def\ergpscm3Hz{\hbox{${\rm\ergps\cm^{-3}\Hz^{-1}\,}$}}
\def\ergpscm3Hzster{\hbox{${\rm\ergps\cm^{-3}\Hz^{-1}\ster^{-1}\,}$}}
\def\gpcm3{\hbox{${\rm\g\cm^{-3}\,}$}}
\def\ergpcm2{\hbox{${\rm\erg\cm^{-2}\,}$}}
\def\ergpcm3{\hbox{${\rm\erg\cm^{-3}\,}$}}
\def\phpscm2{\hbox{${\rm photons\s^{-1}\cm^{-2}\,}$}}

\def\etacar{$\eta\thinspace\rm{Car}\;$}

\def\aap{{\rm A\&A}}
\def\apj{{\rm ApJ}}
\def\apjl{{\rm ApJL}}
\def\apjs{{\rm ApJS}}
\def\aj{{\rm AJ}}
\def\mnras{{\rm MNRAS}}

\def\apss{{\rm Ap\&SS}}

\def\cpc{{\rm Comp.~Phys.~Comm.}}

\def\ijnmf{{\rm Internat.~J.~Numer.~Methods~Fluids}}

\title[Numerical heat conduction in hydrodynamical models]{Numerical
  heat conduction in hydrodynamical models of colliding hypersonic
  flows}

\author[E.~R.~Parkin \& J.~M.~Pittard]
       {E.~R.~Parkin$^{1,2}$\thanks{E-mail:parkin@astro.ulg.ac.be} \&
         J.~M.~Pittard$^{2}$\\ $^{1}$Institut d'Astrophysique et de
         G\'{e}ophysique, Universit\'{e} de Li\`{e}ge, 17, All\'{e}e
         du 6 Ao\^{u}t, B5c, B-4000 Sart Tilman, Belgium
         \\ $^{2}$School of Physics and Astronomy, The University of
         Leeds, Woodhouse Lane, Leeds LS2 9JT, UK}

\begin{document}

\date{Accepted 2010 April 20. Received 2010 April 20; in original form
  2010 March 01}

\pagerange{\pageref{firstpage}--\pageref{lastpage}} \pubyear{2009}

\maketitle

\label{firstpage}

\begin{abstract}
Hydrodynamical models of colliding hypersonic flows are presented
which explore the dependence of the resulting dynamics and the
characteristics of the derived X-ray emission on numerical conduction
and viscosity. For the purpose of our investigation we present models
of colliding flow with plane-parallel and cylindrical
divergence. Numerical conduction causes erroneous heating of gas
across the contact discontinuity which has implications for the rate
at which the gas cools. We find that the dynamics of the shocked gas
and the resulting X-ray emission are strongly dependent on the
contrast in the density and temperature either side of the contact
discontinuity, these effects being strongest where the postshock gas
of one flow behaves quasi-adiabatically while the postshock gas of the
other flow is strongly radiative.

Introducing additional numerical viscosity into the simulations has
the effect of damping the growth of instabilities, which in some cases
act to increase the volume of shocked gas and can re-heat gas via
sub-shocks as it flows downstream. The resulting reduction in the
surface area between adjacent flows, and therefore of the amount of
numerical conduction, leads to a commensurate reduction in spurious
X-ray emission, though the dynamics of the collision are compromised.

The simulation resolution also affects the degree of numerical
conduction. A finer resolution better resolves the interfaces of high
density and temperature contrast and although numerical conduction
still exists the volume of affected gas is considerably
reduced. However, since it is not always practical to increase the
resolution, it is imperative that the degree of numerical conduction
is understood so that inaccurate interpretations can be avoided. This
work has implications for the dynamics and emission from astrophysical
phenomena which involve high Mach number shocks.

\end{abstract}

\begin{keywords}
hydrodynamics - methods:numerical - conduction - Shock waves - X-rays:general -
ISM:jets and outflows - stars:outflows
\end{keywords}

\section{Introduction}
\label{sec:intro}

Colliding hypersonic flows occur in a number of astrophysical
environments and over a wide range of scales, e.g. massive young
stellar objects \citep{Parkin:2009b}, astrophysical jets
\citep{Falle:1991, Shang:2006, Bonito:2007, Sutherland:2007},
colliding wind binary systems \citep[CWBs,][]{Stevens:1992,
  Pittard:2009}, wind-blown bubbles around evolved stars
\citep[see][and references there-in]{Arthur:2007}, SNe
\citep[e.g.][]{Tenorio-Tagle:1991, Dwarkadas:2007}, and the cumulative
outflows from young star clusters \citep[][]{Canto:2000,
  Rockefeller:2005, Wunsch:2008, Rodriquez-Gonzalez:2008,
  Reyes-Iturbide:2009} and starburst galaxies
\citep[][]{Strickland:2000, Tenorio-Tagle:2003, Tang:2009}. Flow
collisions can be subject to turbulent motions, the growth of linear
and non-linear instabilities in boundary layers, and in some cases a
global instability of the shocked gas \citep[i.e. radiative
  overstability, ][]{Chevalier:1982}. The combination of these effects
leads to complex scenarios for which numerical hydrodynamics has
proved to be a useful investigatory tool.

However, in the discretization of the governing equations of
hydrodynamics, additional terms are introduced which are purely
numerical in origin. Depending on the order of the scheme, the
appearance of these terms acts to disperse or dissipate the solution,
and therefore terms such as ``numerical dispersion'', ``numerical
diffusion'' or ``artificial viscosity'' are often used to describe
them. The undesirable effects of numerical diffusion are minimized as
one uses higher order schemes, though all schemes are only first order
accurate near discontinuites such as shocks (where flow variables as
well as the perpendicular velocity component, $v_{\rm p}$, are
discontinuous) and contact discontinuities (where there is a density
and/or temperature jump but $v_{\rm p}$ is unchanged). Contact
discontinuities, and interfaces between different fluids, create
special problems for multi-dimensional hydrodynamic codes. Unlike
shocks, which contain a self-steepening mechanism, contact
discontinuities spread diffusively during a calculation, and continue
to broaden as the calculation progresses \citep[see
  e.g. ][]{Robertson:2010}. Some schemes employ an algorithm known as
a contact discontinuity steepener to limit this diffusion
\citep[e.g.][]{Fryxell:2000}. However, their use remains
controversial, since the algorithm is based on empirical values with
no physical or mathematical basis, and requires some care, since under
certain circumstances it can produce incorrect results
(i.e. ``staircasing'', Blondin, private communication).

Purely numerical effects are most prevalent when there are large
density and temperature contrasts. Unfortunately, these frequently
occur in practice, as when radiative cooling is effective cold dense
regions of gas can form. Such regions are also inherently unstable,
and compressed interface layers may be fragmented resulting in cold
dense clumps/filaments residing next to hot tenous gas. When modelling
such phenomena, the numerical transfer of heat from hot to cold cells
can change the behaviour of the shocked gas. In particular, hot cells
on one side of the contact discontinuity can reduce the net cooling
rate of denser gas in adjacent cells on the other side of the contact
discontinuity, and vice-versa. A further concern comes when one
derives synthetic emission from the simulation output. For instance,
due to the $\rho^2$ dependence of {\it thermal} emission, artificial
heating caused by numerical conduction can cause dramatic differences
in the spectral hardness and the magnitude of the integrated
luminosity.

The goal of this work is to provide both a qualitative and
quantitative analysis of the effects of numerical conduction and
viscosity on the dynamics and observables from colliding flows as a
function of the density and temperature constrast between the
postshock gas. For the purposes of our investigation we have performed
hydrodynamic simulations of colliding flows in plane-parallel and
cylindrical geometries. In both scenarios the influence of efficient
radiative cooling and powerful instabilities cause cold dense
layers/clumps to reside next to hot rarefied gas. We show that the
calculated X-ray emission from the postshock gas is strongly dependent
on the parameters of the opposing flows.

The remainder of this paper is structured as follows: in
\S~\ref{sec:model} we give a description of the hydrodynamics code and
details of the X-ray emission calculations. In \S~\ref{sec:results} we
present model descriptions and results, in \S~\ref{sec:discussion} a
discussion, and we close with conclusions in \S~\ref{sec:conclusions}.

\section{Numerical method}
\label{sec:model}

\subsection{The hydrodynamics code}

The collision of hypersonic flows is modelled by numerically solving
the time-dependent equations of Eulerian hydrodynamics in a 2D
cartesian coordinate system. The relevant equations for mass,
momentum, and energy conservation are:
\begin{eqnarray}
\frac{\partial\rho}{\partial t} + \nabla \cdot \rho {\bf u} &  =  & 0, \\
\frac{\partial\rho{\bf u}}{\partial t} + \nabla\cdot\rho{\bf uu} + \nabla P & = & 0,\\
\frac{\partial\rho E}{\partial t} + \nabla\cdot[(\rho E + P){\bf u}] & =& \left(\frac{\rho}{m_{\rm H}}\right)^{2}\Lambda(T).
\end{eqnarray}

\noindent Here $E = e + \frac{1}{2}|{\bf u}|^{2}$, is the total gas
energy, $e$ is the internal energy, ${\bf u}$ is the gas velocity,
$\rho$ is the mass density, $P$ is the pressure, $T$ is the
temperature, and $m_{\rm H}$ is the mass of hydrogen. We use the ideal
gas equation of state, $P = (\gamma - 1)/\rho e$, where the adiabatic
index $\gamma = 5/3$, to close the equations.

The radiative cooling is calculated from the \textsc{MEKAL} thermal
plasma code \citep{Kaastra:1992,Mewe:1995} distributed in
\textsc{XSPEC} (v11.2.0). The temperature of the pre-shock flows is
assumed to be maintained at $\approx 10^{4}\;$K through
photoionization heating. Gas in the shocked region which rapidly cools
is prevented from cooling below this temperature.

The simulations presented in this work were performed using the
\textsc{FLASH} version 3.1.1 hydrodynamics code \citep{Fryxell:2000}
which uses the piecewise-parabolic method \cite{Colella:1984} to solve
the hydrodynamic equations. This code has been designed to operate
with a block-structured AMR grid \citep[e.g.][]{Berger:1989} using the
\textsc{PARAMESH} package \citep{MacNeice:2000} under the
message-passing interface (MPI) architecture. In the models presented
in this work the AMR uses square blocks consisting of $8^2$
cells. Details relating to the adopted resolution and size of the
simulation domain for the colliding (plane-parallel) laminar flow and
cylindrically diverging colliding wind binary models are given in
\S\S~\ref{subsec:slab_model} and \ref{subsec:cwb_model},
respectively. Contact discontinuity steepening was used in all
calculations, with the standard parameters\footnote{Further tests in
  which $\eta_1, \eta_2$, and $\epsilon$ were varied revealed that
  within the limit that the dynamics of the simulation remained
  reasonably unaffected, the numerical conduction effects that we
  discuss were not significantly reduced.} \citep[$\eta_1=20,
  \eta_2=0.05, \epsilon=0.01$, ][]{Fryxell:2000}. A customized unit
has been implemented into the {\sc FLASH} code for optically thin
radiative cooling using the explicit method described in
\cite{Strickland:1995}. An advected scalar variable is included in the
hydrodynamics calculations to distinguish between the flows.

\subsection{X-ray emission}
\label{subsec:xray_emission}

To calculate the intrinsic X-ray emission from the simulation we
assume solar abundances and use emissivities for optically thin gas in
collisional ionization equilibrium obtained from look-up tables
calculated from the \textsc{MEKAL} plasma code containing 200
logarithmically spaced energy bins in the range 0.1-10 keV, and 101
logarithmically spaced temperature bins in the range
$10^{4}-10^{9}\;$K. The advected scalar is used to separate the X-ray
emission contributions from each flow\footnote{Further tests in which
  a cut is placed on the fluid dye variable when calculating the X-ray
  emission from each fluid component did not prove successful in
  significantly reducing the contamination of the results by numerical
  conduction. The main trend from using a cut was that the softer of
  the two emission components becomes harder and vice-versa.}.

\section{Results}
\label{sec:results}

We have performed two sets of simulations to examine the effect of
numerical conduction on the gas dynamics and the derived X-ray
characteristics of colliding flows. The first scenario is that of
colliding plane-parallel flow and the second is of the wind-wind
collision in a massive stellar binary system.

In many of the models presented in this work the postshock gas has a
temperature which places it near a local minimum in the cooling
function, $\Lambda(T)$, at a value of $\approx 10^{-23}\;{\rm
  erg~cm^{3}~s^{-1}}$. This fact can be utilised to determine
approximate values for the cooling time, $t_{\rm cool}$, and the
cooling length, $l_{\rm cool}$, of postshock gas:

\begin{equation}
  t_{\rm cool}  = \frac{3 k T}{8 n \Lambda (T)}  \approx 10^{-10} \frac{v_{8}^2}{\rho}\;{\rm s}, \label{eqn:tcool_slab}
\end{equation}
\begin{equation}
  l_{\rm cool}  \simeq v \cdot t_{\rm cool}  \approx 10^{-2} \frac{v_{8}^3}{\rho}\;{\rm cm}, \label{eqn:lcool_slab}
\end{equation}

\noindent where $T$ is the postshock gas temperature in K, $v_8$ is
the gas velocity in units of $10^{8}\;{\rm cm~s^{-1}}$, $n$ is the gas
number density in ${\rm cm}^{-3}$, and $\rho$ the gas density in
g~cm$^{-3}$. Note that $v$, $n$, and $\rho$ are the values immediately
preshock.

In models of CWBs the geometry of the colliding flows allows one to
define an escape time for postshock flow leaving the system, $t_{\rm
  esc}$. The strength of cooling in these systems can then be
characterised by a dimensionless cooling parameter which is the ratio
of the cooling time to the postshock flow time \citep{Stevens:1992},

\begin{equation}
  \chi = \frac{t_{\rm cool}}{t_{\rm esc}} = \frac{v_{8}^{4}d_{12}}{\dot{M}_{-7}}, \label{eqn:chi}
\end{equation}

\noindent where $v_8$ is the gas velocity in units of $10^{8}\;{\rm
  cm~s^{-1}}$, $d_{12}$ is the binary separation in units of
$10^{12}\;$cm, and $\dot{M}_{-7}$ is the stellar wind mass-loss rate
in units of $10^{-7}\Msolpyr$. Values of $\chi \gg 1$ are
representative of adiabatic gas, whereas $\chi \ltsimm 1$ indicates
that the postshock gas is radiative and will cool rapidly. In the
following we use the subscripts 1 and 2 to indicate the cooling
parameter for the postshock wind~1 and 2 material, respectively.

Using the Rankine-Hugoniot shock jump conditions we can estimate the
postshock gas temperature, $T$, and a corresponding energy for
(thermal) X-ray emission from the postshock gas as,

\begin{equation}
  kT \simeq 1.17 v_{8}^{2}\;{\rm keV}.
\label{eqn:energy}
\end{equation}

\begin{figure*}
  \begin{center}
    \begin{tabular}{ccc}
\resizebox{45mm}{!}{\includegraphics{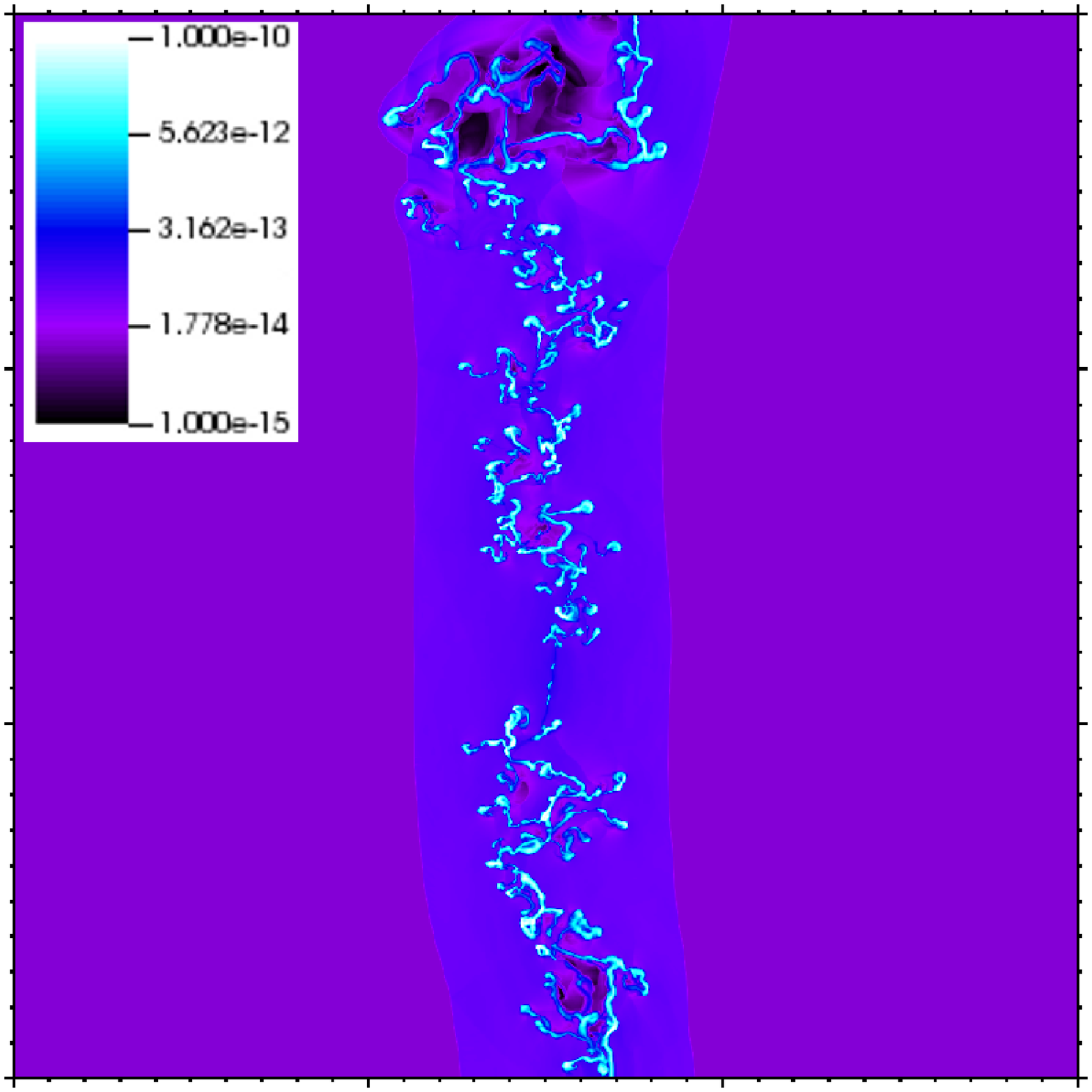}} &
\resizebox{45mm}{!}{\includegraphics{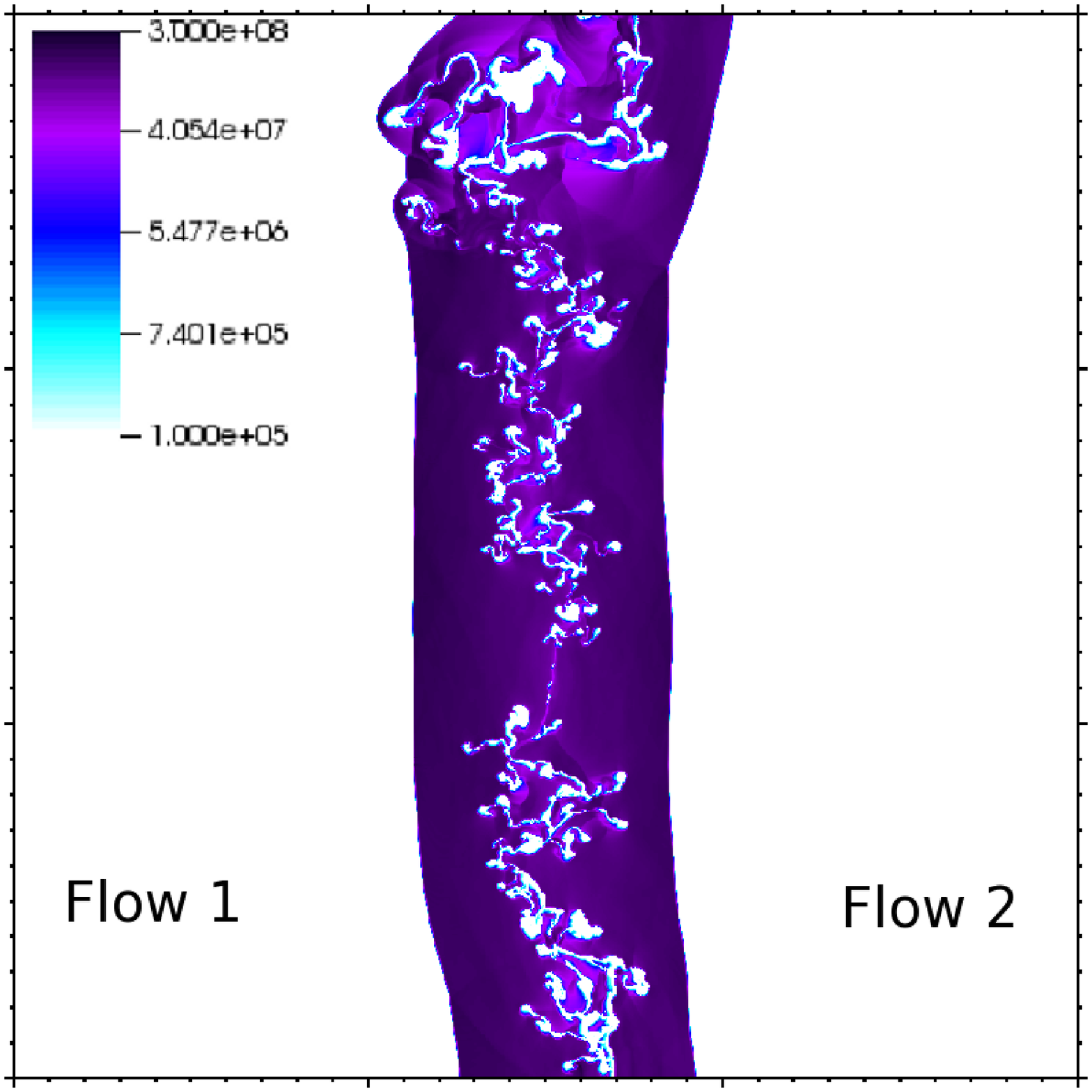}} &
\resizebox{45mm}{!}{\includegraphics{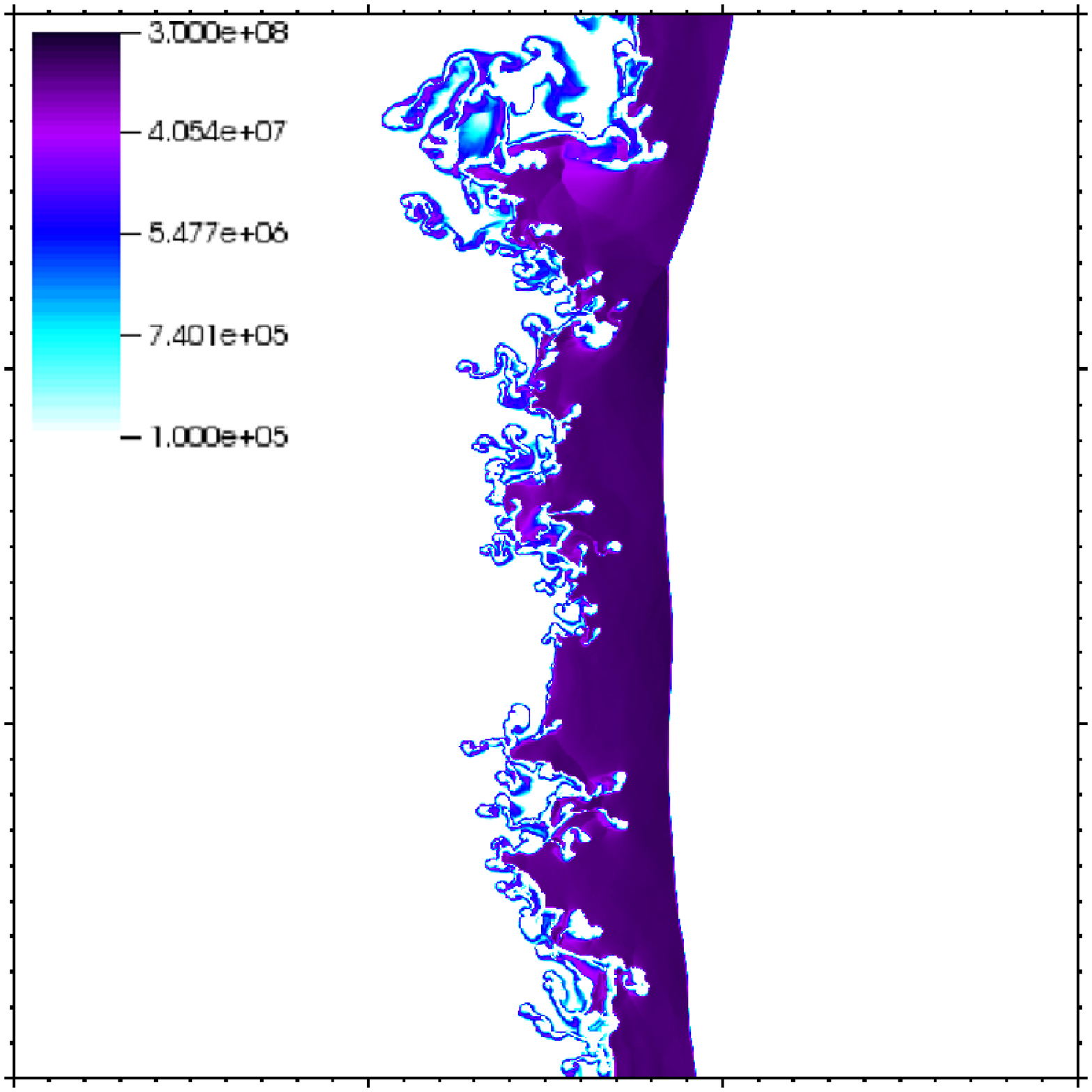}} \\

\resizebox{45mm}{!}{\includegraphics{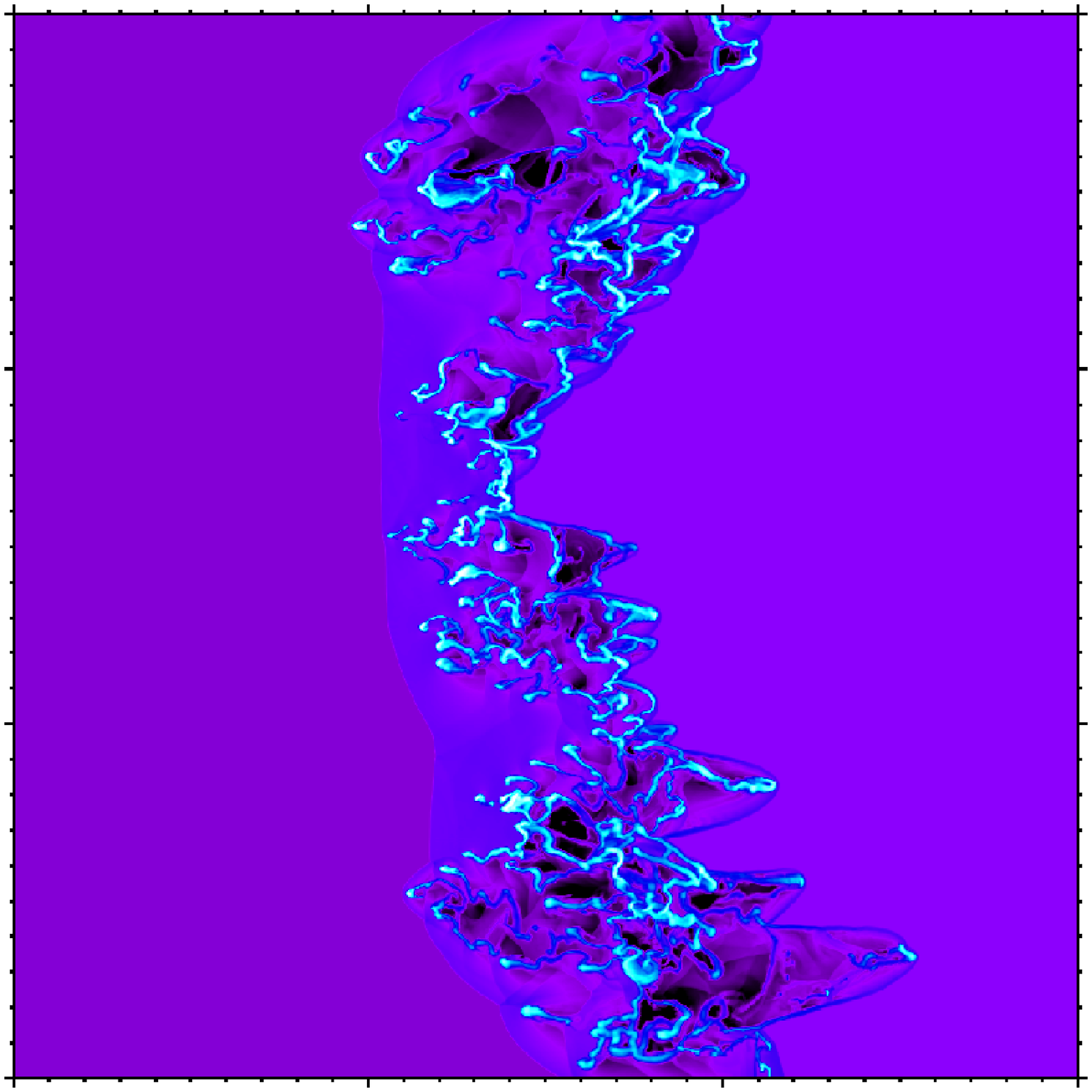}} &
\resizebox{45mm}{!}{\includegraphics{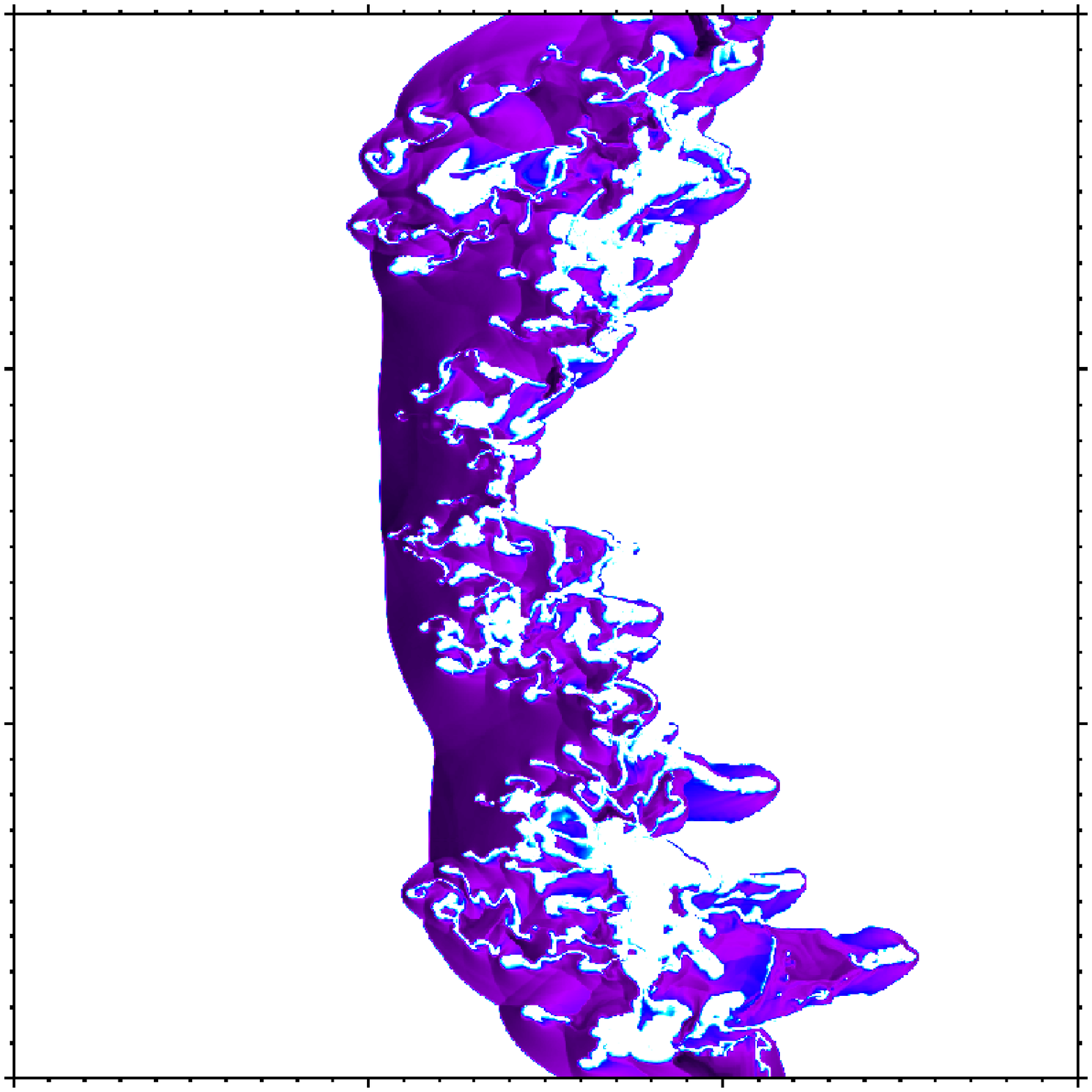}} &
\resizebox{45mm}{!}{\includegraphics{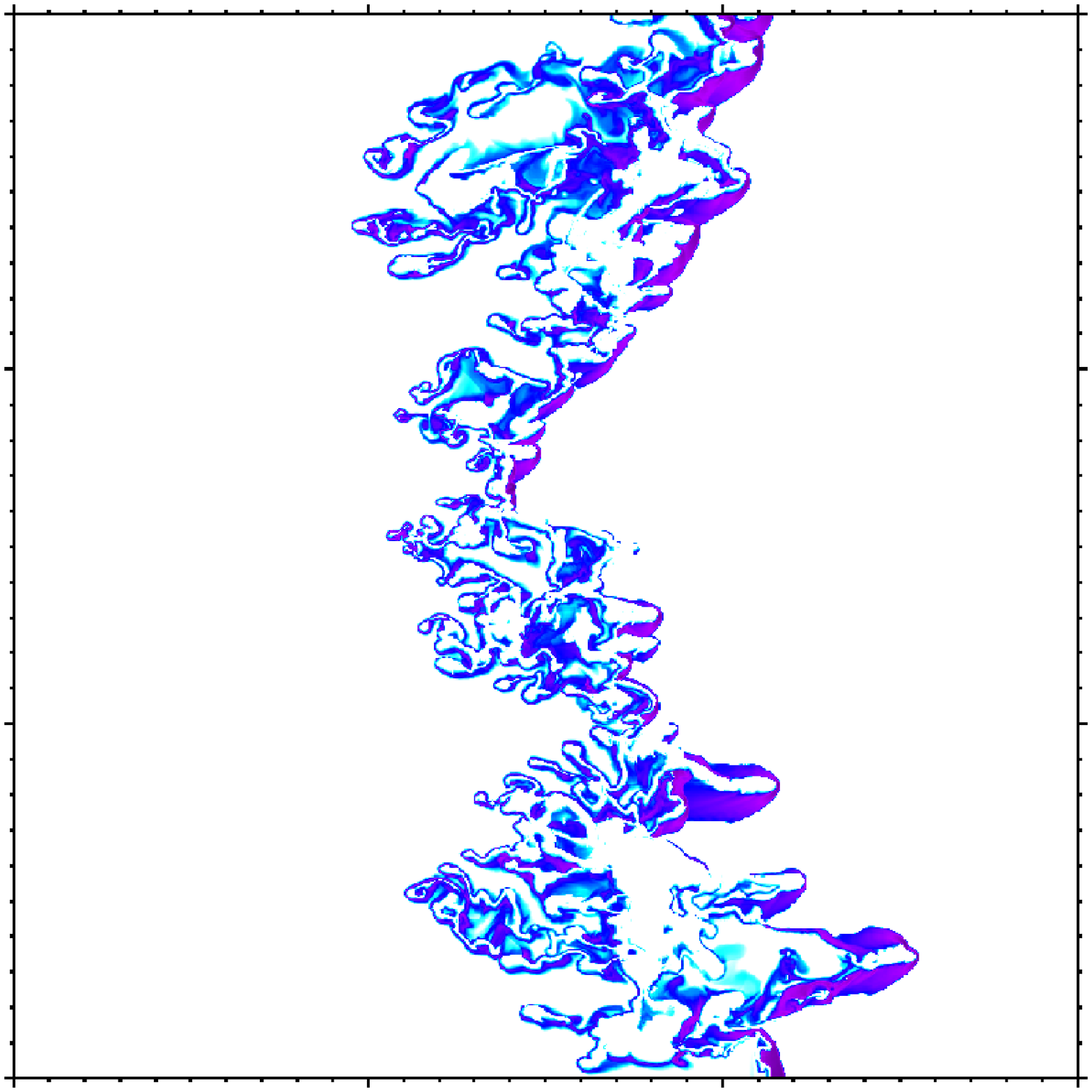}} \\

\resizebox{45mm}{!}{\includegraphics{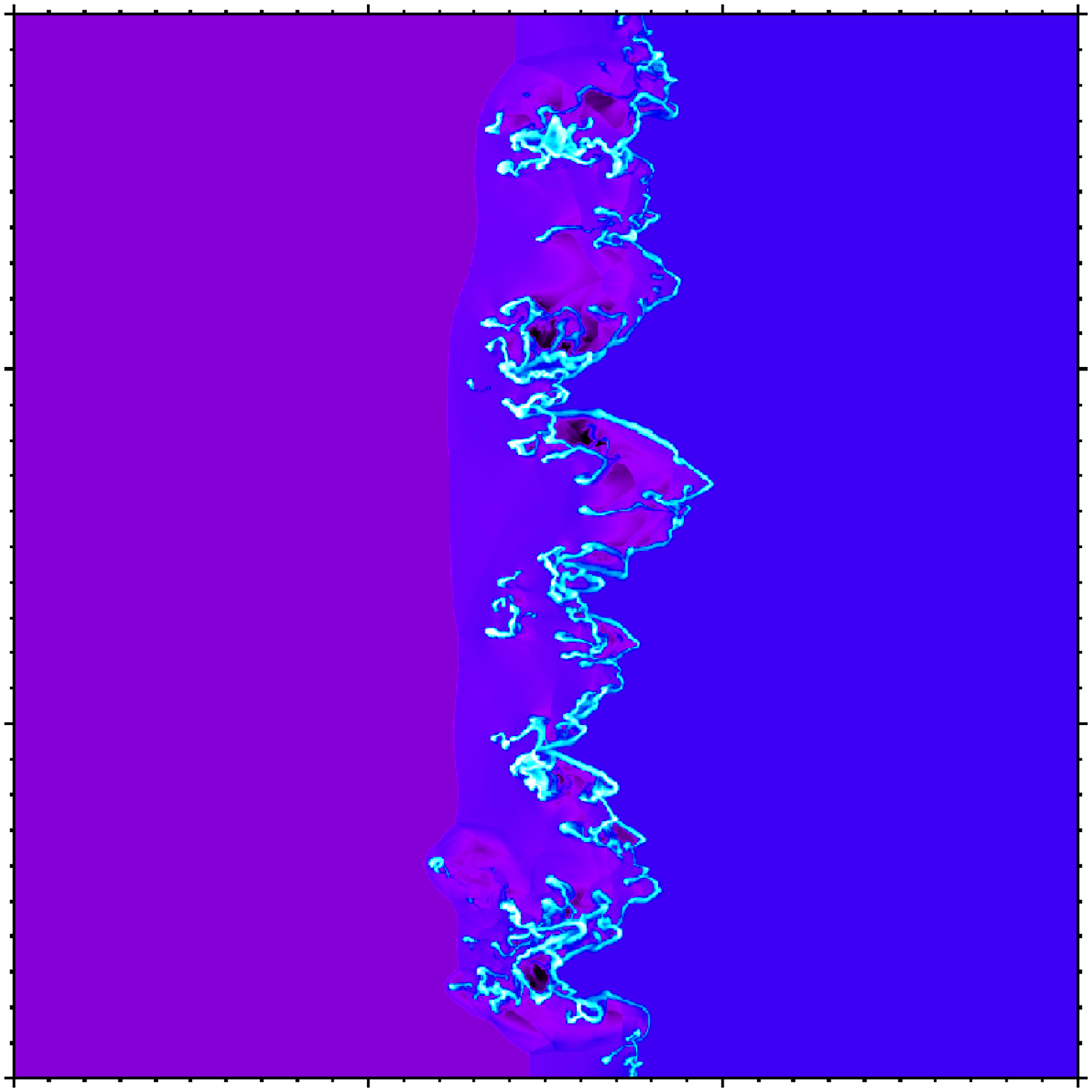}} &
\resizebox{45mm}{!}{\includegraphics{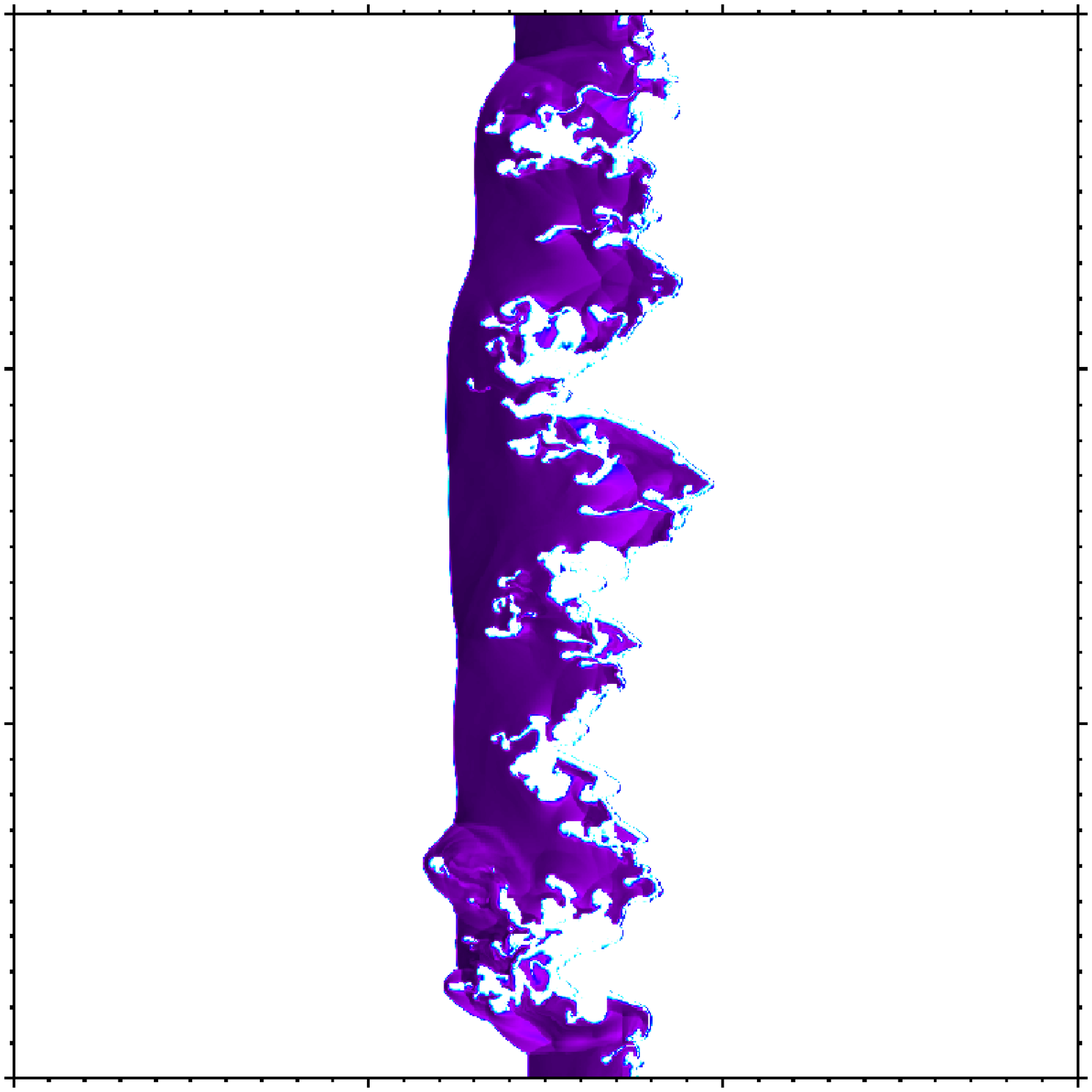}} &
\resizebox{45mm}{!}{\includegraphics{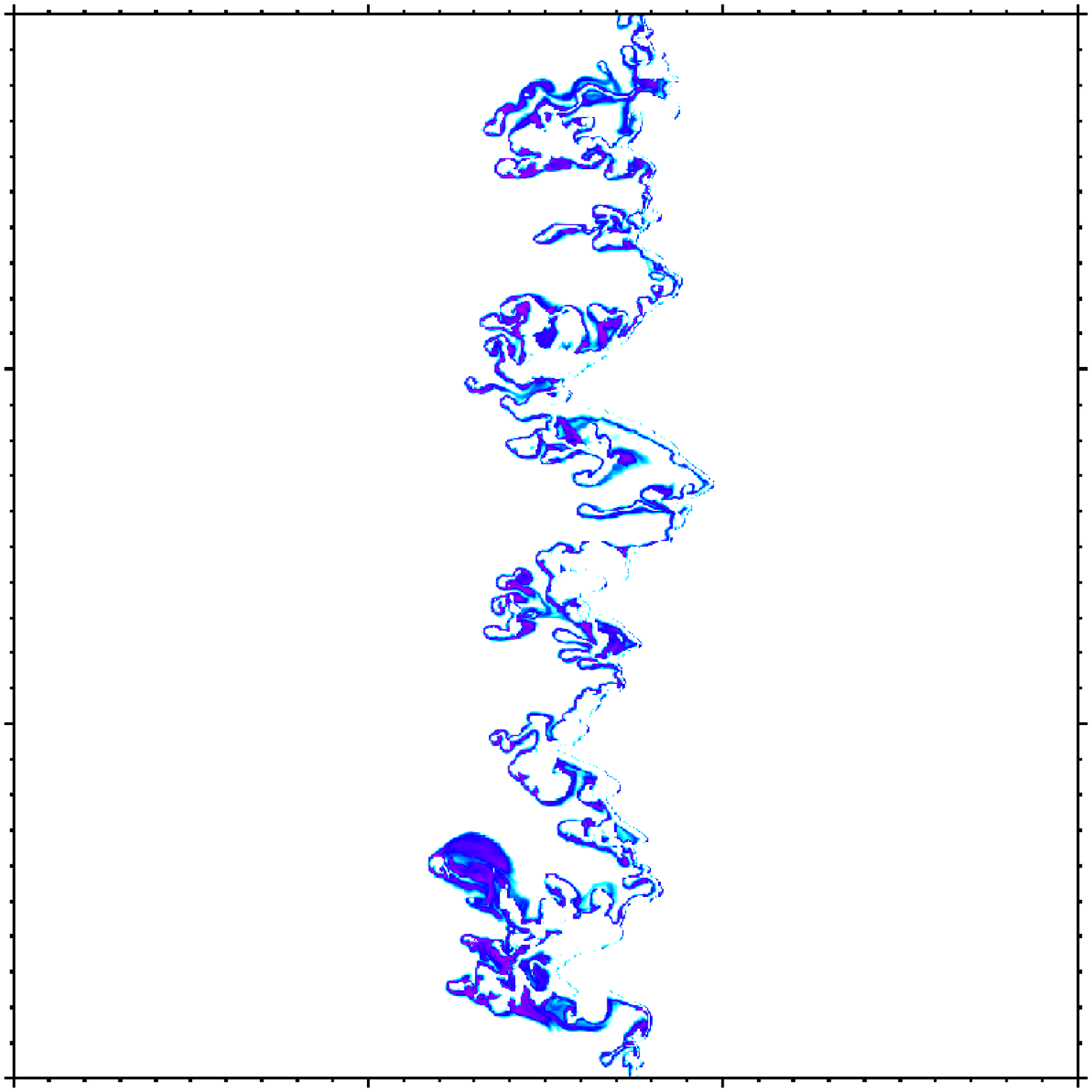}} \\

\resizebox{45mm}{!}{\includegraphics{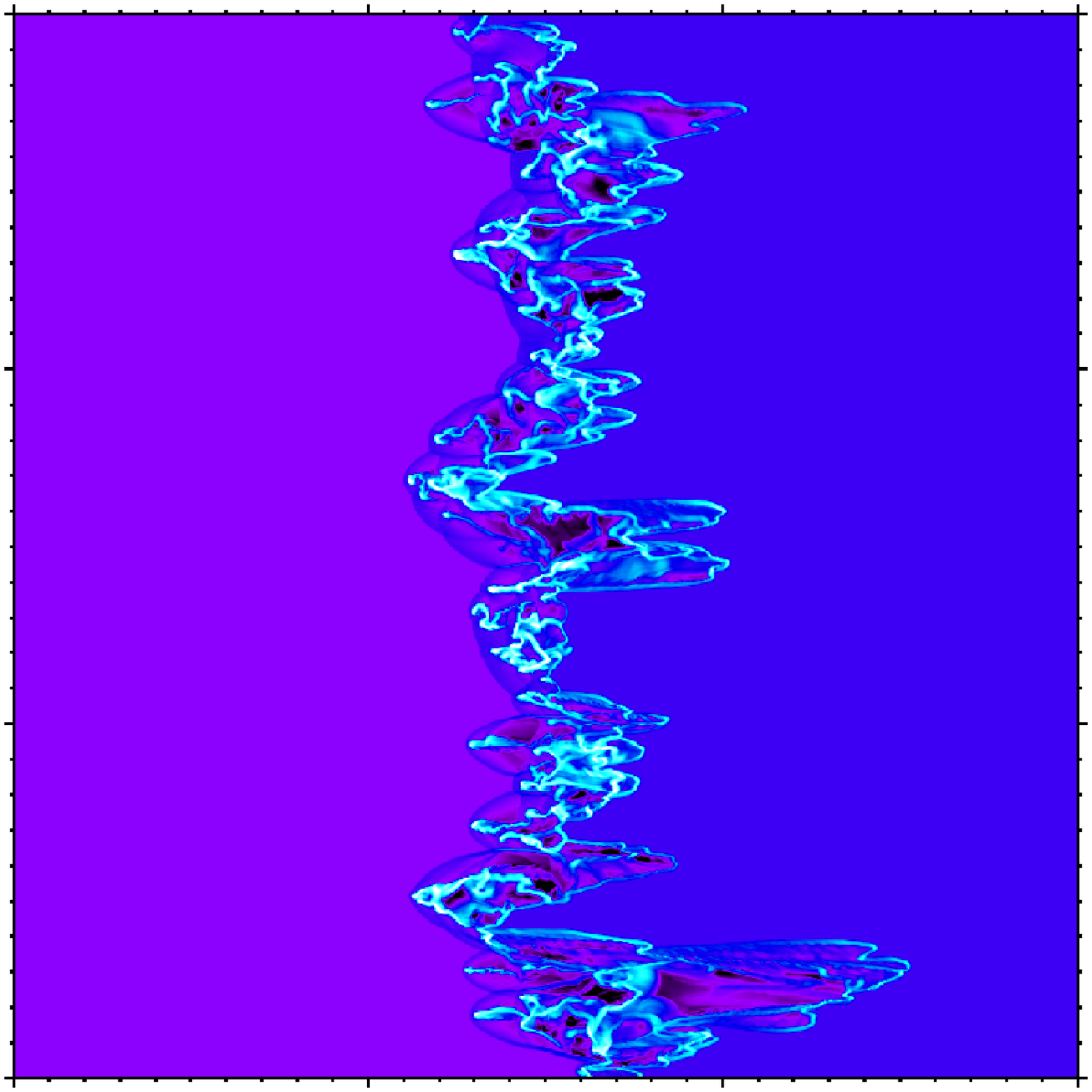}} &
\resizebox{45mm}{!}{\includegraphics{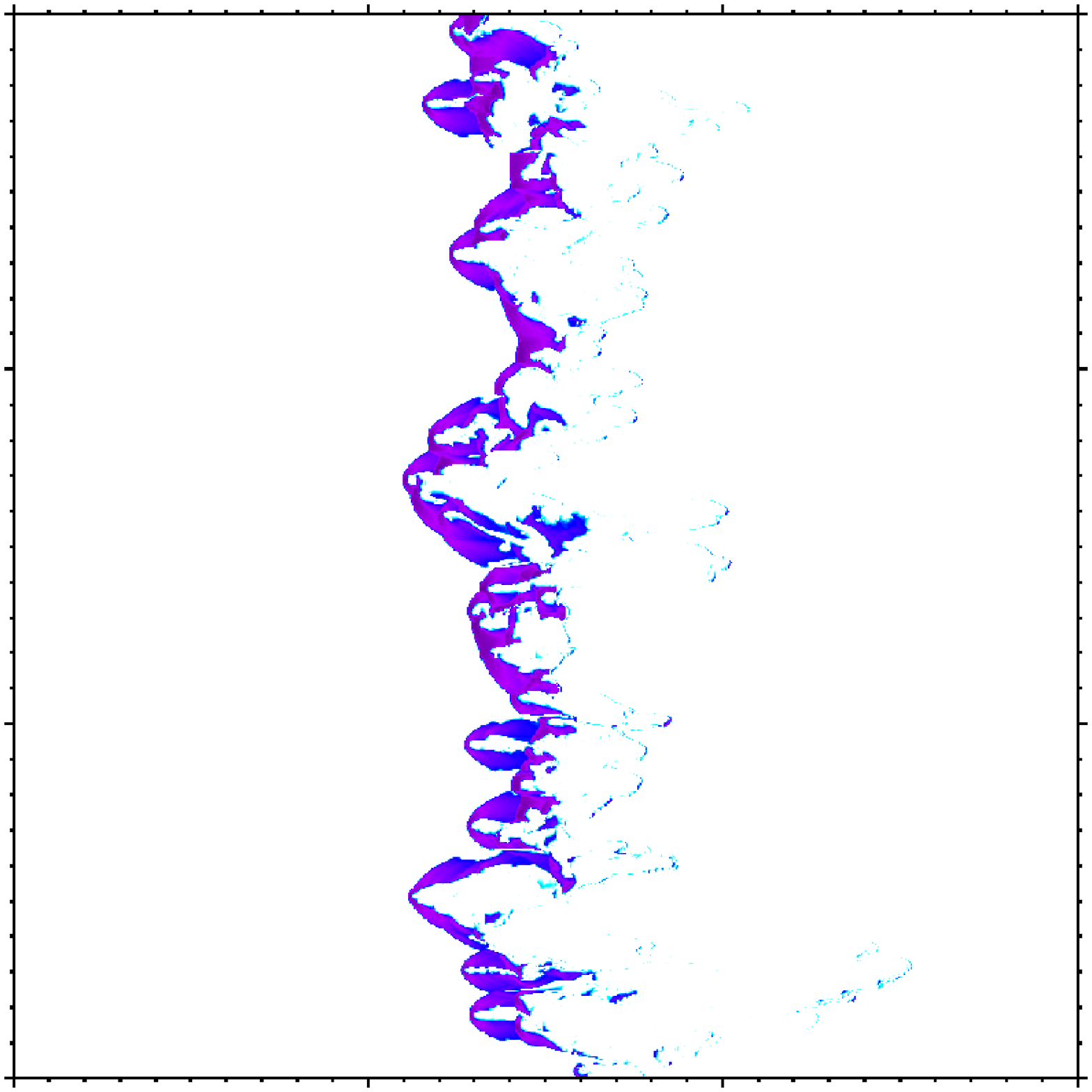}} &
\resizebox{45mm}{!}{\includegraphics{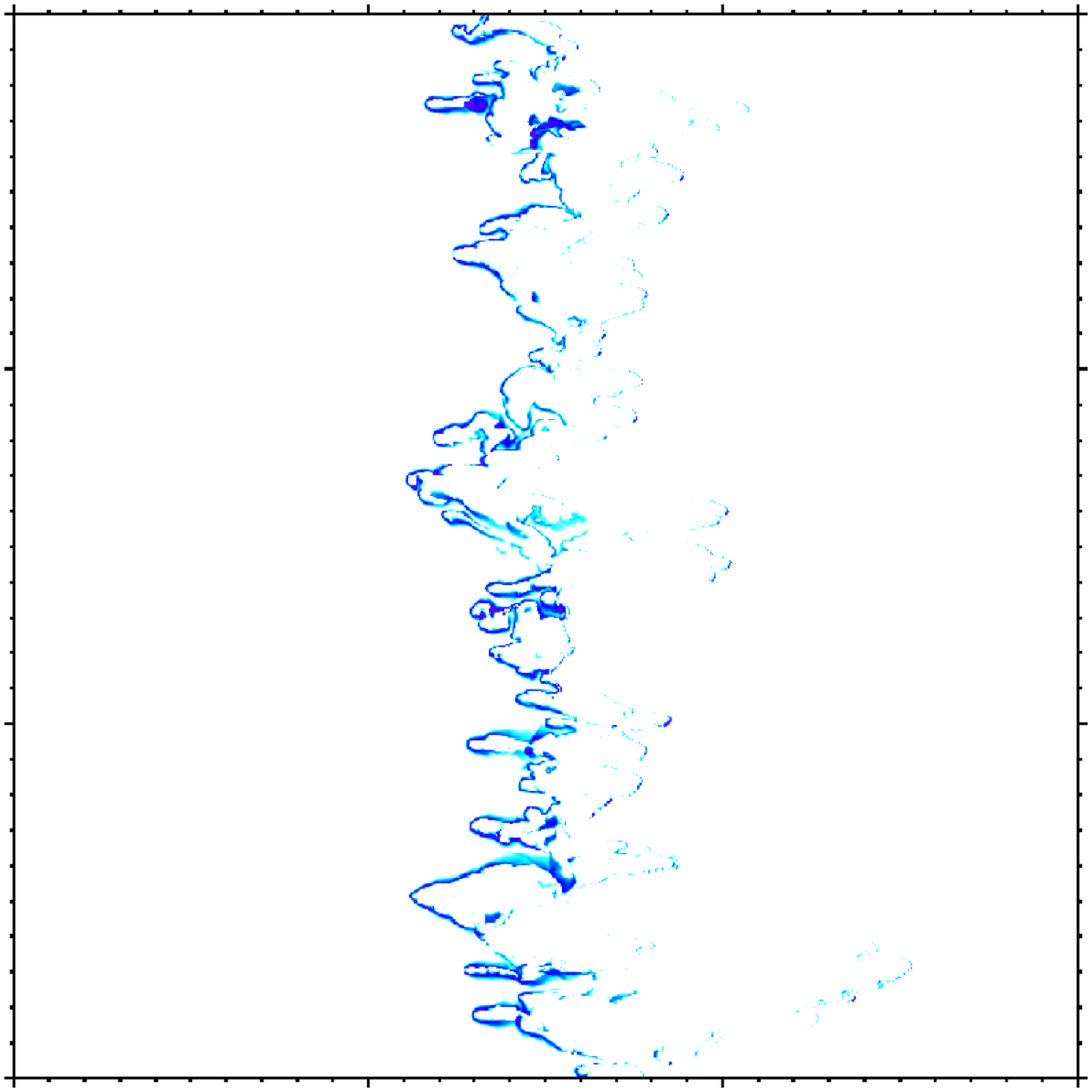}} \\

\resizebox{45mm}{!}{\includegraphics{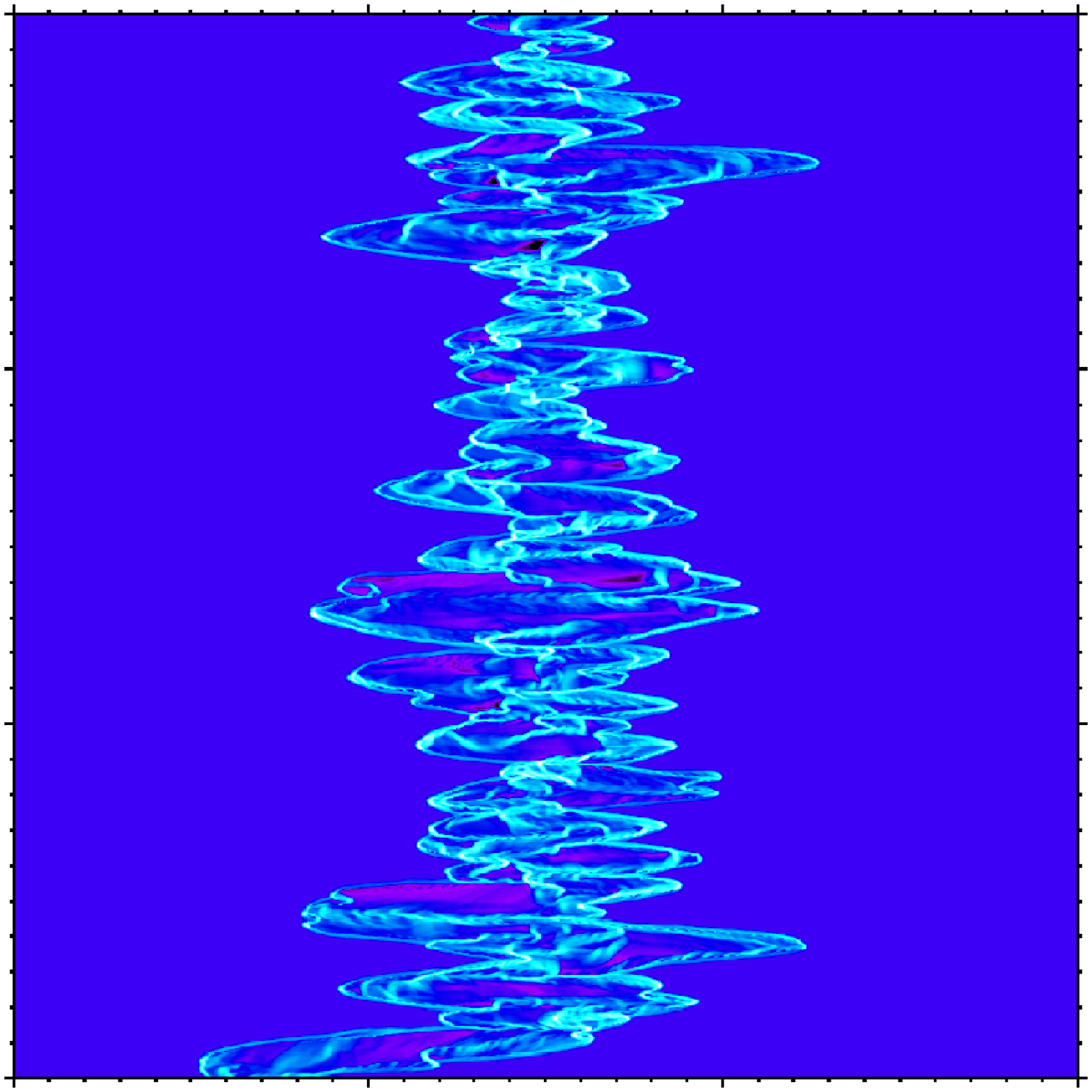}} &
\resizebox{45mm}{!}{\includegraphics{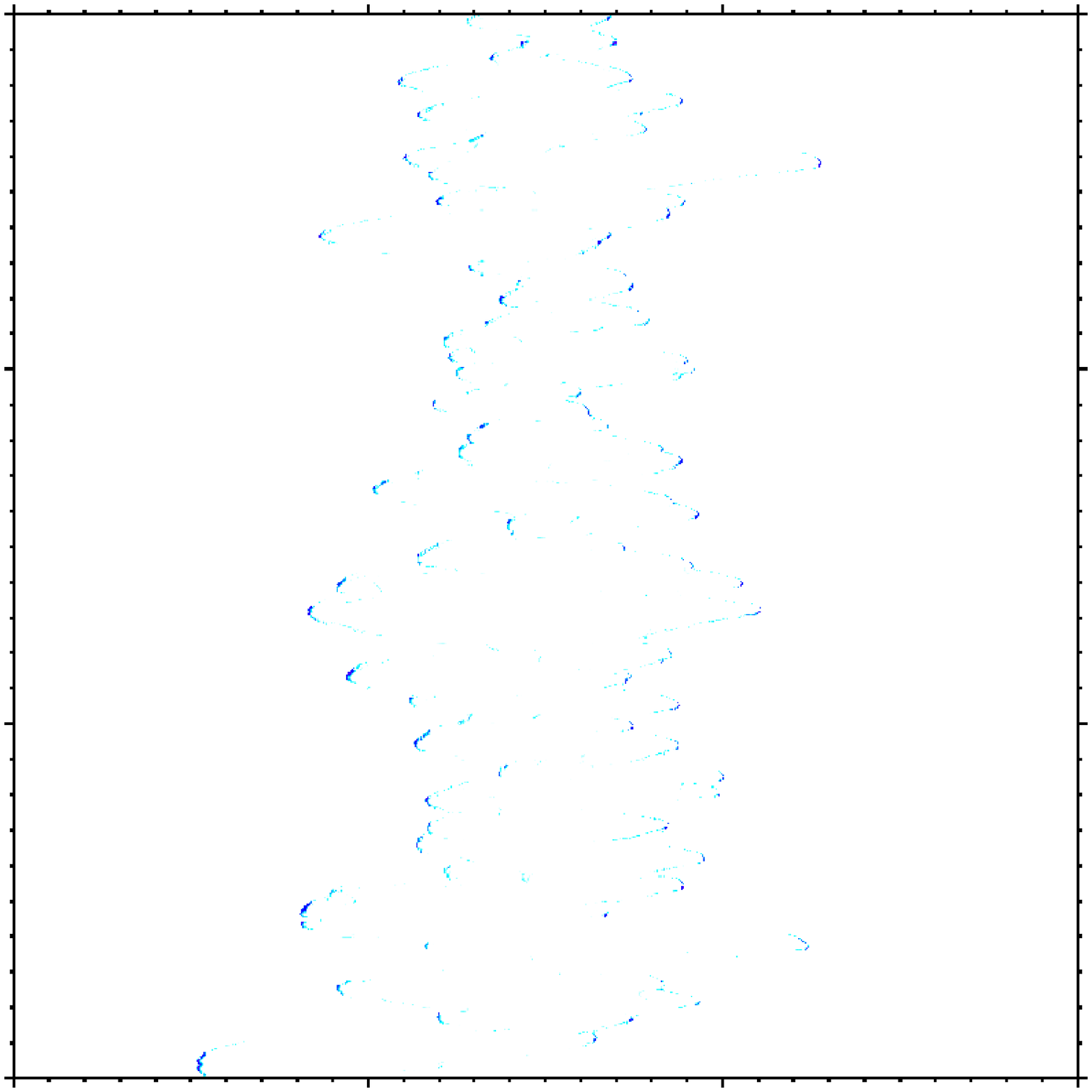}} &
\resizebox{45mm}{!}{\includegraphics{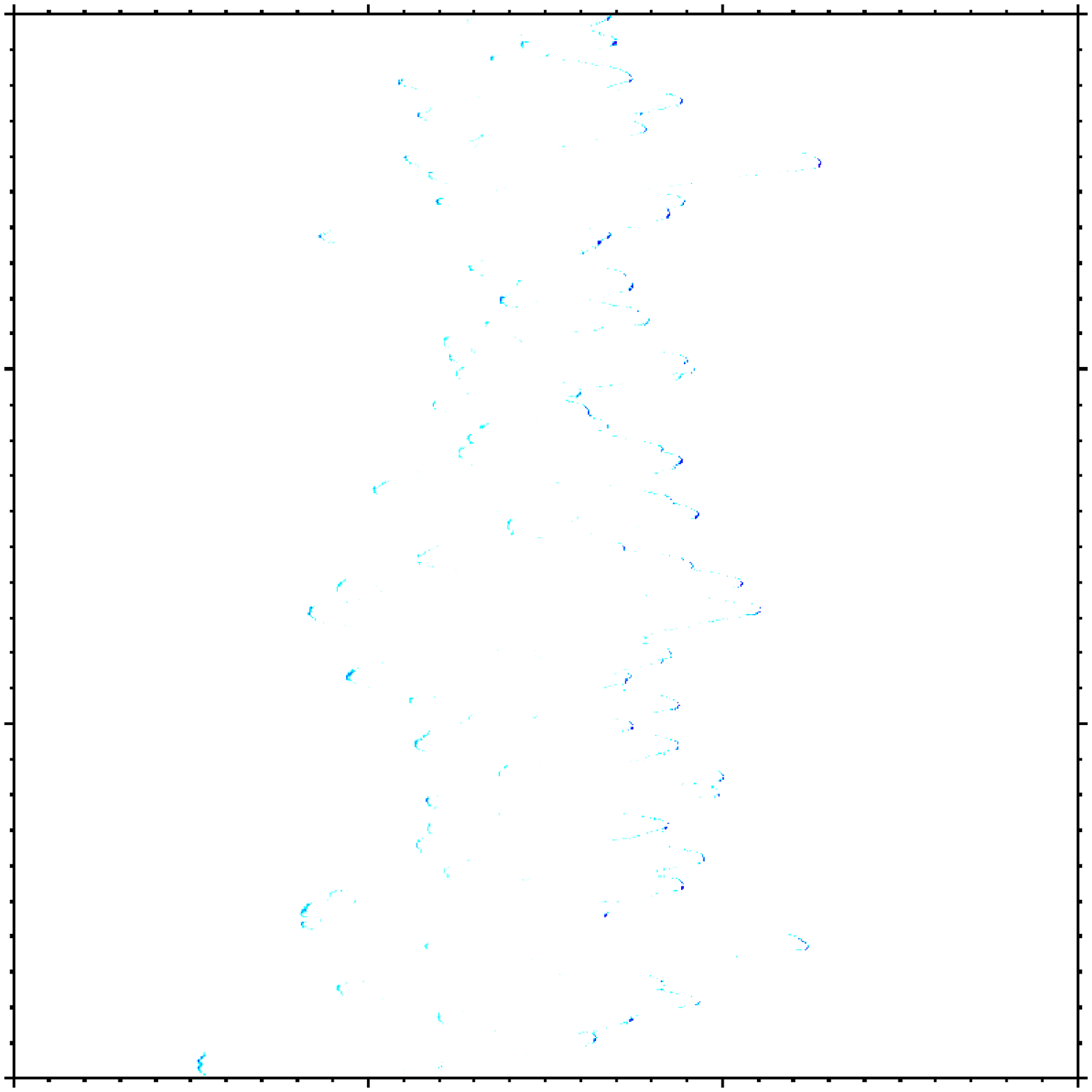}} \\

    \end{tabular}
    \caption{Simulation snapshots showing density (left column),
      temperature (middle column), and flow 2 temperature (right
      column). Models shown from top to bottom: SLAB-A, SLAB-B,
      SLAB-C, SLAB-D, and SLAB-E. Model parameters are listed in
      Table~\ref{tab:slab_model_parameters}. Large tick marks
      correspond to a distance of $1\times10^{13}\;$cm.}
    \label{fig:slab_model_images}
  \end{center}
\end{figure*}
\subsection{Colliding plane-parallel flows}
\label{subsec:slab_model}

We introduce the effects of numerical conduction with a model of
colliding plane-parallel flows, relevant to a wide variety of
astrophysical phenomena. Each flow is of constant density and
velocity, and they are separated by a vertical discontinuity through
the centre of the grid. The simulation domain extends to $x_{\rm
  max}=y_{\rm max}=3\times10^{13}\;$cm and is covered with a coarse
grid consisting of $x \times y = 15 \times 15\;$blocks, with each
block consisting of $8\times8$ cells. We allow for 3 nested levels of
refinement which gives a minimum cell size of $3.125\times10^{10}\;$cm
and an effective resolution of $960\times960\;$cells. The left and
right hand boundaries are inflow conditions for flow 1 and 2,
respectively, while the upper and lower boundaries employ zero
gradients. The flows are hypersonic, so their ram pressure dominates
over their thermal pressure. The simulations are allowed to run for
$10^{7}\;$s which is sufficiently long to allow the postshock gas to
cool and for the initial conditions to flow off the grid. We note that
previous models of colliding plane-parallel flows typically focus on
the {\it supersonic} regime \citep[e.g. ][]{Walder:1998, Walder:2000,
  Folini:2006, Heitsch:2006, Heitsch:2008, Niklaus:2009,
  Banerjee:2009} which is more relevant to star formation. The region
of parameter space examined here explores the {\it hypersonic} regime,
a situation which is more relevant to systems such as CWBs, SNe,
cluster/galactic winds, and jets. The simulation results highlight the
turbulent nature of hypersonic flow collisions, which deserve a
detailed investigation. However, for the purposes of this work we
provide a brief description of the dynamics and defer a more indepth
analysis to a later paper.

\begin{table*}
\begin{center}
\caption[]{Parameters pertaining to the colliding laminar flow
  simulations. The cooling time, $t_{\rm cool}$, and the cooling
  length, $l_{\rm cool}$, are calculated from
  Eqs.~\ref{eqn:tcool_slab} and \ref{eqn:lcool_slab}, respectively.}
\begin{tabular}{lllllllll}
\hline
Model  & $\rho_{1}$ & $v_{1}$ & $t_{\rm cool 1}$ & $l_{\rm cool 1}$ & $\rho_{2}$ & $v_{2}$ & $t_{\rm cool 2}$ & $l_{\rm cool 2}$ \\
 & (g~cm$^{-3}$) & $(10^{8}\;{\rm cm~s^{-1}})$ & (s) & (cm) & (g~cm$^{-3}$) & $(10^{8}\;{\rm cm~s^{-1}})$ & (s) & (cm) \\
\hline
SLAB-A & $1.1\times10^{-14}$  & 3 & $8.18\times10^{4}$ & $2.45\times10^{13}$ & $1.1\times10^{-14}$  & 3 & $8.18\times10^{4}$ & $2.45\times10^{13}$ \\
SLAB-B & $1.1\times10^{-14}$  & 3 & $8.18\times10^{4}$ & $2.45\times10^{13}$ & $2.5\times10^{-14}$  & 2 & $1.60\times10^{4}$ & $3.20\times10^{12}$ \\
SLAB-C & $1.1\times10^{-14}$  & 3 & $8.18\times10^{4}$ & $2.45\times10^{13}$ & $1.0\times10^{-13}$  & 1 & $1.00\times10^{3}$ & $1.00\times10^{11}$ \\
SLAB-D & $2.5\times10^{-14}$  & 2 & $1.60\times10^{4}$ & $3.20\times10^{12}$ & $1.0\times10^{-13}$  & 1 & $1.00\times10^{3}$ & $1.00\times10^{11}$ \\
SLAB-E & $1.0\times10^{-13}$  & 1 & $1.00\times10^{3}$ & $1.00\times10^{11}$ & $1.0\times10^{-13}$  & 1 & $1.00\times10^{3}$ & $1.00\times10^{11}$ \\
\hline
\label{tab:slab_model_parameters}
\end{tabular}
\end{center}
\end{table*}

We have run a series of models where the densities and velocities of
the flows have been varied while maintaining equal ram-pressure for
the flows (Table~\ref{tab:slab_model_parameters}) and keeping the
resolution constant. Through models SLAB-A to SLAB-C we decrease
(increase) the preshock gas velocity (density) of flow 2. Following
this, in models SLAB-C to SLAB-E the preshock parameters of flow 2 are
kept fixed while the velocity (density) of flow 1 is decreased
(increased). As such, in model SLAB-A the postshock gas of both flows
is adiabatic, whereas in model SLAB-E the postshock gas is strongly
radiative. The preshock temperature of the flows is $10^{4}\;$K,
giving Mach numbers of 67, 134, and 300 for preshock velocities of
1000, 2000, and 3000 km s$^{-1}$, respectively. The postshock gas is
allowed to cool back to $10^{4}\;$K. In the following, we discuss the
effect of these variations on the gas dynamics, and then present the
results of X-ray calculations which provide a quantitative analysis of
the effect of numerical conduction on observable quantities.

Model SLAB-A consists of two hypersonic flows with identical preshock
parameters ($v_1=v_2=3\times10^{8}\;{\rm cm~s^{-1}}$, $\rho_1 = \rho_2
= 1.1\times10^{-14}\;{\rm g~cm^{-3}}$). As the flows collide shocks
are generated which heat gas to $T\gtsimm10^{8}\;$K. This gas slowly
cools so that at later times a thin dense shell of cold ($T\simeq
10^{4}\;$K) gas is formed near the contact discontinuity separating
the flows.  This is Rayleigh-Taylor (RT) unstable, and fingers of
material from the dense shell begin to protrude into the hot tenous
gas (Fig.~\ref{fig:slab_model_images}, top panels). At early times in
the simulation the shock front of both flows oscillates dramatically
\citep[this is the radiative over-stability, e.g.][]{Chevalier:1982,
  Imamura:1984, Walder:1996, Pittard:2005, Mignone:2005}. By the time
of the snapshot shown in Fig.~\ref{fig:slab_model_images} these
oscillations have been damped by waves which reflect back from the
central cold dense layer out of phase with the original shock wave,
disrupting the coherence between waves in the oscillating shock, and
causing it to deteriorate into an approximate steady state
\citep{Strickland:1995}. This effect is enhanced by the distortion of
the central cold dense gas layer \citep{Walder:1996}.

In models SLAB-B and SLAB-C the velocity (density) of flow 2 is
decreased (increased) while the ram pressure of the preshock flow
remains the same. As in model SLAB-A the postshock gas of both flows
demonstrates over-stability. As the preshock velocity (density) of
flow~2 decreases (increases) the frequency of oscillations and the
width of the layer of postshock gas increase and decrease
respectively, as $t_{\rm cool}$ and $l_{\rm cool}$ both decrease. The
thin dense shell of gas which separates the shocks becomes noticeably
more structured than in model SLAB-A. At the end of the simulations
there is a complicated spatial distribution of gas in which cold and
dense material, lower density shock heated gas, and hot, rarefied gas
produced by the vigorous action of instabilities all reside adjacent
to one another (Fig.~\ref{fig:slab_model_images}). The hottest
postshock gas of flow~2 occurs just behind the shock, as is also the
case in model SLAB-A.

In model SLAB-C, the postshock gas of flow~2 now cools rapidly to form
a cold dense shell, and unlike models SLAB-A and SLAB-B the width of
the region of postshock gas is not resolved. Of particular note is the
fact that the maximum temperature attained by flow 2 gas in model
SLAB-C is $\simeq 6.7\times 10^{7}\;$K, which is substantially higher
than the expected value of $T\simeq 1.4\times10^{7}\;$K (from the
preshock velocity). Examining the location of flow~2 gas with
$T>10^5\;$K (Fig.~\ref{fig:slab_model_images}) we see that the hottest
postshock gas now occurs at the contact discontinuity between the
postshock gas of flows 1 and 2 in contrast to what is seen in models
SLAB-A and SLAB-B. This is clearly the effect of numerical conduction
rearing its ugly head.

\begin{figure}
  \begin{center}
    \begin{tabular}{c}
      \resizebox{80mm}{!}{\includegraphics{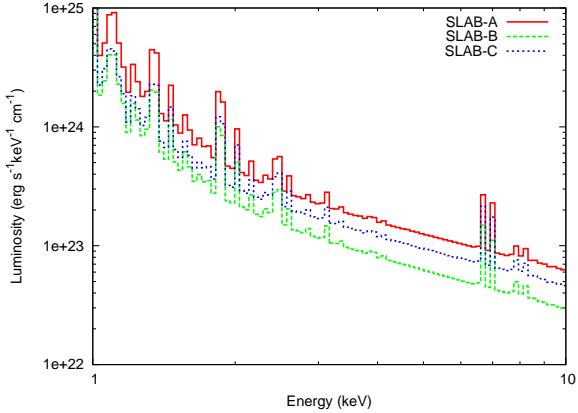}} \\
    \end{tabular}
    \caption{X-ray spectra calculated from flow 1 in models SLAB-A,
      SLAB-B, and SLAB-C. Model parameters are noted in
      Table~\ref{tab:slab_model_parameters}. Only flow 2 parameters
      are varied between the models.}
    \label{fig:slab_spec}
  \end{center}
\end{figure}

\begin{figure}
  \begin{center}
    \begin{tabular}{c}
      \resizebox{80mm}{!}{\includegraphics{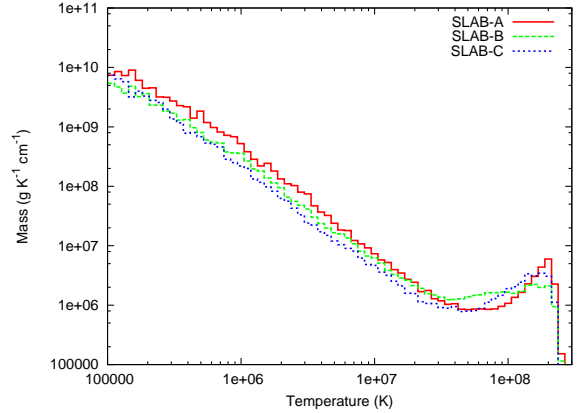}} \\
    \end{tabular}
    \caption{Distribution of flow~1 mass as a function of postshock
      gas temperature in models SLAB-A, SLAB-B, and SLAB-C. The
      differences between the distributions at $T\ltsimm
      4\times10^{7}\;$K are due to differing degrees of numerical
      conduction, whereas at $T\gtsimm 4\times10^{7}\;$K they are
      related to the orientation of the shock with respect to the
      upstream flow.}
    \label{fig:slab_mvst}
  \end{center}
\end{figure}

In models SLAB-A to SLAB-C we have examined the effect of
progressively increasing the density and temperature constrast between
the postshock gas whilst keeping the parameters of flow 1 fixed so as
to have hot, and largely adiabatic, postshock gas on at least one side
of the contact discontinuity. In models SLAB-C to SLAB-E we now look
at the opposite situation in which we keep highly radiative gas on at
least one side of the contact discontinuity while reducing the
preshock velocity of the flow on the other side of the contact
discontinuity until its postshock gas also becomes highly
radiative. Through an intermediary model (SLAB-D) we arrive at model
SLAB-E, in which two equal flows with $v_1=v_2=1\times10^{8}\;{\rm
  cm~s^{-1}}$, $\rho_1 = \rho_2 = 1\times10^{-13}\;{\rm g~cm^{-3}}$
produce highly radiative postshock gas. Through this sequence of
models the decrease (increase) in the preshock velocity (density) of
flow 1 causes a corresponding decrease in the cooling time and cooling
length of its postshock gas. Examining progressive trends we note that
through models SLAB-C to SLAB-E the reduced cooling length in the
postshock gas causes the width of the region of postshock gas of flow
1 to become narrower and pertubations to the cold dense layer increase
in ferocity (Fig.~\ref{fig:slab_model_images}). In models SLAB-C and
SLAB-D the distortion of the cold dense layer is due to a combination
of thin-shell \citep{Vishniac:1983}, RT, and Kelvin-Helmholtz (KH)
instabilities. In model SLAB-E the shocks which bound the cold dense
layer are isothermal with large amplitude oscillations resulting from
the non-linear thin shell instability \citep[NTSI,
][]{Vishniac:1994}. It is interesting to note that in model SLAB-E
almost all of the postshock gas cools rapidly and collapses into the
dense layer at $T\ltsimm10^{5}\;$K. However, there is a small amount
of gas at $T\simeq1.6\times10^{7}\;$K, higher than the
$1.36\times10^{7}\;$K expected from the flow velocity. We attribute
this difference to instability driven oscillations moving upstream
into the incoming flow, and which thus cause higher preshock
velocities in the frame of the shock.

\begin{figure}
  \begin{center}
    \begin{tabular}{c}
      \resizebox{80mm}{!}{\includegraphics{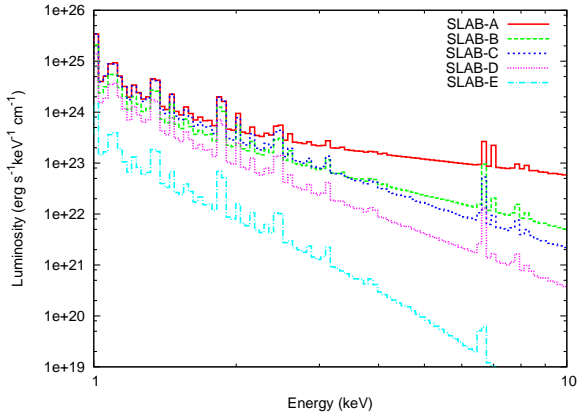}} \\
    \end{tabular}
    \caption{X-ray spectra calculated from flow 2 in models SLAB-C,
      SLAB-D, and SLAB-E. Model parameters are noted in
      Table~\ref{tab:slab_model_parameters}.}
    \label{fig:slab_spec2}
  \end{center}
\end{figure}

To highlight the impact of numerical conduction on the inferred
observables we examine the 1-10 keV X-ray spectra calculated from the
models (see \S~\ref{subsec:xray_emission} for details). In
Fig.~\ref{fig:slab_spec} the X-ray spectra calculated from flow 1 in
models SLAB-A, SLAB-B, and SLAB-C are shown. Comparing the spectra
from models SLAB-A and SLAB-B we see that as $v_2$ is decreased the
normalization of the spectrum decreases. However, a further reduction
in $v_{2}$ between models SLAB-B and SLAB-C actually produces a higher
normalization in the spectrum of model SLAB-C. To explain these
differences one must examine the temperature plots in
Fig.~\ref{fig:slab_model_images} and the mass-weighted temperature
distributions in Fig.~\ref{fig:slab_mvst}. The lower normalization in
the model SLAB-B spectrum compared to that of models SLAB-A and SLAB-C
is due to a lower amount of gas at $\gtsimm10^{8}\;$K in the former,
which is evidently due to the orientation of the shock with respect to
the upstream flow - in models SLAB-A and SLAB-C more of the shock
front is normal to the preshock flow, with more severe distortions in
model SLAB-B. The differences between the spectra for models SLAB-A
and SLAB-C are the direct result of increased numerical conduction in
model SLAB-C - the higher surface area for interactions in model
SLAB-C results in a larger amount of energy being extracted from hot
flow~1 gas by cold flow~2 gas in the temperature range $T\simeq10^7 -
4\times10^{7}\;$K.

We now turn our attention to the effect that numerical conduction has
on the spectrum of the colder, denser postshock gas. In models SLAB-C,
SLAB-D, and SLAB-E the preshock parameters of flow~2 are
identical. Therefore, intuitively one would expect that the spectrum
calculated from this gas should also be identical. Yet, examining
Fig.~\ref{fig:slab_spec2} we see that this is clearly not the
case. Comparing the spectra calculated for models SLAB-C and SLAB-E,
the numerical conduction of heat in the former results in a higher
normalization to the spectrum ($\sim2\;$dex). This represents a cause
for concern as it is clear from the flow~2 temperature plot in
Fig.~\ref{fig:slab_model_images} that the volume of gas contaminated
by numerical conduction in model SLAB-C is relatively small, yet the
impact on the derived X-ray spectrum is significant.

\subsection{Colliding winds in a binary star system}
\label{subsec:cwb_model}

Colliding stellar winds in binary systems are particularly useful for
the study of the effect of instabilities on the global flow dynamics
and resulting emission characteristics. This is in part due to the
inherent geometry of the flows; close to the line-of-centres between
the stars the stellar winds collide head on, postshock gas at the
stagnation point is then accelerated in the tangential direction, and
there can be considerable shear between the flows either side of the
contact discontinuity. In addition, the importance of radiative
cooling in some systems can bring about density and temperature
contrasts of many orders of magnitude between adjacent gas.

The model consists of two stars with hypersonic, isotropic stellar
winds which collide at their terminal speeds. We restrict our
investigation to 2D and neglect any orbital motion effects
\citep[e.g.][]{Lemaster:2007, Parkin:2008, Pittard:2009}. Normally, 2D
simulations of colliding stellar winds assume axis-symmetry, and an
$rz$ grid is used. However, such simulations can be susceptible to the
``carbuncle'' instability \citep[see e.g. ][]{Pittard:1998} which is
purely numerical in origin \citep{Quirk:1994}. To avoid this, we have
instead adopted slab-symmetry, in which the divergence of the flow
goes as $r^{-1}$, and placed the `stars' in the centre of the
grid. The grid extends to $x = \pm 1.5\times10^{15}\;$cm, $y =
0-2.25\times10^{15}\;$cm with star 1 centered at ($0$,
$5\times10^{14}\;$cm) and star 2 centered at (0,
$1\times10^{15}\;$cm), giving a binary separation, $d_{\rm sep}=
5\times10^{14}\;$cm. For the models presented in \S\S~\ref{subsec:nd}
and \ref{subsec:av} the grid is initialized with $x \times y = 32
\times 24$ blocks. We allow for 2 nested levels of refinement, where
adjacent levels differ in resolution by a factor of two. This provides
an effective resolution on the finest grid of
$1024\times768\;$cells. The grid resolution is varied in
\S~\ref{subsec:sim_res} where an investigation into resolution effects
is performed.

\begin{figure*}
  \begin{center}
    \begin{tabular}{ccc}
\resizebox{45mm}{!}{\includegraphics{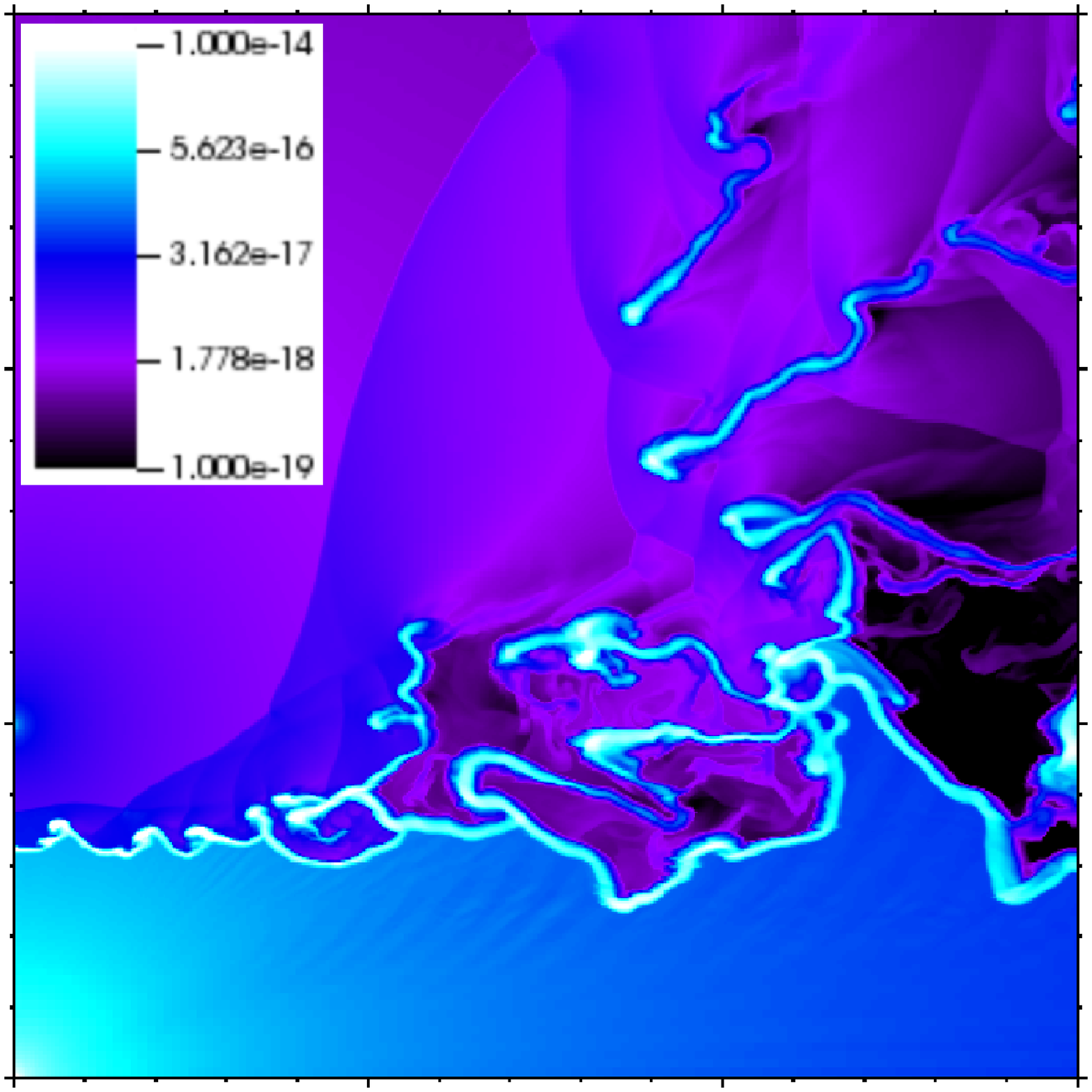}} &
\resizebox{45mm}{!}{\includegraphics{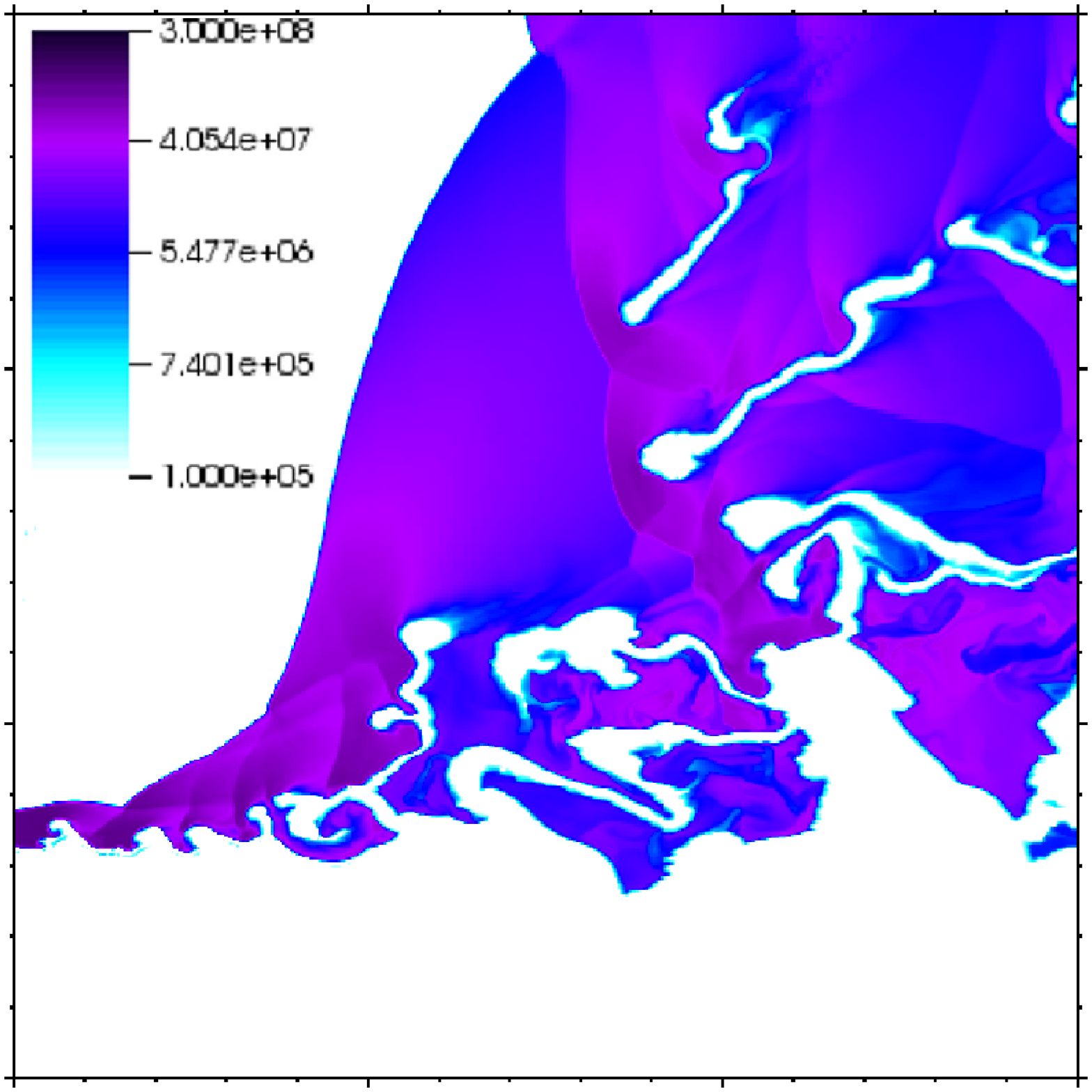}} &
\resizebox{45mm}{!}{\includegraphics{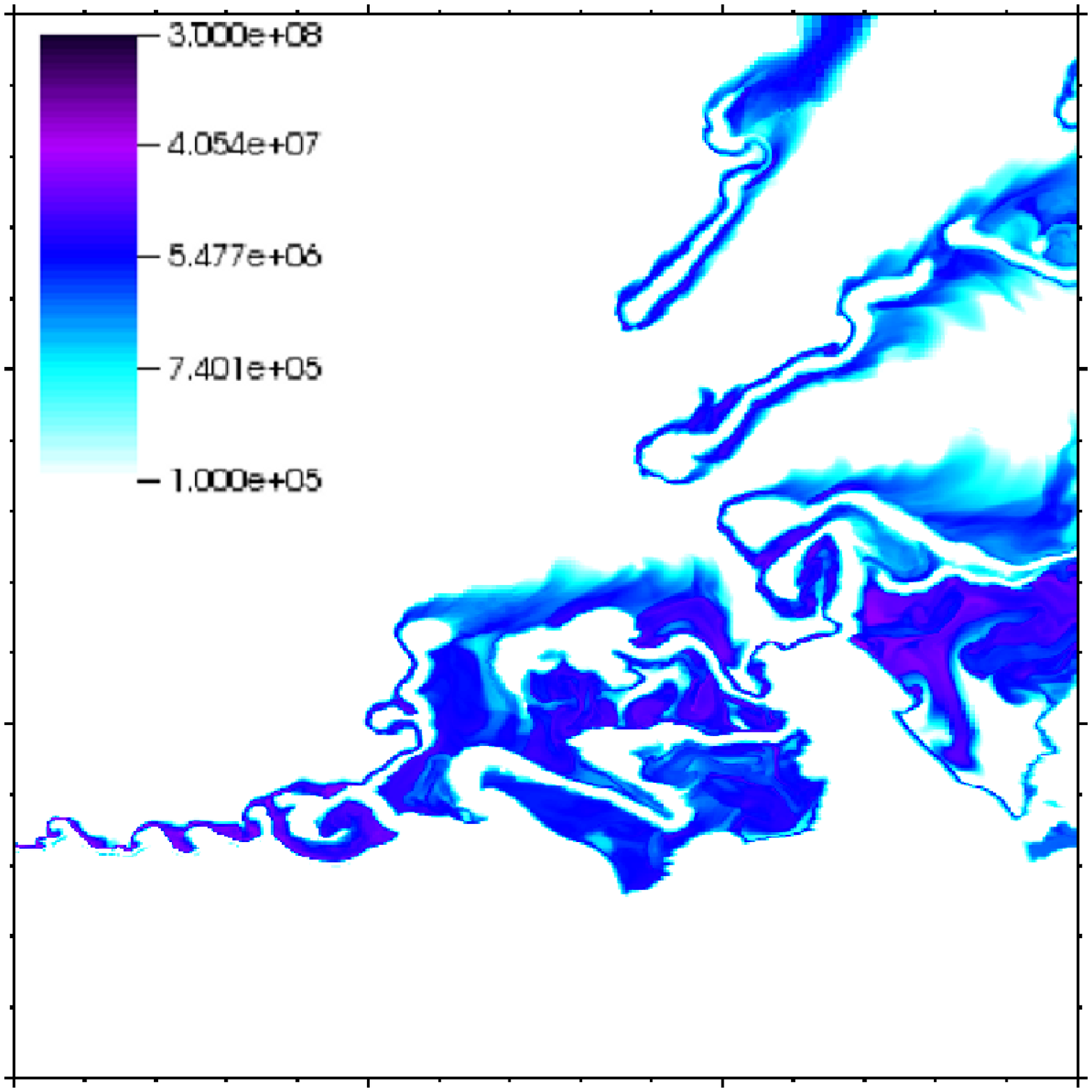}} \\

\resizebox{45mm}{!}{\includegraphics{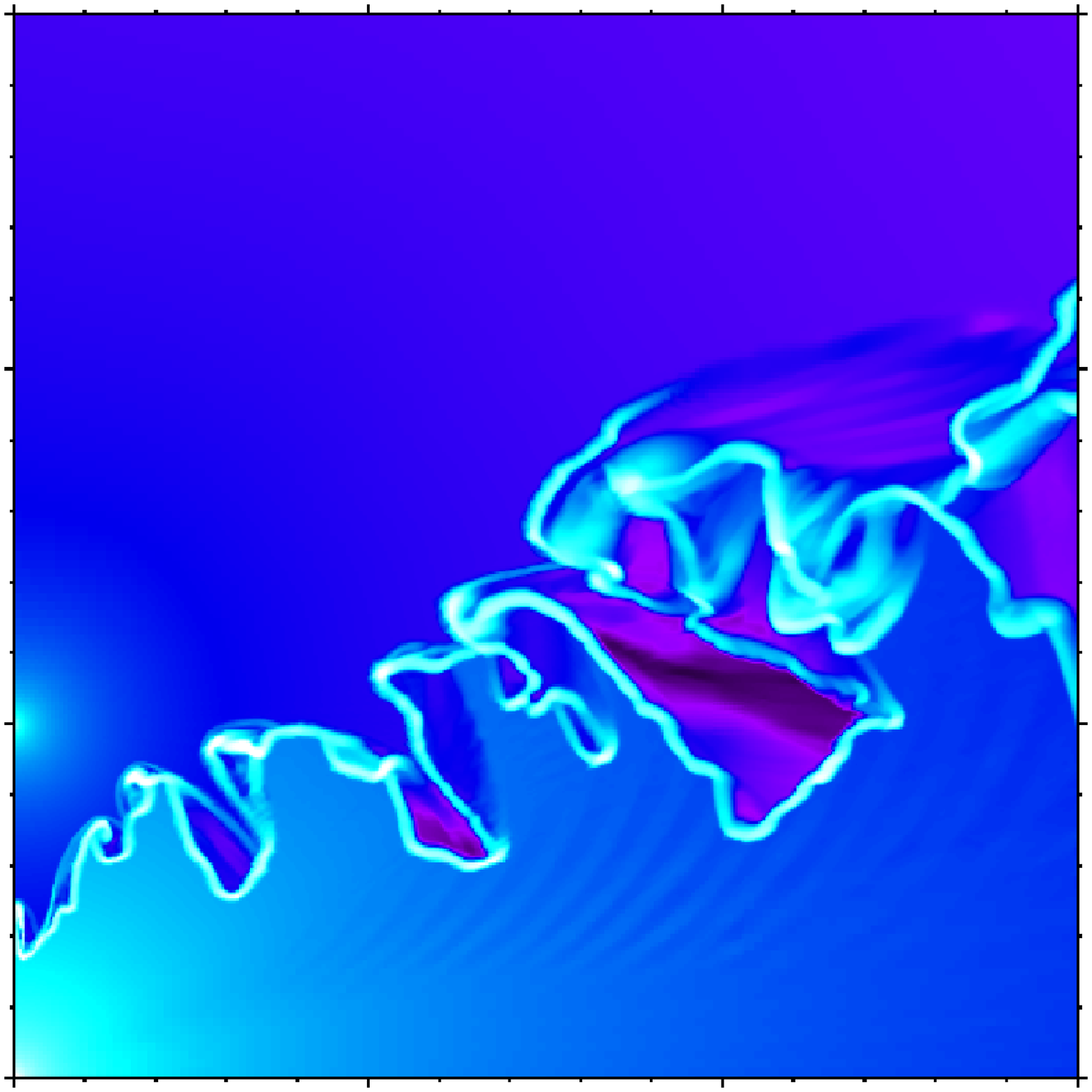}} &
\resizebox{45mm}{!}{\includegraphics{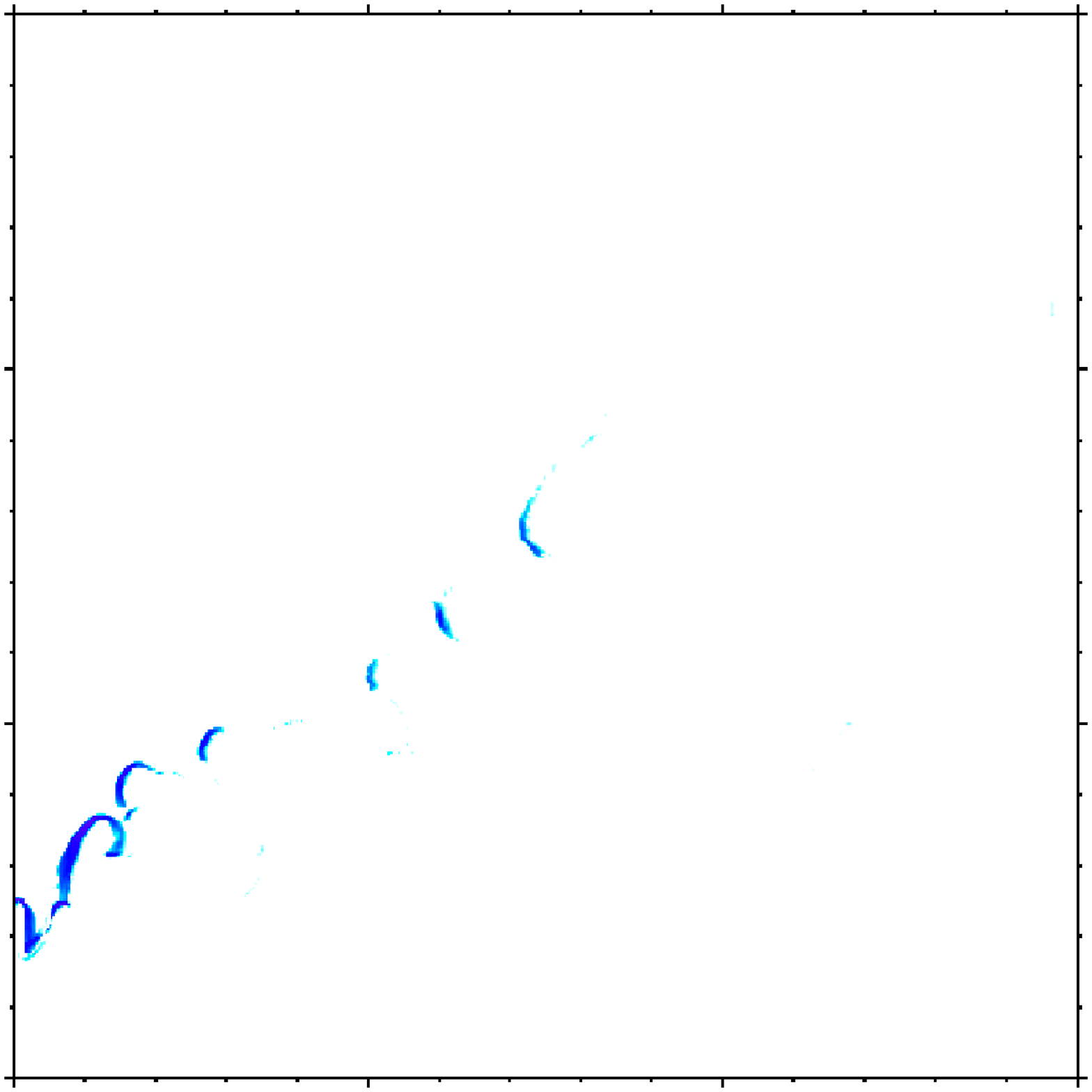}} &
\resizebox{45mm}{!}{\includegraphics{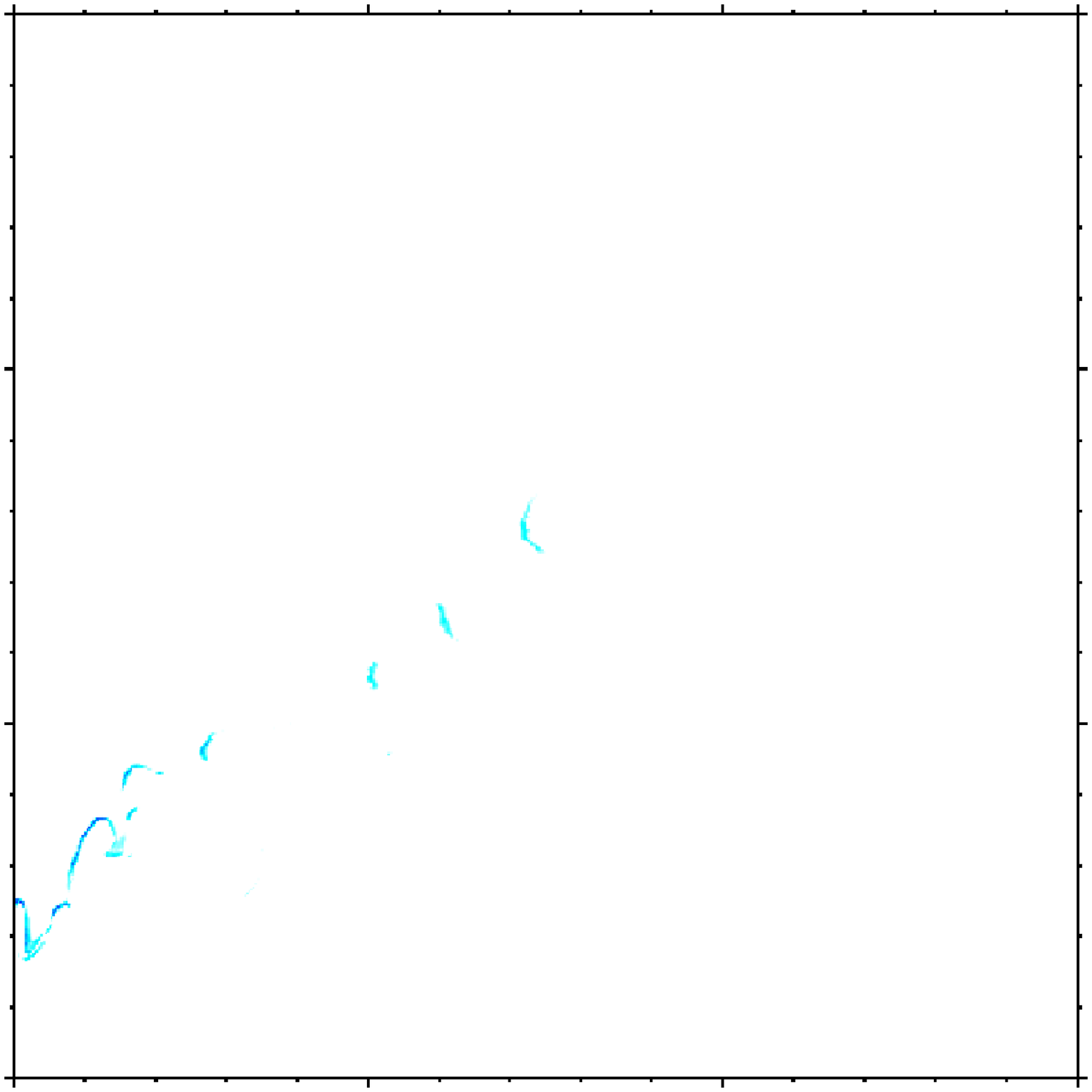}} \\

\resizebox{45mm}{!}{\includegraphics{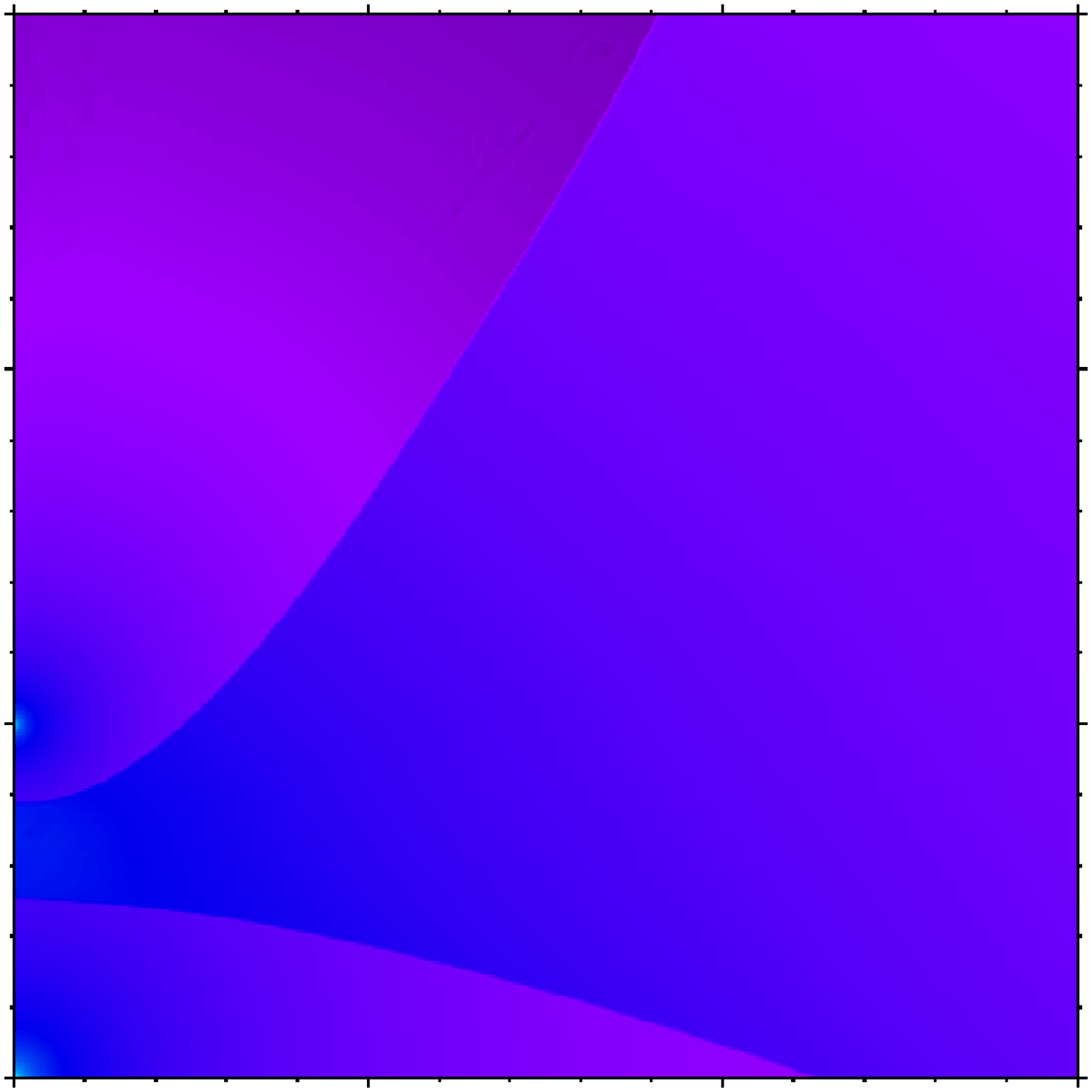}} &
\resizebox{45mm}{!}{\includegraphics{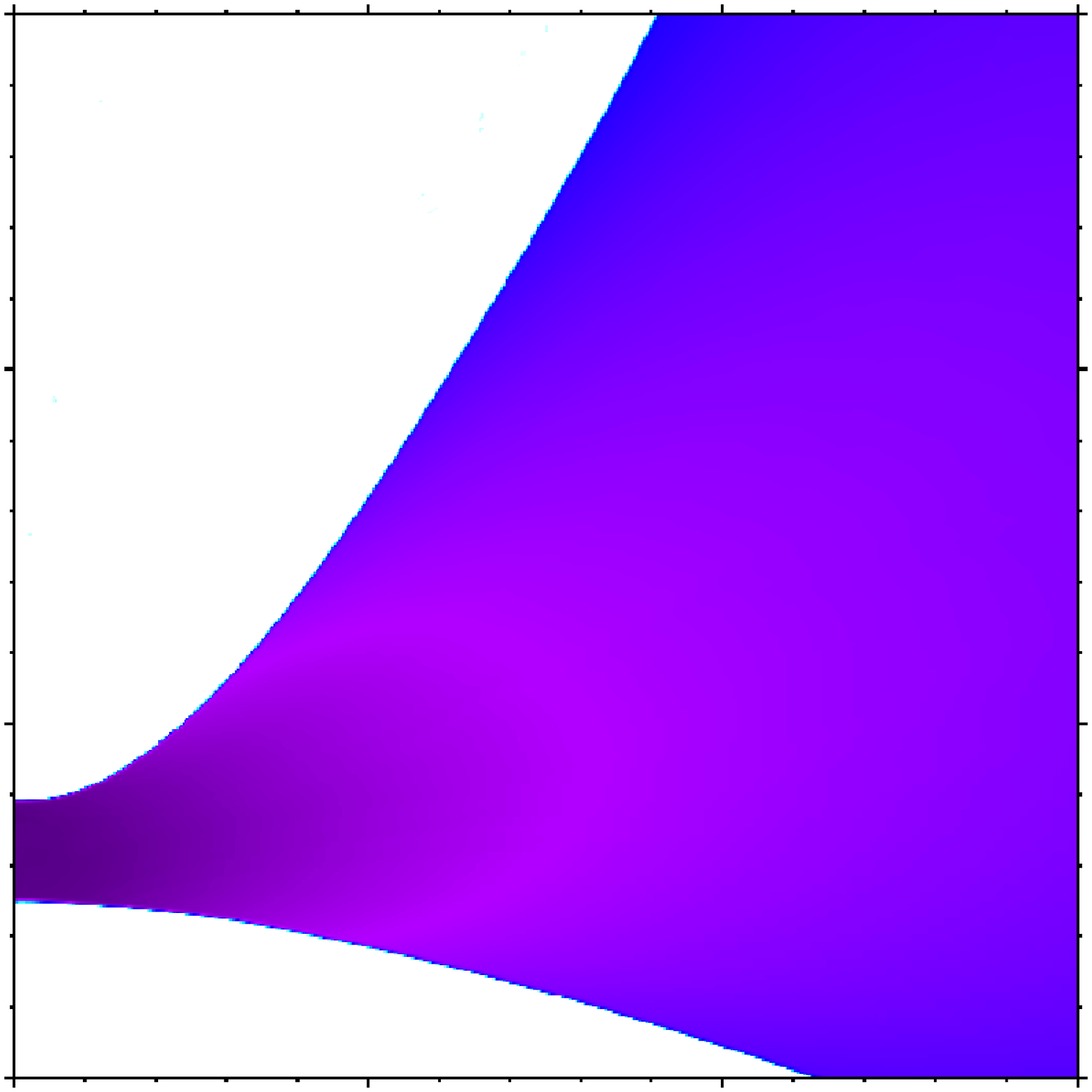}} &
\resizebox{45mm}{!}{\includegraphics{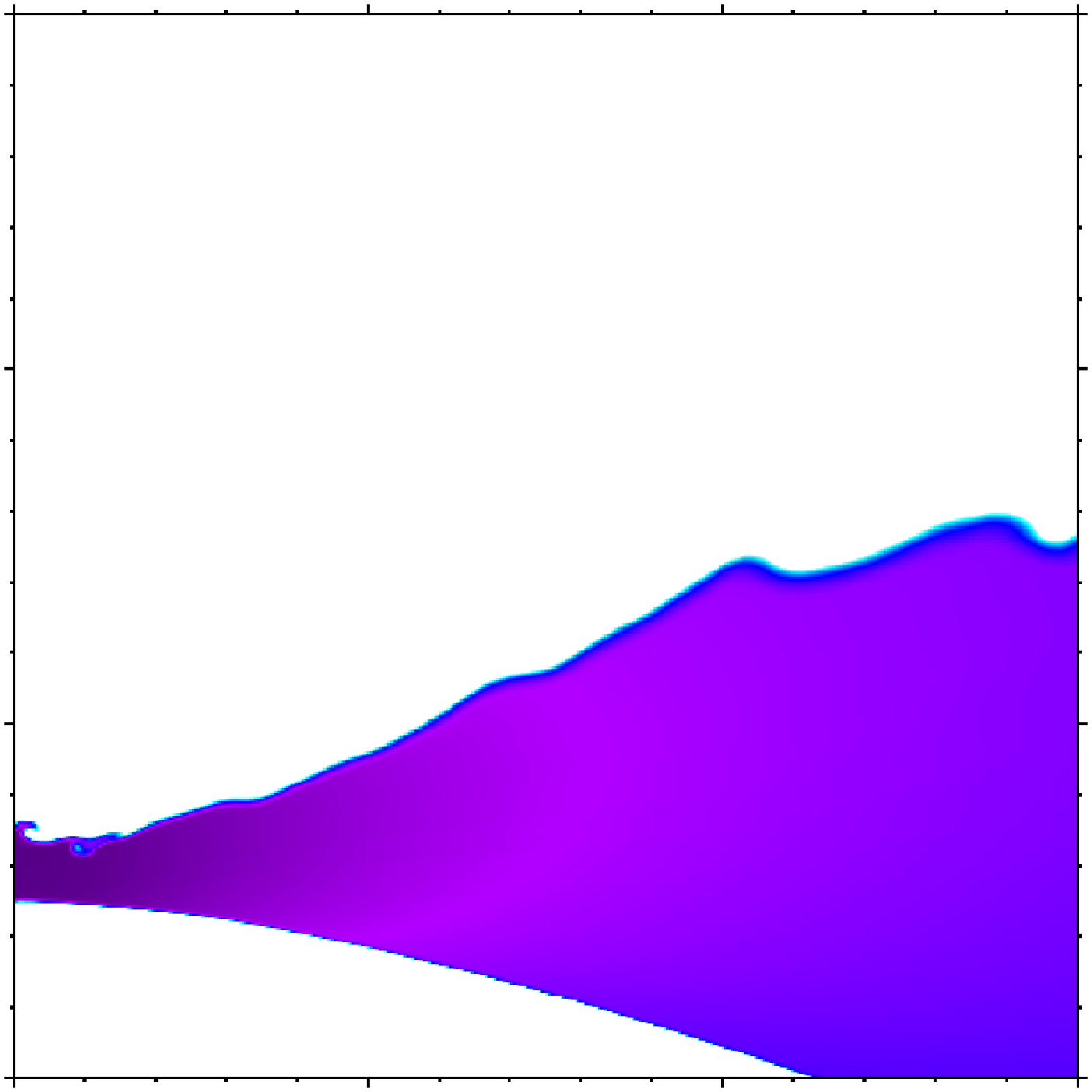}} \\

\resizebox{45mm}{!}{\includegraphics{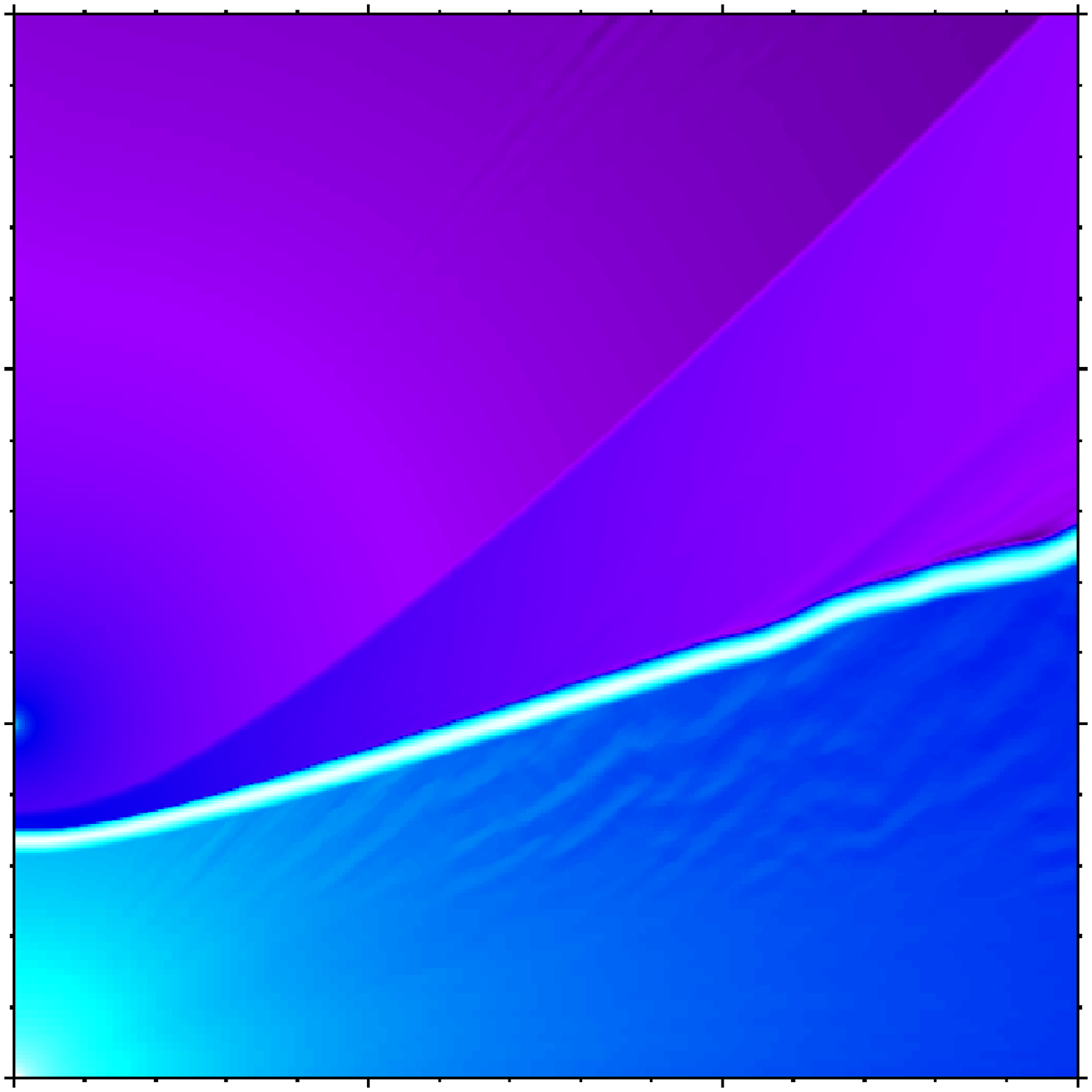}} &
\resizebox{45mm}{!}{\includegraphics{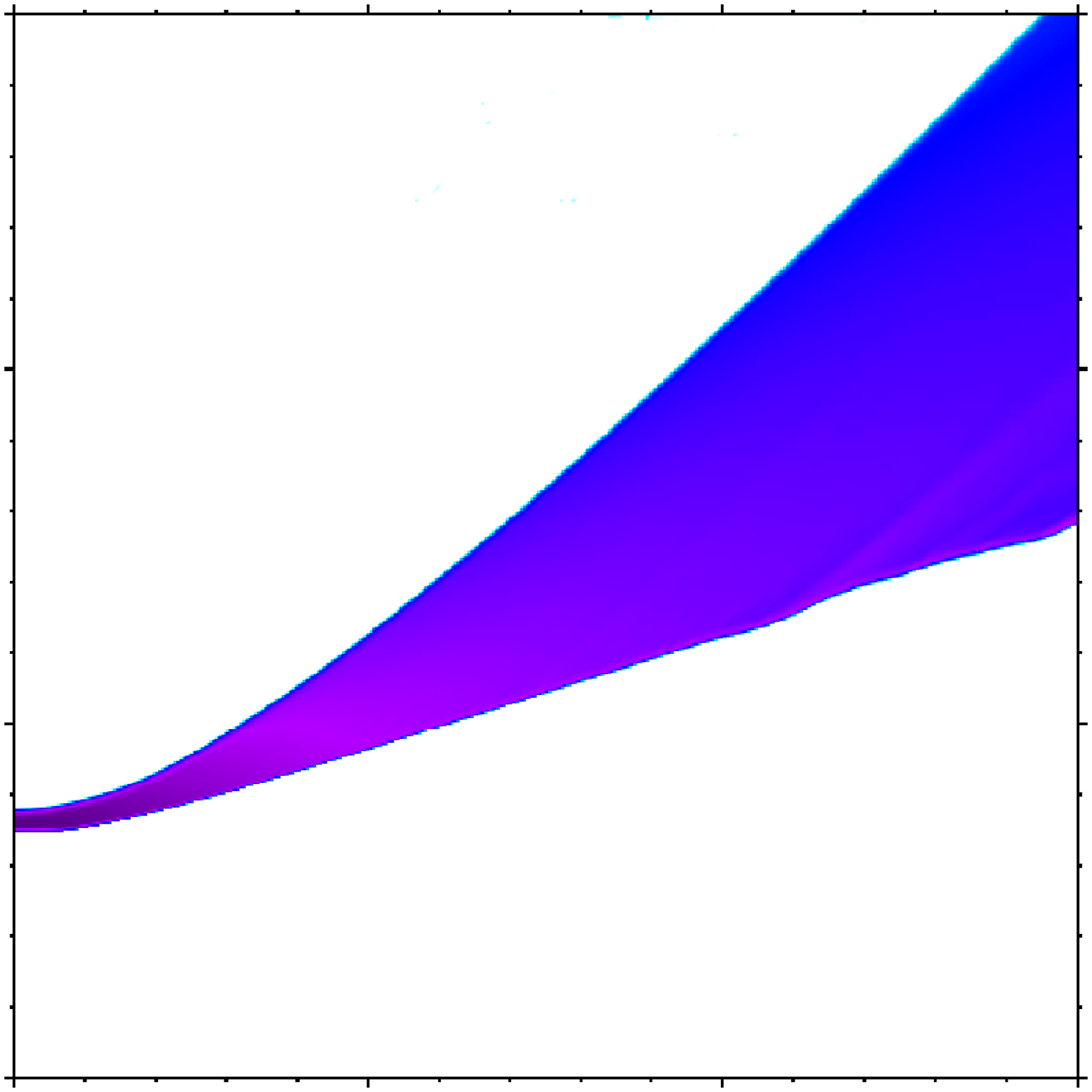}} &
\resizebox{45mm}{!}{\includegraphics{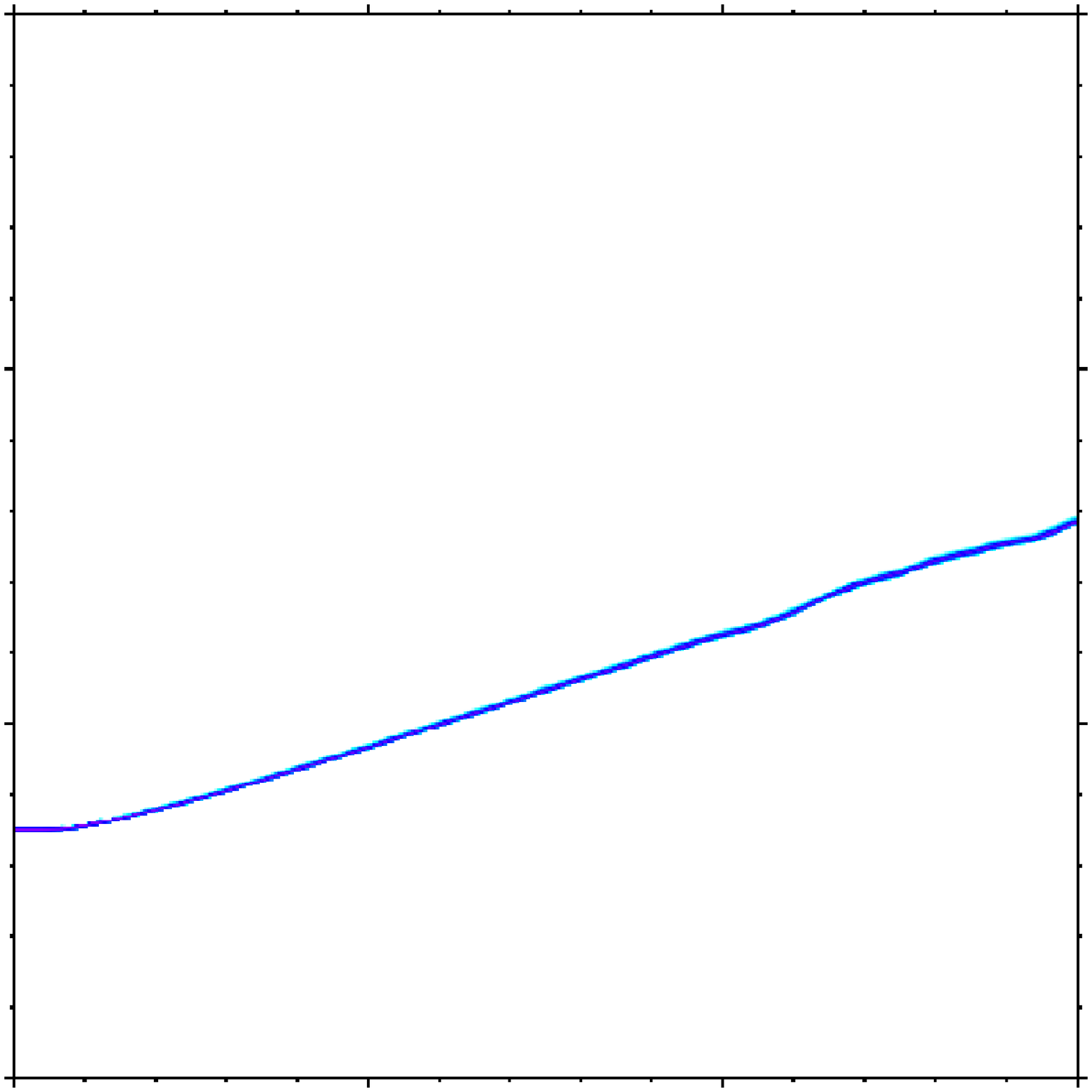}} \\

\resizebox{45mm}{!}{\includegraphics{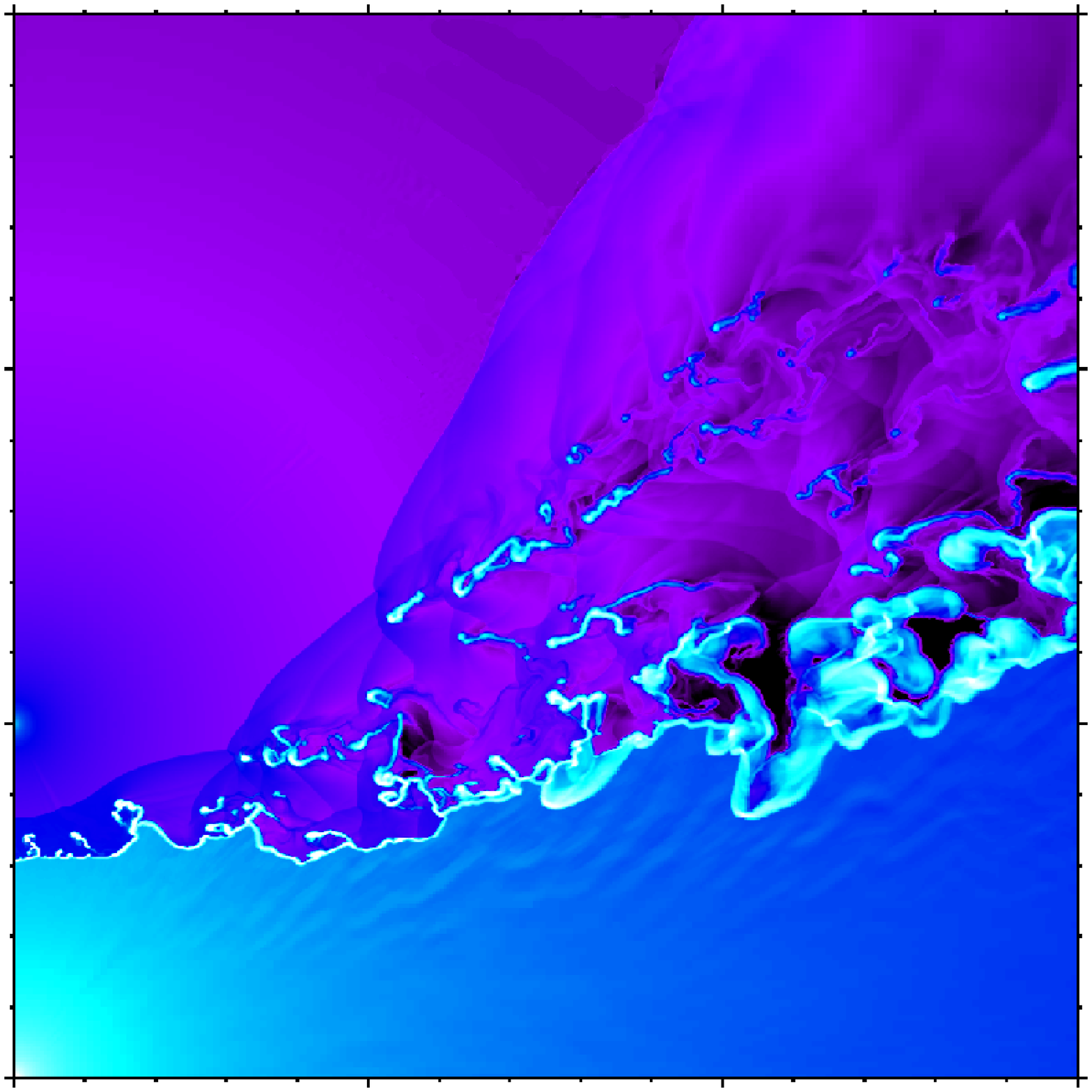}} &
\resizebox{45mm}{!}{\includegraphics{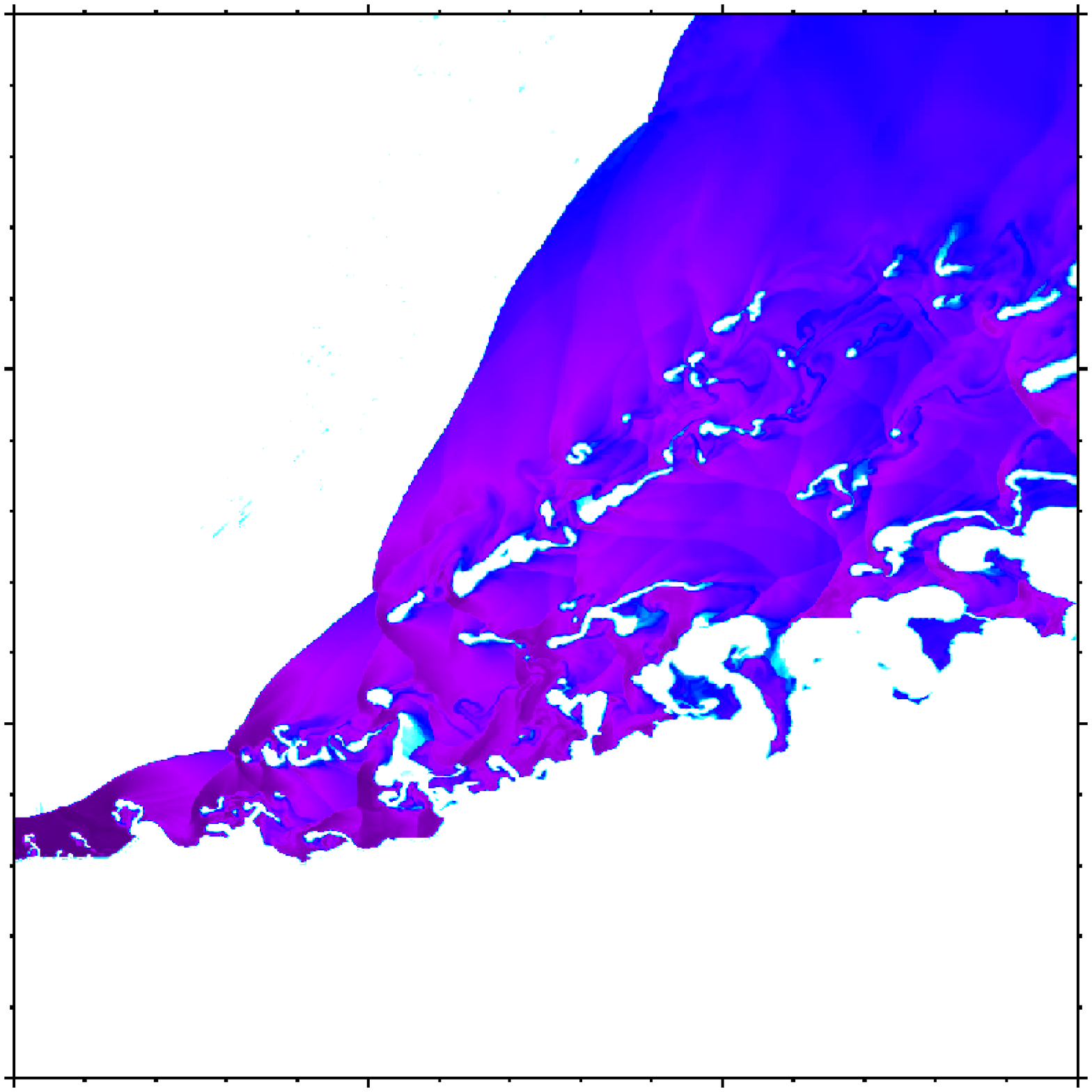}} &
\resizebox{45mm}{!}{\includegraphics{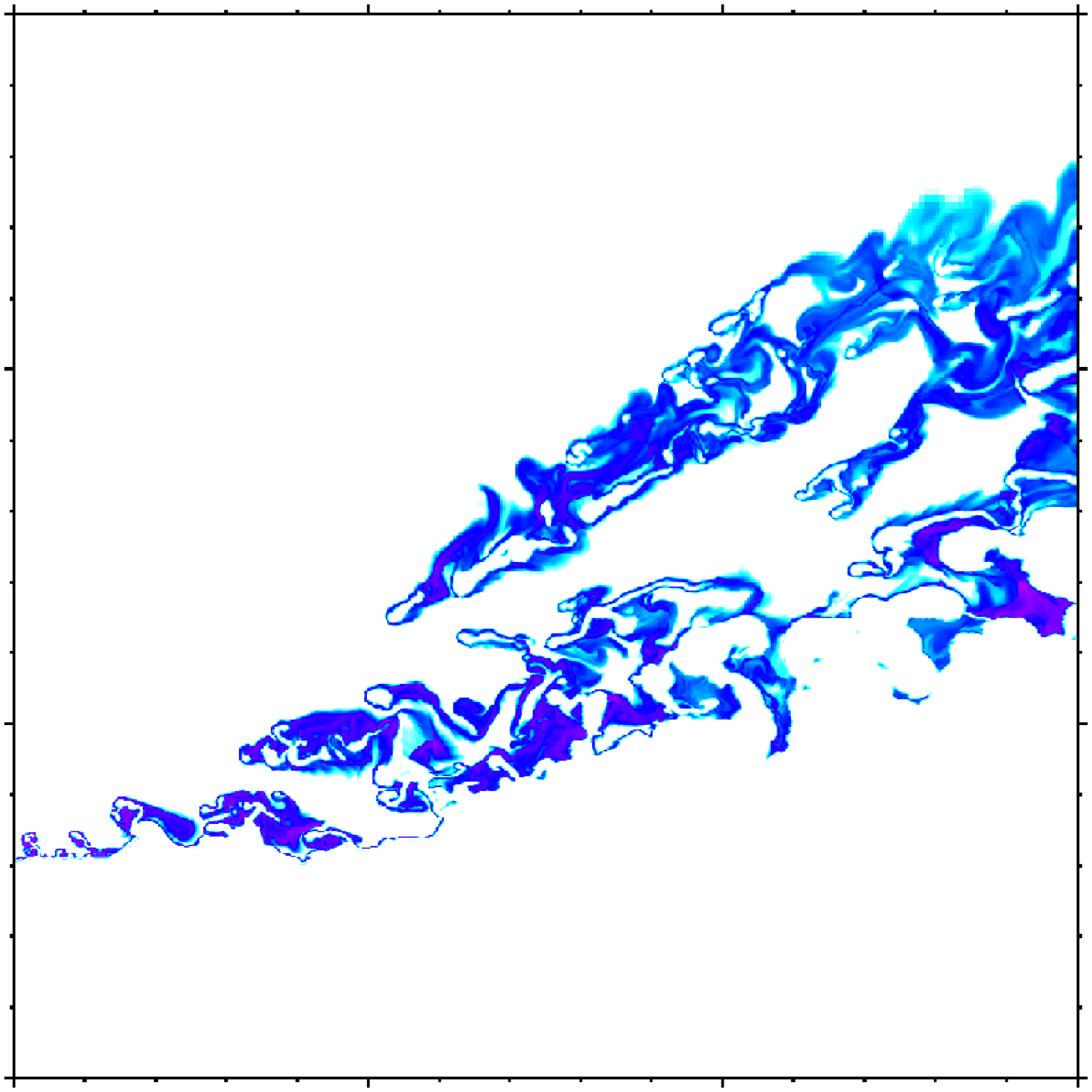}} \\

    \end{tabular}
    \caption{Simulation snapshots showing density (left column),
      temperature (middle column), and wind 1 temperature (right
      column). Models shown from top to bottom: CWB-A, CWB-B, CWB-C,
      CWB-Avisc, and CWB-A+. Wind 1 is flowing from the bottom left
      corner, while wind 2 occupies the top half of each plot. These
      calculations were performed in slab-symmetry (see text for
      details). Model parameters are listed in
      Table~\ref{tab:model_parameters}. Large tick marks correspond to
      a distance of $5\times10^{14}\;$cm. For illustrative purposes,
      only a subsection of the computational grid is shown, with the
      line of centres between the stars actually running through the
      middle of the grid.}
    \label{fig:model_images}
  \end{center}
\end{figure*}

To initiate the stellar winds appropriate hydrodynamic variables
(i.e. $\rho, P, {\bf v}$) are mapped into cells residing within a
radial distance of the respective star of $5\times10^{12}\;$cm. Model
parameters are noted in Table~\ref{tab:model_parameters}. The
unshocked winds are initialized with $T=10^{4}\;$K. Noting that in 2D
cartesian geometry the divergence of the flow goes as $r^{-1}$ we
calculate the stellar wind density profiles using the following
equations,

\begin{equation}
\rho_1 = \frac{\dot{M}_1\alpha}{2 \pi r_1 v_1},
\end{equation}

\begin{equation}
\rho_2 =  \frac{\dot{M}_2\alpha}{2 \pi r_2 v_2\sqrt{\eta}},
\end{equation}

\noindent where

\begin{equation}
\alpha = \frac{(1 + \sqrt{\eta})}{2 d_{\rm sep}}.
\end{equation}

\noindent $r$ and $v$ are the radial distance from, and stellar wind
velocity adopted for, the respective star, and $\eta~ (= \dot{M}_2
v_2/\dot{M}_1 v_1)$ is the wind momentum ratio which is kept constant
at a value of 0.2 in all models. In the above equations we are
effectively fixing the stagnation point wind density in our 2D
cartesian geometry simulations to the equivalent attained from a 2D
axis-symmetric or 3D simulation. This approach has the advantage that
the preshock gas densities only differ slightly from those of an
equivalent simulation with an $r^{-2}$ divergence (i.e. 3D, or 2D
axis-symmetric), and as such radiative cooling is similar.

For our fiducial model we choose parameters similar to those
determined by \cite{Parkin:2009} for the massive star binary system
$\eta\;$Car. In the context of CWBs, \etacar presents us with some
unique problems when it comes to modelling its wind-wind collision
which we do not find when modelling other long-period binaries
(e.g. WR\,140, $\iota$~Ori). These problems arise from the fact that
the slow and dense primary wind, once shocked, cools rapidly into a
dense sheet, while the shocked secondary wind behaves largely
adiabatically \citep{Pittard:2002,Parkin:2009}, which produces large
jumps in the flow variables across the contact discontinuity, with
cold dense gas adjacent to hot rarefied gas. Inevitably there is some
numerical conduction of heat across the interface\footnote{To
  circumvent this problem sophisticated models of CWBs have been
  developed which, at the cost of providing limited information about
  the flow dynamics, avoid details of mixing in the postshock gas
  \citep{Antokhin:2004, Parkin:2008}.}.

In the following sections we present results for simulations performed
with low and high numerical viscosity, and also the results of
resolution tests. In each instance we first discuss the gas dynamics
and then present the results from X-ray calculations.

\begin{table*}
\begin{center}
\caption[]{Parameters pertaining to the CWB simulations of
  \S~\ref{subsec:cwb_model}. The wind momentum ratio, $\eta$, is kept
  constant in all simulations.}
\begin{tabular}{llllllll}
\hline
Model &  Comment & $\dot{M}_{1}$ & $v_{1}$ & $\chi_{1}$ & $\dot{M}_{2}$ & $v_{2}$ & $\chi_{2}$ \\
 &  & $(10^{-7}\Msolpyr)$ & $(10^{8}\;{\rm cm~s^{-1}})$ &  & $(10^{-7}\Msolpyr)$ & $(10^{8}\;{\rm cm~s^{-1}})$ &  \\
\hline
CWB-A &  & 3000 & 0.5 & 0.01 & 100 & 3 & 405 \\
CWB-B &  $\chi_2 \downarrow$ & 3000 & 0.5 & 0.01 & 300 & 1 & 1.67 \\
CWB-C &  $\chi_1 \uparrow$ & 500 & 3 & 81 & 100 & 3 & 405 \\
\hline
\label{tab:model_parameters}
\end{tabular}
\end{center}
\end{table*}

\begin{figure}
  \begin{center}
    \begin{tabular}{c}
      \resizebox{60mm}{!}{\includegraphics{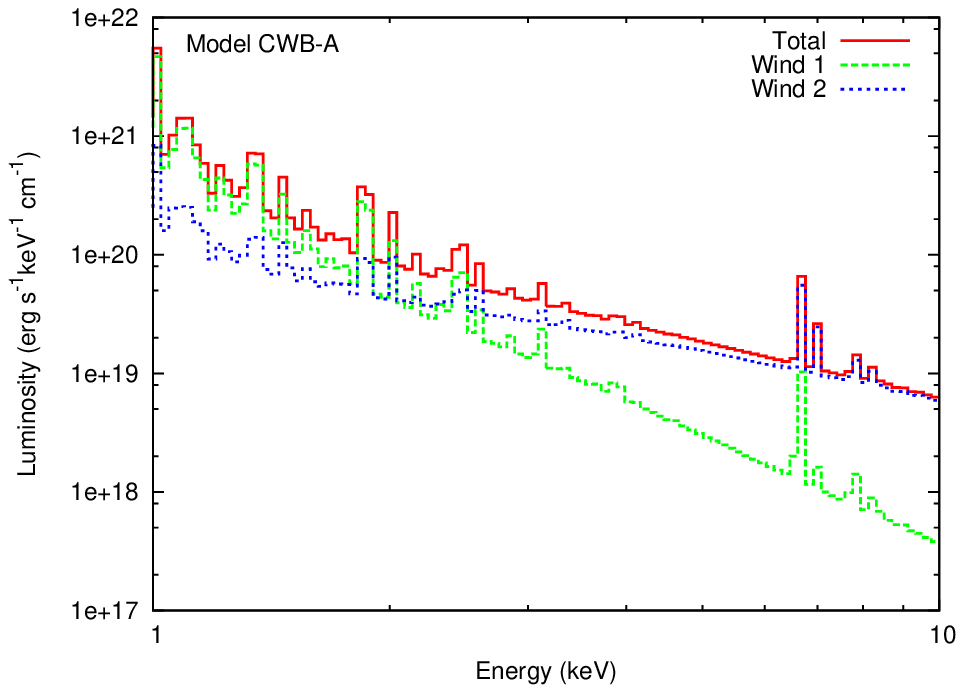}} \\
      \resizebox{60mm}{!}{\includegraphics{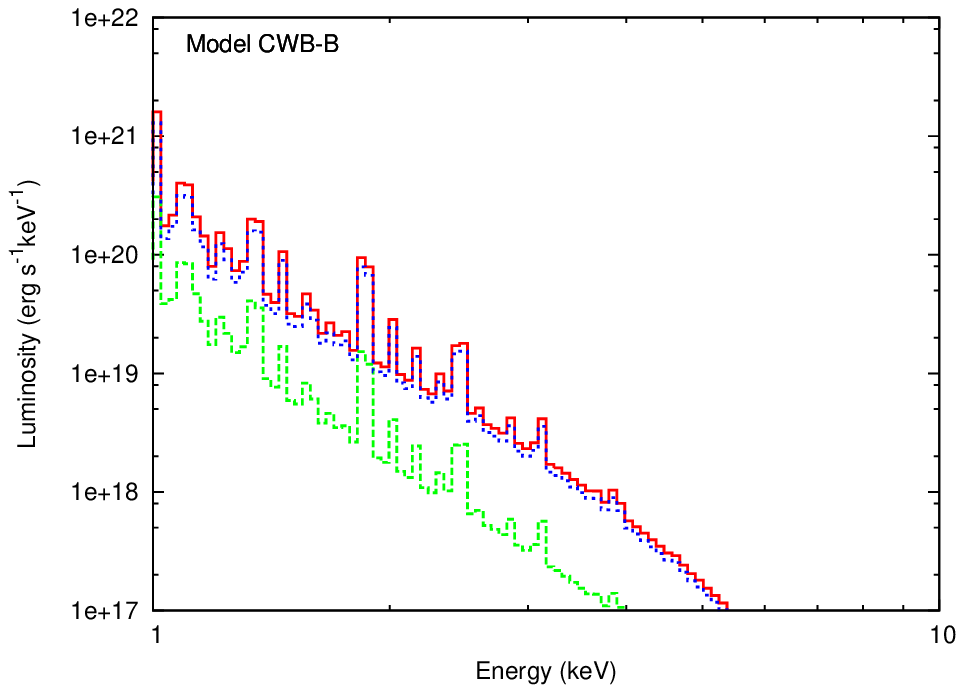}} \\
      \resizebox{60mm}{!}{\includegraphics{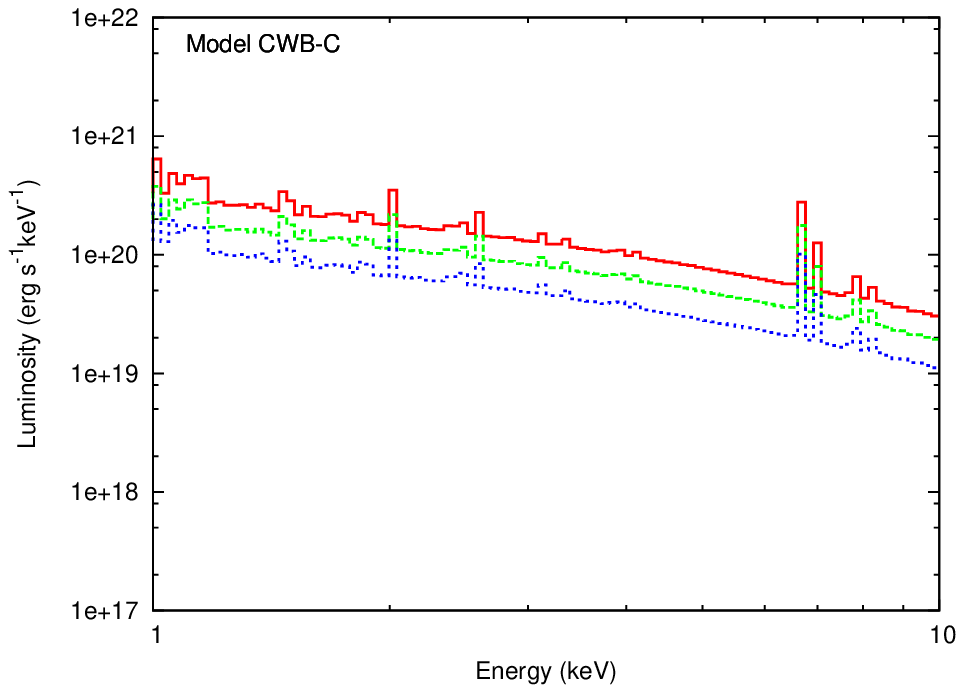}} \\
      \resizebox{60mm}{!}{\includegraphics{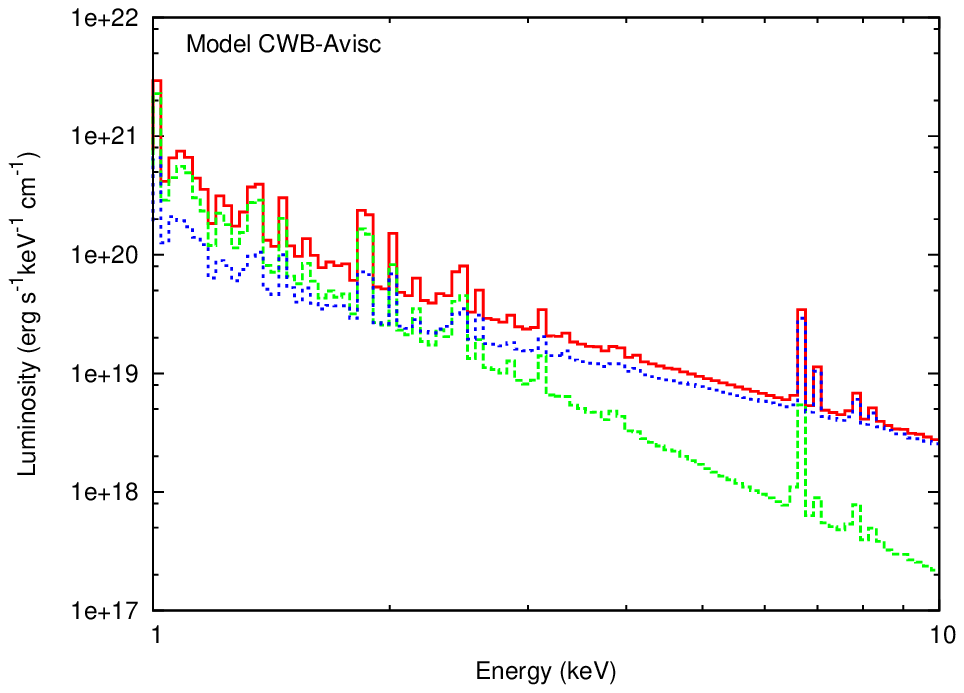}} \\
      \resizebox{60mm}{!}{\includegraphics{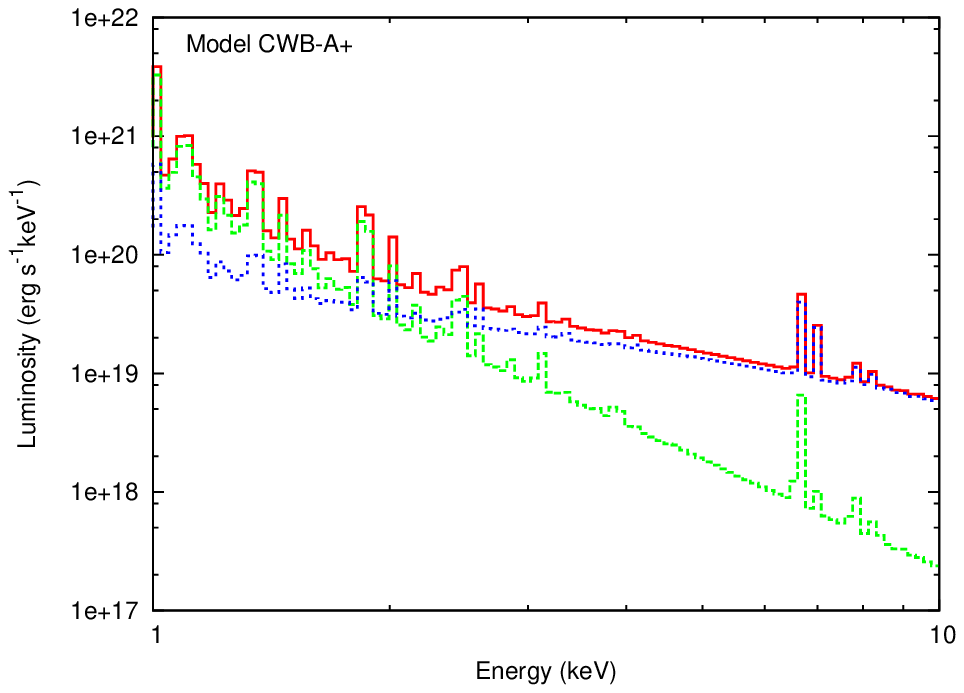}} \\
    \end{tabular}
    \caption{X-ray spectra calculated from the CWB simulations noted
      in Table~\ref{tab:model_parameters}. The total (solid red line),
      wind 1 (dashed green line), and wind 2 (dotted blue line)
      emission contributions are shown. From top to bottom: CWB-A,
      CWB-B, CWB-C, CWB-Avisc, and CWB-A+. Corresponding integrated
      luminosities are listed in Table~\ref{tab:luminosities}.}
    \label{fig:model_spectra}
  \end{center}
\end{figure}

\subsubsection{Low numerical viscosity}
\label{subsec:nd}

In model CWB-A wind~1 is highly radiative ($\chi_{1} = 0.01$, see
Eq.~\ref{eqn:chi}) and wind~2 is largely adiabatic ($\chi_{2} \simeq
400$). The postshock wind~1 gas cools rapidly and collapses to form a
thin dense shell (Fig.~\ref{fig:model_images}). The rapid onset of RT,
KH, and thin-shell instabilities fragments this shell, causing clumps
of wind~1 material to become interspersed with the hot postshock
wind~2 material. The fragmentation of the interface changes the
surface area between hot and cold gas, and increases the total amount
of numerical conduction in the calculation. Postshock wind~2 material
reaches peak temperatures of $\gtsimm 10^{8}\;$K, consistent with the
value expected from Eq.~\ref{eqn:energy}. However, wind~1 gas reaches
temperatures of $\sim 10^{7}\;$K, much higher than the expected
$T\simeq3\times10^{6}\;$K (see the wind~1 temperature plot in
Fig.~\ref{fig:model_images}). As can be seen, the postshock region is
turbulent, which complicates deciphering heating through flow
collisions and spurious heating brought about by numerical
conduction. In \S~\ref{subsec:av} we elucidate the relative
contributions of these two mechanisms by performing tests with
additional numerical viscosity to suppress the growth of
instabilities.

The intrinsic X-ray spectrum from model CWB-A is dominated by emission
from the shocked wind~1 material at $E\ltsimm 2\;$keV and by wind~2
material above this energy (Fig.~\ref{fig:model_spectra}). The
difference in the spectral slopes does not readily relate to the
contrast in preshock velocities. The preshock velocity of wind~1
($v_{1} = 500\;{\rm km s^{-1}}$) corresponds to an energy of $\simeq
0.3\;$keV (using Eq.~\ref{eqn:energy}). Therefore, we would not expect
emission extending to energies of $\simeq10\;$keV at the magnitude
observed in Fig.~\ref{fig:model_spectra}. The total luminosity is
dominated by postshock wind~1 gas (see Table~\ref{tab:luminosities})
which produces considerable soft X-ray emission.

Lowering the velocity of wind~2 to $v_2 = 1\times10^{8}\;{\rm
  cm~s^{-1}}$ (model CWB-B), reduces the cooling parameter to a value
which corresponds to gas which is highly radiative
($\chi_{2}=1.67$). Examining Fig.~\ref{fig:model_images} we see that
such a change causes a dramatic difference to the flow dynamics. The
structure of postshock gas is now dominated by large amplitude
oscillations caused by the NTSI \citep[e.g.][]{Stevens:1992,
  Vishniac:1994}. Heat conduction from wind 2 into wind 1 is lessened
in model CWB-B compared to model CWB-A due to a lower temperature
gradient at the interface between the winds and this has a
catastrophic effect on the resulting emission
(Fig.~\ref{fig:model_spectra}). The spectrum and total luminosity are
now dominated by emission from wind~2, and there is a substantial
reduction in the wind~1 emission at all energies. This is unsurprising
when one compares the wind~1 temperature plots for models CWB-A and
CWB-B; for model CWB-B there is very little wind~1 gas at $T\gtsimm
10^{6}\;$K.

We have explored above the effect that the thermal properties
(e.g. temperature) of the wind~2 postshock gas has on the resulting
emission from wind~1, but what effect does wind~1 have on wind~2 gas?
In model CWB-C the parameters of wind~1 have been modified so as to
produce an adiabatic postshock flow ($\chi_1 = 81$). Compared to
models CWB-A and CWB-B the WCR now appears relatively smooth with only
small perturbations of the contact discontinuity due to velocity shear
between the postshock winds driving a KH instability
(Fig.~\ref{fig:model_images}). Examining the spectra, we see that the
slope of both the wind~1 and 2 spectra are now identical, as expected
for identical preshock velocities. A small increase in the
normalization of the spectrum and integrated luminosity calculated
from wind~2 is seen between models CWB-C and CWB-A, consistent with
heat being conducted out of wind~2 by the neighbouring cold gas of
wind~1 in model CWB-A.

\begin{figure}
  \begin{center}
    \begin{tabular}{c}
      \resizebox{80mm}{!}{\includegraphics{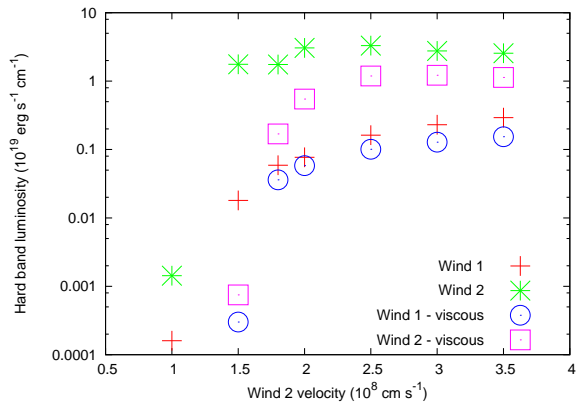}} \\
    \end{tabular}
    \caption{Hard band (7-10 keV) luminosity calculated from wind~1
      and 2 as a function of $v_2$ for models CWB-A and CWB-Avisc. For
      these model calculations the wind~1 parameters were set to those
      of model~CWB-A ($v_1 = 5\times10^{7}\;{\rm cm~s^{-1}}$) and
      wind~2 parameters were varied accordingly while maintaining a
      constant wind momentum ratio. In the ideal situation where there
      is no numerical heat conduction between wind~1 and wind~2, the
      hard band luminosity of wind~1 would be negligible.}
    \label{fig:hr1}
  \end{center}
\end{figure}

In summary, models CWB-A, CWB-B, and CWB-C demonstrate that
differences in the density and postshock cooling rate across the
contact discontinuity are important for the dynamics and the
calculated X-ray emission. To examine how the contamination of the
X-ray emission varies with the wind~2 parameters we have performed
further tests which are shown in Fig.~\ref{fig:hr1}. As the preshock
velocity of wind~2, $v_2$, and therefore $\chi_2$, is increased the
integrated luminosity of wind~1 and 2 in the hard band (7-10 keV) also
increases. There is a clear jump in the hard band luminosity which
occurs between $v_2\simeq1$ and $1.5\times10^8\;{\rm cm~s^{-1}}$, and
corresponds to the transition of wind~2 from a radiative to an
essentially adiabatic postshock flow. When both winds are similar in
character (i.e. radiative in this case) there appears to be little
heat conduction between them, while large differences in the wind
speeds and postshock temperatures result in significant numerical heat
conduction, as shown by the rise in the hard band luminosity from
wind~1 as $v_{2}$ increases.

\begin{table}
\begin{center}
\caption[]{Integrated 1-10 keV X-ray luminosities calculated for the
  simulated models. $L_{\rm TOT}$, $L_{1}$, and $L_{2}$ are the total,
  wind~1, and wind~2 luminosities (in units of erg s$^{-1}$
  cm$^{-1}$), respectively. Models CWB-Avisc and CWB-A+ are variants
  of model CWB-A with higher numerical viscosity (\S~\ref{subsec:av})
  and higher simulation resolution (\S~\ref{subsec:sim_res}),
  respectively.}
\begin{tabular}{llll}
\hline
Model & $L_{\rm TOT}$& $L_{1}$ & $L_{2}$ \\  
\hline
CWB-A & $9.50\times10^{20}$ & $6.42\times10^{20}$ & $3.08\times10^{20}$ \\
CWB-B & $2.42\times10^{20}$ & $4.44\times10^{19}$ & $1.98\times10^{20}$ \\
CWB-C & $9.52\times10^{20}$ & $5.86\times10^{20}$ & $3.66\times10^{20}$ \\
CWB-Avisc & $4.14\times10^{20}$ & $2.48\times10^{20}$ & $1.66\times10^{20}$ \\
CWB-A+ & $7.84\times10^{20}$ & $5.20\times10^{20}$ & $2.64\times10^{20}$ \\
\hline
\label{tab:luminosities}
\end{tabular}
\end{center}
\end{table}

\subsubsection{High artificial viscosity}
\label{subsec:av}

Hydrodynamical codes typically allow the user to modify the magnitude
of additional diffusion-like terms which act as artificial viscosity
in the governing flow equations. Intuitively, one would expect the
addition of artificial viscosity to increase the level of heat
conduction across a unit area of the interface between cold and hot
gas. However, viscosity also suppresses instabilities, and thus
reduces the total area for interactions. Whether the total amount of
numerical heat conduction increases or decreases with increasing
artificial viscosity will depend on the relative strength of these
effects.

Low numerical viscosity in model CWB-A results in the growth of
instabilities and thus a turbulent wind-wind collision region
(WCR). The inclusion of additional numerical viscosity in model
CWB-Avisc surpresses the growth of such instabilities, and in
comparison to model CWB-A the WCR is much smoother
(Fig.~\ref{fig:model_images}). The volume occupied by shocked wind 2
material is significantly smaller in model CWB-Avisc than in model
CWB-A. There appears to be two reasons for this. The reduction in
volume far downstream of the apex of the WCR is caused by the lack of
secondary shocks in this region of the flow, which in model CWB-A
result from the intrusion of dense clumps of gas from wind~1 into the
shocked wind~2 gas. However, there is also a significant reduction in
the width of postshock wind~2 gas at the apex of the WCR which is
evidence for the enhanced numerical transport of heat out of the gas
by the artificial viscosity.

\begin{figure}
  \begin{center}
    \begin{tabular}{c}
      \resizebox{80mm}{!}{\includegraphics{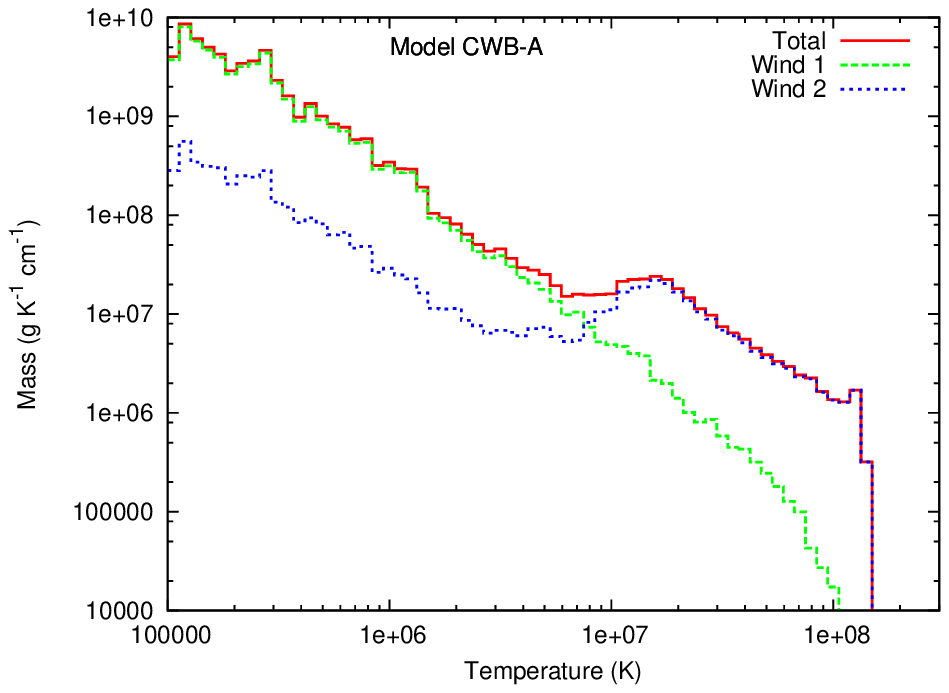}} \\
      \resizebox{80mm}{!}{\includegraphics{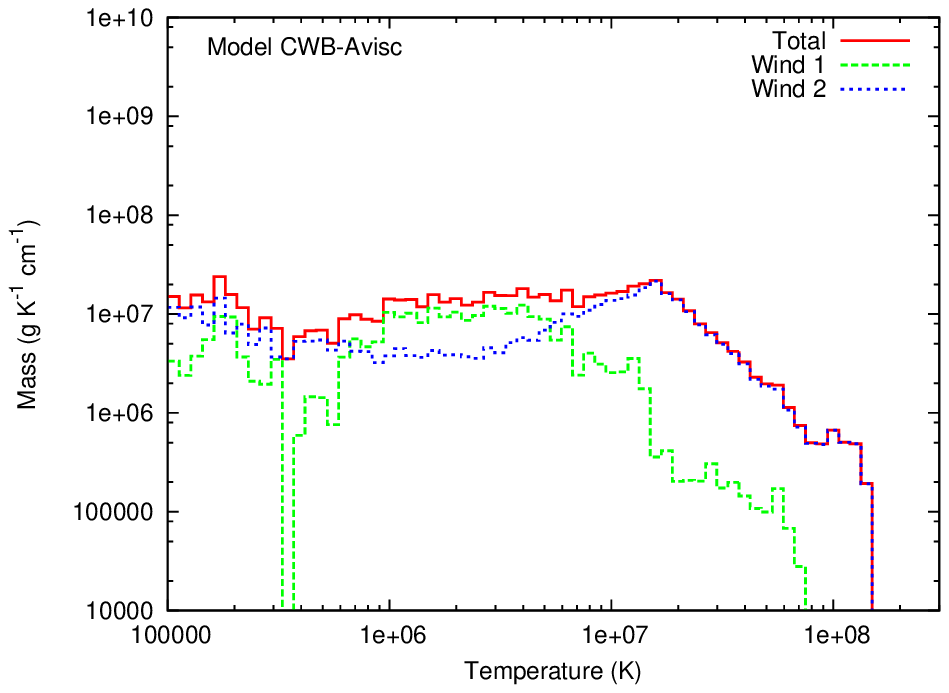}} \\
     \end{tabular}
    \caption{Distribution of mass as a function of postshock gas
      temperature in model CWB-A (upper panel) and model CWB-Avisc
      (lower panel).}
    \label{fig:av_mass}
  \end{center}
\end{figure}

A comparison of the X-ray calculations for models CWB-A and CWB-Avisc
reveals that the latter is of lower luminosity and spectral hardness
(see Table~\ref{tab:luminosities} and
Fig.~\ref{fig:model_spectra}). This can be understood by the fact that
in model CWB-A the higher level of interaction between postshock wind
1 and 2 gas (i.e. slowly moving dense clumps being struck by higher
speed, hot, rarefied flow) involves collisions which heat wind 1 gas
to soft X-ray emitting temperatures and also act to heat wind 2 gas,
whilst in model CWB-Avisc heat is conducted out of the hottest gas
near the apex by the increased artificial viscosity. This point is
highlighted by a comparison of the distribution of mass as a function
of temperature for models CWB-A and CWB-Avisc
(Fig.~\ref{fig:av_mass}). For model CWB-Avisc there is almost three
orders of magnitude less mass at $T\simeq10^{5}\;$K compared to model
CWB-A. In addition, the distribution of wind 2 material in the range
$T=10^{7}-10^{8}\;$K drops-off more rapidly for model CWB-Avisc than
for model CWB-A. These results confirm that i) the increased surface
area for interactions in model CWB-A compared to model CWB-Avisc
results in collisions which heat additional (mainly downstream) wind 2
material to $T\sim 10^{8}\;$K through further shocks while also
heating wind~1 material to $T<10^{7}\;$K through friction and enhanced
heat conduction (though the mass-weighted temperature is actually
reduced - see below), and ii) enhanced heat conduction near the apex
of the WCR reduces the temperature of the wind~2 gas in model
CWB-Avisc.

\begin{figure}
  \begin{center}
    \begin{tabular}{c}
      \resizebox{80mm}{!}{\includegraphics{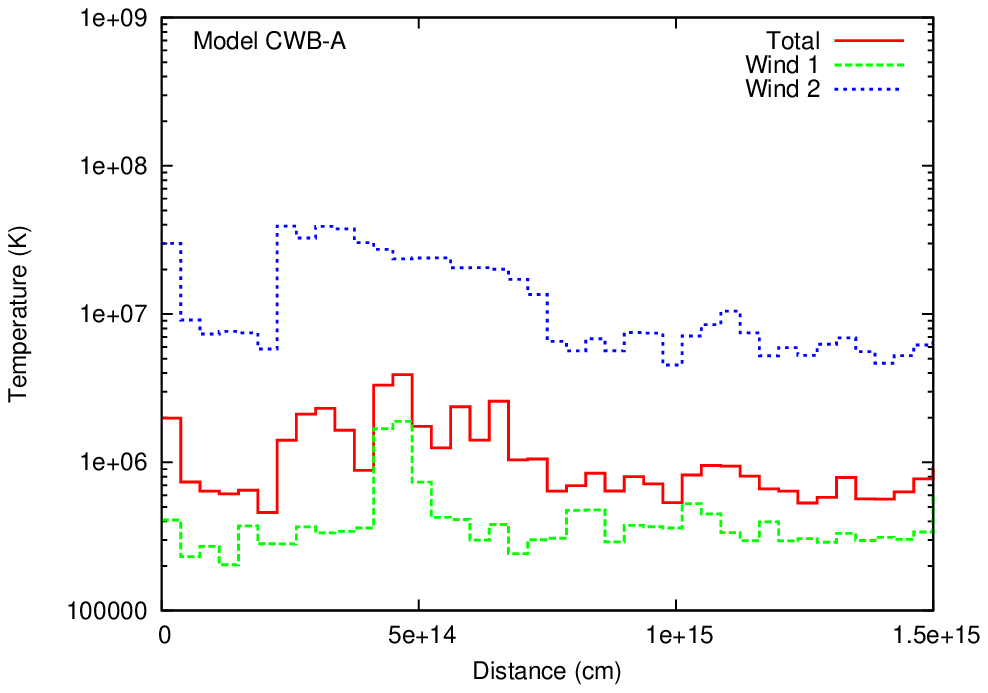}} \\
      \resizebox{80mm}{!}{\includegraphics{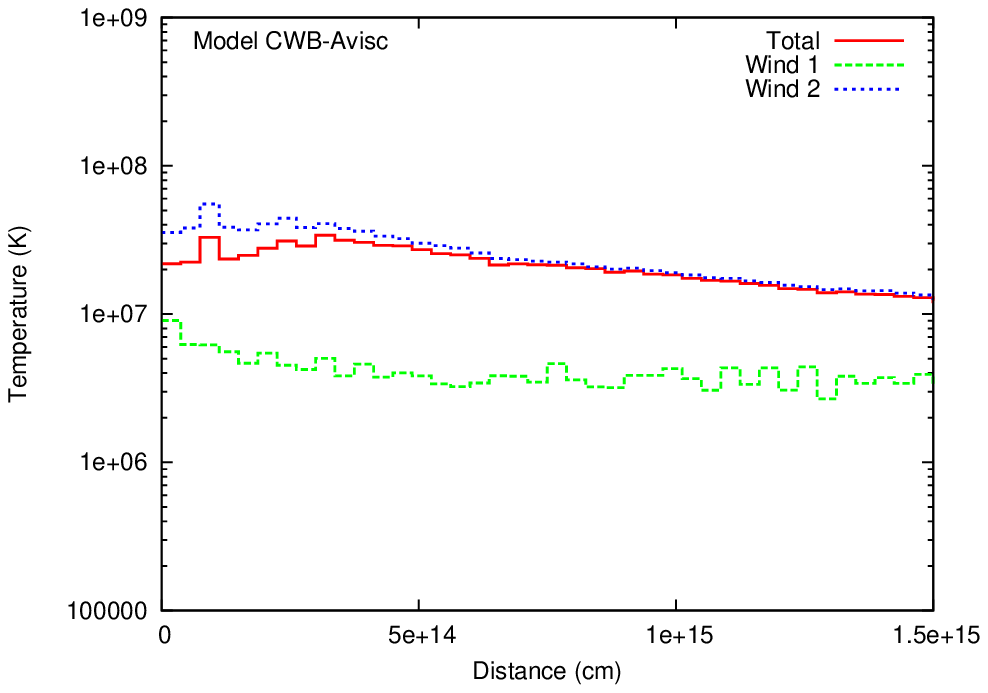}} \\
     \end{tabular}
    \caption{Mass-weighted postshock gas temperature as a function of
      distance from the apex of the WCR in model CWB-A (upper panel)
      and model CWB-Avisc (lower panel). When calculating the
      distributions a temperature cut of $T\ge 10^{5}\;$K was used.}
    \label{fig:av_temp}
  \end{center}
\end{figure}

In model CWB-A the instabilities help to re-heat both wind~1 and 2 gas
as it flows downstream resulting in recurring spikes in the
mass-weighted postshock gas temperature as one moves to larger
distances from the shock apex (Fig.~\ref{fig:av_temp}). However,
despite this additional heating, as the gas flows off the grid the
{\it mass-weighted} wind~2 temperature for model CWB-A is in fact
lower than that of model CWB-Avisc. This would suggest that although
the volume of postshock gas in model CWB-Avisc is smaller in size than
that of model CWB-A, the enhanced mixing at the CD in model CWB-A
produces a lower {\it mass-weighted temperature} than the smooth flow
of model CWB-Avisc. This does not, however, lead to a harder flow~2
spectrum from model CWB-Avisc, because the hottest wind~2 gas in the
models occurs closer to the apex of the WCR (see
Fig.~\ref{fig:model_spectra} and Table~\ref{tab:luminosities}). At the
apex of the WCR the temperature of postshock wind 2 gas is lower in
model CWB-Avisc than in model CWB-A, consistent with the expected
extra numerical conduction introduced by additional artificial
viscosity, which ultimately modifies the gas temperature and pressure.

The wind~1 distributions in Fig.~\ref{fig:av_temp} provide further
evidence for enhanced numerical conduction as the wind~1 temperature
in model CWB-Avisc is clearly higher than in model CWB-A. However, it
should be noted that in the calculations presented in
Fig.~\ref{fig:av_temp} a cut of $T=10^{5}\;$K was made, and thus the
distributions shown are for mass \textit{above} this limit - in model
CWB-Avisc all postshock wind~1 gas at $T> 10^{5}\;$K has an average
temperature of $\simeq 4\times10^6\;$K whereas in model CWB-A this
value is $\simeq7\times10^5\;$K. Examining the location of gas at
these mean temperatures we see that in model CWB-Avisc it resides in a
thin layer at the contact discontinuity whereas in model CWB-A the
turbulent WCR causes this gas to outline wind-clump interactions as
well as highlight the heavily distorted contact discontinuity (see the
wind~1 temperature plots in Fig.~\ref{fig:model_images}). Although
there is more wind~1 gas at $T>10^5\;$K in model CWB-A as a result of
vigorous mixing (see Fig.~\ref{fig:av_mass}), in model CWB-Avisc
heating at the CD via numerical conduction actually causes the {\it
  mass-weighted temperatures} to be higher, resulting in a slightly
harder wind~1 spectrum (see Fig.~\ref{fig:model_spectra}).

Focusing now on the inferred hard band (7-10 keV) luminosity, we have
performed a series of simulations with high artificial viscosity and
with the wind~1 parameters fixed, but with different parameters for
wind~2, the results of which are plotted in Fig.~\ref{fig:hr1}. At
$v_{2}\gtsimm 2.5\times10^{8}\;{\rm cm~s^{-1}}$ ($\chi_2\gtsimm 163$)
the wind~1 and wind~2 data points from the low and high viscosity
calculations correlate well. However, at $v_{2} \ltsimm
2\times10^{8}\;{\rm cm~s^{-1}}$ ($\chi_2\ltsimm 53$) the wind~1 points
with and without additional viscosity diverge considerably. This
result implies that the effects of numerical conduction are somewhat
($v_2 \gtsimm 2\times10^{8}\;{\rm cm~s^{-1}}$) to significantly ($v_2
\ltsimm 10^{8}\;{\rm cm~s^{-1}}$) reduced by the application of
additional artificial viscosity when instabilities that otherwise
would occur are strongly suppressed, thereby reducing the surface area
between flows. However, the cost of this approach is an unrealistic
description of the dynamics occuring in the WCR, which can effect the
derived observational characteristics (i.e. the spectral hardness of
the wind 2 emission from wind 2) as a result of limiting the presence
of smaller scale downstream shocks.

\subsubsection{Simulation resolution}
\label{subsec:sim_res}

To assess the effect of the simulation resolution on the dynamics of
the wind-wind collision region and the resulting X-ray emission,
calculations have been performed with model CWB-A parameters and with
cell sizes in the range $(1.17 - 11.7)\times10^{12}\;$cm.

At lower resolution the coarser grid essentially acts like viscosity;
the wavelength of resolvable instabilities is larger and the timescale
for the growth of the smallest resolvable instabilities is
longer. This affects the formation of structure on smaller spatial
scales as one tends towards lower resolution (i.e. larger cell
sizes). To the contrary, at higher resolution the formation of
structure is enhanced and to illustrate this point we show a snapshot
of the highest resolution simulation (model CWB-A+; twice the
resolution of model CWB-A) in Fig.~\ref{fig:model_images}. The dense
shell of cold, postshock primary wind fragments into more numerous
clumps of smaller scale which as before pass into the supersonic
stream of postshock wind~2 gas, forming bow shocks and tails. Because
of the smaller size of these clumps the timescale for their
destruction is shorter. This may limit the ability of clumps to
traverse completely out of the WCR, due to orbital motion, as shown by
\cite{Pittard:2009}. Despite the improved simulation resolution,
wind~1 gas is still being heated to temperatures well in excess of
those expected from the preshock wind velocity.

\begin{figure}
  \begin{center}
    \begin{tabular}{c}
      \resizebox{80mm}{!}{\includegraphics{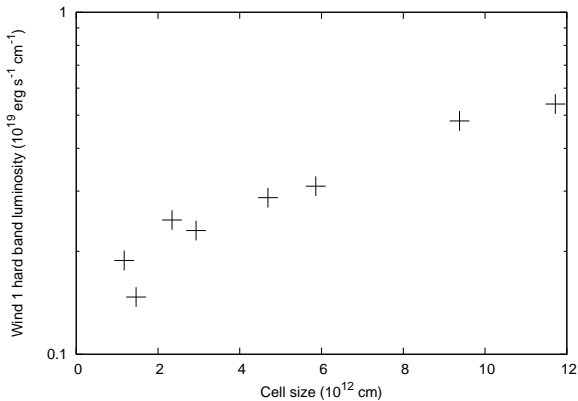}} \\
     \end{tabular}
    \caption{Hard band (7-10 keV) luminosity from wind~1 as a function
      of the grid resolution. Model CWB-A parameters were used for
      these calculations (see \S~\ref{subsec:sim_res}).}
    \label{fig:hr1_res}
  \end{center}
\end{figure}

Comparing the integrated 1-10 keV luminosities from models CWB-A and
CWB-A+ reveals lower values for the latter simulation
(Table~\ref{tab:luminosities}). The explanation for this can be found
in the spectrum for model CWB-A+ (Fig.~\ref{fig:model_spectra}) where
we see that the comparative decrease in the integrated luminosities is
due to a lower normalization in the wind~1 emission at all energies,
and a reduction in soft ($E\ltsimm3\;$keV) emission from wind~2. To
further examine this point we have performed a resolution test, the
results of which are presented in Fig.~\ref{fig:hr1_res}. The general
trend is for the artificial heating of wind~1 (by numerical
conduction) to decrease as the resolution increases (cell size
decreases). Therefore, the increased simulation resolution has the
effect of reducing the net emission calculated from wind~1, and
hardening the spectrum from wind~2.

In conclusion, the degree of fragmentation and the size of the clumps
is dependent on the adopted resolution. Higher resolution simulations
will create smaller clumps and vice-versa. In many (if not most)
astrophysical situations, hydrodynamical codes cannot simulate the
large Reynold's number flows that occur in reality. Thus the
instabilities are not as small as they should be. The increasing
popularity of sub-grid turbulence models may alleviate this problem
\citep[e.g.][]{Pittard:2009b}. Finally we note that magnetic
fields/pressure can limit the compression/density contrast of gas in
the dense shell, and may set a minimum size of clumps, while also
suppressing ablation and yielding longer-living clumps.

\section{Discussion}
\label{sec:discussion}

The numerical conduction of heat between hot and cold cells is an
undesirable side-effect of the approach used to solve the governing
equations of fluid flow. It has implications for a wide range of
phenomena, examples of which include:

\begin{itemize}
\item Colliding flows: the collision of opposing super-sonic flows is
  a common occurance in scenarios ranging from the formation of
  molecular clouds \citep{Folini:2006, Heitsch:2006, Heitsch:2008,
    Niklaus:2009}, the interactions of stellar winds from YSOs
  \citep{Delamarter:2000, Parkin:2009b}, and CWBs \citep{Stevens:1992,
    Pittard:2009} where the preshock winds may also be inhomogeneous
  \citep{Walder:2002, Pittard:2007, Pittard:2009,
    Pittard_Parkin:2010}. Thermal instability and the cooling of gas
  are of key importance to the early stages of star formation, and
  numerical heat conduction will have consequences for simulation
  results.
\item Expanding bubbles: the expansion of a high pressure bubble which
  sweeps up colder surrounding material into a dense shell is relevant
  for wind-blown bubbles around massive stars \citep[see][and
    references there-in]{Arthur:2007} and young stellar clusters
  \citep[][]{Canto:2000, Rockefeller:2005, Wunsch:2008,
    Rodriquez-Gonzalez:2008, Reyes-Iturbide:2009}, expanding
  thin-shells around HII regions \citep{Dale:2009} and SNe
  \citep[e.g.][]{Tenorio-Tagle:1991, Dwarkadas:2007, Ferrand:2010},
  and starburst regions \citep{Strickland:2000, Marcolini:2004,
    Cooper:2008, Strickland:2009, Fujita:2009, Tang:2009}. Numerical
  conduction at the interface between the hotter gas in the interior
  of the bubble, SNR, or HII region, and the surrounding cold dense
  shell, leads to the evaporation of material from the shell into the
  bubble, and will change the emission and dynamics of these objects.
\item Flow-clump interactions: the interaction of a fast (and often
  hot) diffuse flow and a clumpy medium is a common occurance in
  astronomy, occuring in such diverse objects as PNe
  \citep[e.g. ][]{Meaburn:1998, Matsuura:2009} and starburst
  superwinds \citep[e.g. ][]{Strickland:2000b, Cecil:2002}. A large
  body of numerical simulations of this interaction now exists
  \citep[e.g.][]{Klein:1994, MacLow:1994, Gregori:1999, Gregori:2000,
    Mellema:2002, Falle:2002, Fragile:2004, Fragile:2005,
    Melioli:2004, Melioli:2005, Orlando:2005, Orlando:2006,
    Orlando:2008, Pittard:2005b, Pittard:2009b, Dyson:2006,
    Tenorio-Tagle:2006, vanLoo:2007, vanLoo:2010, Shin:2008,
    Pittard:2009b, Cooper:2009, Yirak:2009}. Observational signatures
  such as H$\alpha$, OVI, and X-ray emission from starburst regions
  \citep[e.g. ][]{Westmoquette:2007} and superwinds
  \citep[e.g. ][]{Strickland:2000b, Cecil:2002} will be sensitive to
  the level of heat conduction between the hot and cold phases,
  artificial or otherwise.
\end{itemize}

The X-ray spectra calculated from the colliding plane-parallel flow
models presented in \S~\ref{subsec:slab_model} show that large
deviations can arise from numerical conduction. For example, emission
from flow~2 in model SLAB-C is 100 times greater than that in model
SLAB-E. Numerical heat conduction is also seen in the CWB models. Of
course, in reality some physical mixing/diffusion may occur. In this
work we have not considered thermal electron conduction which, due to
the large temperature gradients present in the simulations, may be
important. For instance, the transport of heat across the contact
discontinuity by thermal electrons can ``evaporate'' cold dense
postshock gas \citep{Myasnikov:1998, Zhekov:1998}, affecting the flow
dynamics and the derived X-ray emissivities. However, depending on the
strength and orientation of the magnetic field the efficiency of
thermal electron conduction may be significantly hindered
\citep[e.g. ][]{Orlando:2008}. Therefore, to accurately account for
the effect of {\it thermal electron} conduction requires the inclusion
of the relevant electron transport physics and magnetic fields in the
simulations. With this in mind, it is important also to understand the
level at which numerical conduction acts at.

The current investigation has focused on grid-based hydrodynamics
(GH), of which there are numerous codes available to the community
\citep{Fryxell:2000, Norman:2000, Teyssier:2002, O'Shea:2004,
  Mignone:2007, Stone:2008}. This choice is in part justified by the
finding that GH is considerably better at handling strong shocks,
contact discontinuities, and instabilities than SPH
\citep{Agertz:2007, Tasker:2008}, although recent developments to the
SPH scheme have improved its ability to model these phenomena
\citep[e.g.][]{Rosswog:2007, Price:2008, Read:2009, Kawata:2009,
  Saitoh:2009, Cha:2010}. In light of the growing literature of code
comparisons which aim to elucidate differences between simulation
results from GH and SPH codes \citep[e.g.][]{Frenk:1999, Agertz:2007,
  Commercon:2008, Tasker:2008, Wadsley:2008, Kitsionas:2008}, it would
be very interesting to repeat the current investigation with a
state-of-the-art SPH code.

\section{Conclusions}
\label{sec:conclusions}

We have presented hydrodynamic models of colliding hypersonic flows
with the aim of examining the effects of numerical conduction on the
simulation dynamics and the derived X-ray characteristics. The
conduction of heat occurs across flow discontinuities due to diffusive
terms introduced in the discretization of the governing flow
equations. A key conclusion from this work is that the magnitude of
the numerical heat conduction is strongly related to the density (and
temperature) contrast between adjacent gas. X-ray calculations
performed on the simulation results show that significant changes to
spectra can occur by numerical conduction alone. Further tests
performed with additional artificial viscosity reveal a complicated
relationship between the flow dynamics, the magnitude of numerical
conduction, and the resulting X-ray emission. For instance, the
inherent instability of the collision regions of hypersonic flows
naturally enhances the interface area between the flows, which in turn
enhances the level of numerical conduction. Introducing sufficient
viscosity to damp the growth of instabilities can reduce these
effects, but the additional diffusion introduced into the fluid
equations may increase the level of numerical heat conduction where
the interface is relatively stable (e.g. near the apex of the
wind-wind collision region in a colliding winds binary
system). Finally, we note that while enhancing the resolution of the
simulation increases the growth of small scale instabilities, and thus
the area of the interface between the hot and cold phases, the overall
effect of numerical conduction is reduced.

In the present work we have highlighted a fundamental problem
encountered when using grid-based hydrodynamics to model fluids where
high density and temperature contrasts are present - conditions which
can be found in a multitude of astrophysical phenomena. Unfortunately,
there is no simple fix. The brute-force approach to resolving this
problem would be to employ higher simulation resolution, though this
is not always a realistic option. 

\subsection*{Acknowledgements}
This work was supported in part by a Henry Ellison Scholarship from
the University of Leeds, and by a PRODEX XMM/Integral contract
(Belspo). JMP gratefully acknowledges funding from the Royal Society
and previous discussions with Robin Williams which instigated this
work. The software used in this work was in part developed by the
DOE-supported ASC/Alliance Center for Astrophysical Thermonuclear
Flashes at the University of Chicago.


\begin{thebibliography}{}

\bibitem[\protect\citeauthoryear{{Agertz}, {Moore}, {Stadel}, {Potter},
  {Miniati}, {Read}, {Mayer}, {Gawryszczak}, {Kravtsov}, {Nordlund}, {Pearce},
  {Quilis}, {Rudd}, {Springel}, {Stone}, {Tasker}, {Teyssier}, {Wadsley} \&
  {Walder}}{{Agertz} et~al.}{2007}]{Agertz:2007}
{Agertz}, O., {et~al.} 2007, \mnras, 380, 963

\bibitem[\protect\citeauthoryear{{Antokhin}, {Owocki} \& {Brown}}{{Antokhin}
  et~al.}{2004}]{Antokhin:2004}
{Antokhin}, I.~I., {Owocki}, S.~P., \& {Brown}, J.~C. 2004, \apj, 611, 434

\bibitem[\protect\citeauthoryear{{Arthur}}{{Arthur}}{2007}]{Arthur:2007}
{Arthur}, S.~J. 2007, {Wind-Blown Bubbles around Evolved Stars}.
Springer Dordrecht, pp 183--+

\bibitem[\protect\citeauthoryear{{Banerjee}, {V{\'a}zquez-Semadeni}, 
    {Hennebelle}, \& {Klessen}}{Banerjee et~al.}{2009}]{Banerjee:2009}
{Banerjee}, R., {V{\'a}zquez-Semadeni}, E., {Hennebelle}, P., \& 
{Klessen}, R.~S. 2009, \mnras, 398, 1082

\bibitem[\protect\citeauthoryear{{Berger} \& {Oliger}}{{Berger} \&
  {Oliger}}{1989}]{Berger:1989}
{Berger}, M.~J. \& {Oliger}, J. 1989, Journal of Computational Physics, 53, 484

\bibitem[\protect\citeauthoryear{{Bonito}, {Orlando}, {Peres}, {Favata} \&
  {Rosner}}{{Bonito} et~al.}{2007}]{Bonito:2007}
{Bonito}, R., {Orlando}, S., {Peres}, G., {Favata}, F., \& {Rosner}, R. 2007,
  \aap, 462, 645

\bibitem[\protect\citeauthoryear{{Cant{\'o}}, {Raga} \&
  {Rodr{\'{\i}}guez}}{{Cant{\'o}} et~al.}{2000}]{Canto:2000}
{Cant{\'o}}, J., {Raga}, A.~C., \& {Rodr{\'{\i}}guez}, L.~F. 2000, \apj, 536,
  896

\bibitem[\protect\citeauthoryear{{Cecil}, {Bland-Hawthorn} \&
  {Veilleux}}{{Cecil} et~al.}{2002}]{Cecil:2002}
{Cecil}, G., {Bland-Hawthorn}, J., \& {Veilleux}, S. 2002, \apj, 576, 745

\bibitem[\protect\citeauthoryear{{Cha}, {Inutsuka} \& {Nayakshin}}{{Cha}
  et~al.}{2010}]{Cha:2010}
{Cha}, S., {Inutsuka}, S., \& {Nayakshin}, S. 2010MNRAS.tmp..83C

\bibitem[\protect\citeauthoryear{{Chevalier} \& {Imamura}}{{Chevalier} \&
  {Imamura}}{1982}]{Chevalier:1982}
{Chevalier}, R.~A. \& {Imamura}, J.~N. 1982, \apj, 261, 543

\bibitem[\protect\citeauthoryear{{Colella} \& {Woodward}}{{Colella} \&
  {Woodward}}{1984}]{Colella:1984}
{Colella}, P. \& {Woodward}, P.~R. 1984, Journal of Computational Physics, 54,
  174

\bibitem[\protect\citeauthoryear{{Commer{\c c}on}, {Hennebelle}, {Audit},
  {Chabrier} \& {Teyssier}}{{Commer{\c c}on} et~al.}{2008}]{Commercon:2008}
{Commer{\c c}on}, B., {Hennebelle}, P., {Audit}, E., {Chabrier}, G., \&
  {Teyssier}, R. 2008, \aap, 482, 371

\bibitem[\protect\citeauthoryear{{Cooper}, {Bicknell}, {Sutherland} \&
  {Bland-Hawthorn}}{{Cooper} et~al.}{2008}]{Cooper:2008}
{Cooper}, J.~L., {Bicknell}, G.~V., {Sutherland}, R.~S., \& {Bland-Hawthorn},
  J. 2008, \apj, 674, 157

\bibitem[\protect\citeauthoryear{{Cooper}, {Bicknell}, {Sutherland} \&
  {Bland-Hawthorn}}{{Cooper} et~al.}{2009}]{Cooper:2009}
{Cooper}, J.~L., {Bicknell}, G.~V., {Sutherland}, R.~S., \& {Bland-Hawthorn},
  J. 2009, \apj, 703, 330

\bibitem[\protect\citeauthoryear{{Dale}, {W{\"u}nsch}, {Whitworth} \& {Palou{\v
  s}}}{{Dale} et~al.}{2009}]{Dale:2009}
{Dale}, J.~E., {W{\"u}nsch}, R., {Whitworth}, A., \& {Palou{\v s}}, J. 2009,
  \mnras, 398, 1537

\bibitem[\protect\citeauthoryear{{Delamarter}, {Frank} \&
  {Hartmann}}{{Delamarter} et~al.}{2000}]{Delamarter:2000}
{Delamarter}, G., {Frank}, A., \& {Hartmann}, L. 2000, \apj, 530, 923

\bibitem[\protect\citeauthoryear{{Dwarkadas}}{{Dwarkadas}}{2007}]{Dwarkadas:20%
07}
{Dwarkadas}, V.~V. 2007, \apss, 307, 153

\bibitem[\protect\citeauthoryear{{Dyson}, {Pittard}, {Meaburn} \&
  {Falle}}{{Dyson} et~al.}{2006}]{Dyson:2006}
{Dyson}, J.~E., {Pittard}, J.~M., {Meaburn}, J., \& {Falle}, S.~A.~E.~G. 2006,
  \aap, 457, 561

\bibitem[\protect\citeauthoryear{{Falle}}{{Falle}}{1991}]{Falle:1991}
{Falle}, S.~A.~E.~G. 1991, \mnras, 250, 581

\bibitem[\protect\citeauthoryear{{Falle}, {Coker}, {Pittard}, {Dyson} \&
  {Hartquist}}{{Falle} et~al.}{2002}]{Falle:2002}
{Falle}, S.~A.~E.~G., {Coker}, R.~F., {Pittard}, J.~M., {Dyson}, J.~E., \&
  {Hartquist}, T.~W. 2002, \mnras, 329, 670

\bibitem[\protect\citeauthoryear{{Ferrand}, {Decourchelle}, {Ballet},
  {Teyssier} \& {Fraschetti}}{{Ferrand} et~al.}{2010}]{Ferrand:2010}
{Ferrand}, G., {Decourchelle}, A., {Ballet}, J., {Teyssier}, R., \&
  {Fraschetti}, F. 2010, \aap, 509, 10

\bibitem[\protect\citeauthoryear{{Folini} \& {Walder}}{{Folini} \&
  {Walder}}{2006}]{Folini:2006}
{Folini}, D. \& {Walder}, R. 2006, \aap, 459, 1

\bibitem[\protect\citeauthoryear{{Fragile}, {Anninos}, {Gustafson} \&
  {Murray}}{{Fragile} et~al.}{2005}]{Fragile:2005}
{Fragile}, P.~C., {Anninos}, P., {Gustafson}, K., \& {Murray}, S.~D. 2005,
  \apj, 619, 327

\bibitem[\protect\citeauthoryear{{Fragile}, {Murray}, {Anninos} \& {van
  Breugel}}{{Fragile} et~al.}{2004}]{Fragile:2004}
{Fragile}, P.~C., {Murray}, S.~D., {Anninos}, P., \& {van Breugel}, W. 2004,
  \apj, 604, 74

\bibitem[\protect\citeauthoryear{{Frenk}, {White}, {Bode}, {Bond}, {Bryan},
  {Cen}, {Couchman}, {Evrard}, {Gnedin}, {Jenkins}, {Khokhlov} \&
  {Klypin}}{{Frenk} et~al.}{1999}]{Frenk:1999}
{Frenk}, C.~S., {et~al.} 1999, \apj, 525, 554

\bibitem[\protect\citeauthoryear{{Fryxell}, {Olson}, {Ricker}, {Timmes},
  {Zingale}, {Lamb}, {MacNeice}, {Rosner}, {Truran} \& {Tufo}}{{Fryxell}
  et~al.}{2000}]{Fryxell:2000}
{Fryxell}, B., {et~al.} 2000, \apjs, 131, 273

\bibitem[\protect\citeauthoryear{{Fujita}, {Martin}, {Low}, {New} \&
  {Weaver}}{{Fujita} et~al.}{2009}]{Fujita:2009}
{Fujita}, A., {Martin}, C.~L., {Low}, M., {New}, K.~C.~B., \& {Weaver}, R.
  2009, \apj, 698, 693

\bibitem[\protect\citeauthoryear{{Gregori}, {Miniati}, {Ryu} \&
  {Jones}}{{Gregori} et~al.}{1999}]{Gregori:1999}
{Gregori}, G., {Miniati}, F., {Ryu}, D., \& {Jones}, T.~W. 1999, \apjl, 527,
  L113

\bibitem[\protect\citeauthoryear{{Gregori}, {Miniati}, {Ryu} \&
  {Jones}}{{Gregori} et~al.}{2000}]{Gregori:2000}
{Gregori}, G., {Miniati}, F., {Ryu}, D., \& {Jones}, T.~W. 2000, \apj, 543, 775

\bibitem[\protect\citeauthoryear{{Heitsch}, {Hartmann}, {Slyz}, {Devriendt} \&
  {Burkert}}{{Heitsch} et~al.}{2008}]{Heitsch:2008}
{Heitsch}, F., {Hartmann}, L.~W., {Slyz}, A.~D., {Devriendt}, J.~E.~G., \&
  {Burkert}, A. 2008, \apj, 674, 316

\bibitem[\protect\citeauthoryear{{Heitsch}, {Slyz}, {Devriendt}, {Hartmann} \&
  {Burkert}}{{Heitsch} et~al.}{2006}]{Heitsch:2006}
{Heitsch}, F., {Slyz}, A.~D., {Devriendt}, J.~E.~G., {Hartmann}, L.~W., \&
  {Burkert}, A. 2006, \apj, 648, 1052

\bibitem[\protect\citeauthoryear{{Imamura}, {Wolff} \& {Durisen}}{{Imamura}
  et~al.}{1984}]{Imamura:1984}
{Imamura}, J.~N., {Wolff}, M.~T., \& {Durisen}, R.~H. 1984, \apj, 276, 667

\bibitem[\protect\citeauthoryear{{Kaastra}}{{Kaastra}}{1992}]{Kaastra:1992}
{Kaastra}, J.~S. 1992, Internal SRON-Leiden Report

\bibitem[\protect\citeauthoryear{{Kawata}, {Okamoto}, {Cen} \&
  {Gibson}}{{Kawata} et~al.}{2009}]{Kawata:2009}
{Kawata}, D., {Okamoto}, T., {Cen}, R., \& {Gibson}, B.~K. 2009, ArXiv e-prints

\bibitem[\protect\citeauthoryear{{Kitsionas}, {Federrath}, {Klessen},
  {Schmidt}, {Price}, {Dursi}, {Gritschneder}, {Walch}, {Piontek}, {Kim},
  {Jappsen}, {Ciecielag} \& {Mac Low}}{{Kitsionas}
  et~al.}{2008}]{Kitsionas:2008}
{Kitsionas}, S., {et~al.} 2008, arXiv:0810.4599

\bibitem[\protect\citeauthoryear{{Klein}, {McKee} \& {Colella}}{{Klein}
  et~al.}{1994}]{Klein:1994}
{Klein}, R.~I., {McKee}, C.~F., \& {Colella}, P. 1994, \apj, 420, 213

\bibitem[\protect\citeauthoryear{{Lemaster}, {Stone} \& {Gardiner}}{{Lemaster}
  et~al.}{2007}]{Lemaster:2007}
{Lemaster}, M.~N., {Stone}, J.~M., \& {Gardiner}, T.~A. 2007, \apj, 662, 582

\bibitem[\protect\citeauthoryear{{Mac Low}, {McKee}, {Klein}, {Stone} \&
  {Norman}}{{Mac Low} et~al.}{1994}]{MacLow:1994}
{Mac Low}, M., {McKee}, C.~F., {Klein}, R.~I., {Stone}, J.~M., \& {Norman},
  M.~L. 1994, \apj, 433, 757

\bibitem[\protect\citeauthoryear{{MacNeice}, {Olson}, {Mobarry}, {deFainchtein}
  \& {Packer}}{{MacNeice} et~al.}{2000}]{MacNeice:2000}
{MacNeice}, P., {Olson}, K.~M., {Mobarry}, C., {deFainchtein}, R., \& {Packer},
  C. 2000, \cpc, 126, 330

\bibitem[\protect\citeauthoryear{{Marcolini}, {Brighenti} \&
  {D'Ercole}}{{Marcolini} et~al.}{2004}]{Marcolini:2004}
{Marcolini}, A., {Brighenti}, F., \& {D'Ercole}, A. 2004, \mnras, 352, 363

\bibitem[\protect\citeauthoryear{{Matsuura}, {Speck}, {McHunu}, {Tanaka},
  {Wright}, {Smith}, {Zijlstra}, {Viti} \& {Wesson}}{{Matsuura}
  et~al.}{2009}]{Matsuura:2009}
{Matsuura}, M., {et~al.} 2009, \apj, 700, 1067

\bibitem[\protect\citeauthoryear{{Meaburn}, {Clayton}, {Bryce}, {Walsh},
  {Holloway} \& {Steffen}}{{Meaburn} et~al.}{1998}]{Meaburn:1998}
{Meaburn}, J., {Clayton}, C.~A., {Bryce}, M., {Walsh}, J.~R., {Holloway},
  A.~J., \& {Steffen}, W. 1998, \mnras, 294, 201

\bibitem[\protect\citeauthoryear{{Melioli} \& {de Gouveia Dal Pino}}{{Melioli}
  \& {de Gouveia Dal Pino}}{2004}]{Melioli:2004}
{Melioli}, C. \& {de Gouveia Dal Pino}, E.~M. 2004, \aap, 424, 817

\bibitem[\protect\citeauthoryear{{Melioli}, {de Gouveia dal Pino} \&
  {Raga}}{{Melioli} et~al.}{2005}]{Melioli:2005}
{Melioli}, C., {de Gouveia dal Pino}, E.~M., \& {Raga}, A. 2005, \aap, 443, 495

\bibitem[\protect\citeauthoryear{{Mellema}, {Kurk} \&
  {R{\"o}ttgering}}{{Mellema} et~al.}{2002}]{Mellema:2002}
{Mellema}, G., {Kurk}, J.~D., \& {R{\"o}ttgering}, H.~J.~A. 2002, \aap, 395,
  L13

\bibitem[\protect\citeauthoryear{{Mewe}, {Kaastra} \& {Liedahl}}{{Mewe}
  et~al.}{1995}]{Mewe:1995}
{Mewe}, R., {Kaastra}, J.~S., \& {Liedahl}, D.~A. 1995, Legacy, 6, 16

\bibitem[\protect\citeauthoryear{{Mignone}}{{Mignone}}{2005}]{Mignone:2005}
{Mignone}, A. 2005, \apj, 626, 373

\bibitem[\protect\citeauthoryear{{Mignone}, {Bodo}, {Massaglia}, {Matsakos},
  {Tesileanu}, {Zanni} \& {Ferrari}}{{Mignone} et~al.}{2007}]{Mignone:2007}
{Mignone}, A., {Bodo}, G., {Massaglia}, S., {Matsakos}, T., {Tesileanu}, O.,
  {Zanni}, C., \& {Ferrari}, A. 2007, \apjs, 170, 228

\bibitem[\protect\citeauthoryear{{Myasnikov} \& {Zhekov}}{{Myasnikov}
    \& {Zhekov}}{1998}]{Myasnikov:1998} {Myasnikov}, A.~V. \&
  {Zhekov}, S.~A. 1998, /mnras, 300, 686

\bibitem[\protect\citeauthoryear{{Niklaus}, {Schmidt} \& {Niemeyer}}{{Niklaus}
  et~al.}{2009}]{Niklaus:2009}
{Niklaus}, M., {Schmidt}, W., \& {Niemeyer}, J.~C. 2009, \aap, 506, 1065

\bibitem[\protect\citeauthoryear{{Norman}}{{Norman}}{2000}]{Norman:2000}
{Norman}, M.~L. 2000, in {Arthur} S.~J.,  {Brickhouse} N.~S.,   {Franco} J.,
  eds, Revista Mexicana de Astronomia y Astrofisica Conference Series Vol.~9 of
  Revista Mexicana de Astronomia y Astrofisica Conference Series, {Introducing
  ZEUS-MP: A 3D, Parallel, Multiphysics Code for Astrophysical Fluid Dynamics}.
pp 66--71

\bibitem[\protect\citeauthoryear{{Orlando}, {Bocchino}, {Peres}, {Reale},
  {Plewa} \& {Rosner}}{{Orlando} et~al.}{2006}]{Orlando:2006}
{Orlando}, S., {Bocchino}, F., {Peres}, G., {Reale}, F., {Plewa}, T., \&
  {Rosner}, R. 2006, \aap, 457, 545

\bibitem[\protect\citeauthoryear{{Orlando}, {Bocchino}, {Reale}, {Peres} \&
  {Pagano}}{{Orlando} et~al.}{2008}]{Orlando:2008}
{Orlando}, S., {Bocchino}, F., {Reale}, F., {Peres}, G., \& {Pagano}, P. 2008,
  \apj, 678, 274

\bibitem[\protect\citeauthoryear{{Orlando}, {Peres}, {Reale}, {Bocchino},
  {Rosner}, {Plewa} \& {Siegel}}{{Orlando} et~al.}{2005}]{Orlando:2005}
{Orlando}, S., {Peres}, G., {Reale}, F., {Bocchino}, F., {Rosner}, R., {Plewa},
  T., \& {Siegel}, A. 2005, \aap, 444, 505

\bibitem[\protect\citeauthoryear{{O'Shea}, {Bryan}, {Bordner}, {Norman},
  {Abel}, {Harkness} \& {Kritsuk}}{{O'Shea} et~al.}{2004}]{O'Shea:2004}
{O'Shea}, B.~W., {Bryan}, G., {Bordner}, J., {Norman}, M.~L., {Abel}, T.,
  {Harkness}, R., \& {Kritsuk}, A. 2004, arXiv:astro-ph/0403044

\bibitem[\protect\citeauthoryear{{Parkin} \& {Pittard}}{{Parkin} \&
  {Pittard}}{2008}]{Parkin:2008}
{Parkin}, E.~R. \& {Pittard}, J.~M. 2008, \mnras, 388, 1047

\bibitem[\protect\citeauthoryear{{Parkin}, {Pittard}, {Corcoran}, {Hamaguchi}
  \& {Stevens}}{{Parkin} et~al.}{2009a}]{Parkin:2009}
{Parkin}, E.~R., {Pittard}, J.~M., {Corcoran}, M.~F., {Hamaguchi}, K., \&
  {Stevens}, I.~R. 2009a, \mnras, 394, 1758

\bibitem[\protect\citeauthoryear{{Parkin}, {Pittard}, {Hoare}, {Wright} \&
  {Drake}}{{Parkin} et~al.}{2009b}]{Parkin:2009b}
{Parkin}, E.~R., {Pittard}, J.~M., {Hoare}, M.~G., {Wright}, N.~J., \& {Drake},
  J.~J. 2009b, \mnras, pp 1372--+

\bibitem[\protect\citeauthoryear{{Pittard}}{{Pittard}}{2007}]{Pittard:2007}
{Pittard}, J.~M. 2007, \apj, 660, L141

\bibitem[\protect\citeauthoryear{{Pittard}}{{Pittard}}{2009}]{Pittard:2009}
{Pittard}, J.~M. 2009, \mnras, 396, 1743

\bibitem[\protect\citeauthoryear{{Pittard} \& {Corcoran}}{{Pittard} \&
  {Corcoran}}{2002}]{Pittard:2002}
{Pittard}, J.~M. \& {Corcoran}, M.~F. 2002, \aap, 383, 636

\bibitem[\protect\citeauthoryear{{Pittard}, {Dobson}, {Durisen}, {Dyson},
  {Hartquist} \& {O'Brien}}{{Pittard} et~al.}{2005}]{Pittard:2005}
{Pittard}, J.~M., {Dobson}, M.~S., {Durisen}, R.~H., {Dyson}, J.~E.,
  {Hartquist}, T.~W., \& {O'Brien}, J.~T. 2005, \aap, 438, 11

\bibitem[\protect\citeauthoryear{{Pittard}, {Dyson}, {Falle} \&
  {Hartquist}}{{Pittard} et~al.}{2005}]{Pittard:2005b}
{Pittard}, J.~M., {Dyson}, J.~E., {Falle}, S.~A.~E.~G., \& {Hartquist}, T.~W.
  2005, \mnras, 361, 1077

\bibitem[\protect\citeauthoryear{{Pittard}, {Falle}, {Hartquist} \&
  {Dyson}}{{Pittard} et~al.}{2009}]{Pittard:2009b}
{Pittard}, J.~M., {Falle}, S.~A.~E.~G., {Hartquist}, T.~W., \& {Dyson}, J.~E.
  2009, \mnras, 394, 1351

\bibitem[\protect\citeauthoryear{{Pittard} \& {Parkin}}{{Pittard} \&
  {Parkin}}{2010}]{Pittard_Parkin:2010}
{Pittard}, J.~M. \& {Parkin}, E.~R. 2010, \mnras, 403, 1657

\bibitem[\protect\citeauthoryear{{Pittard}, {Stevens}, {Corcoran} \&
  {Ishibashi}}{{Pittard} et~al.}{1998}]{Pittard:1998}
{Pittard}, J.~M., {Stevens}, I.~R., {Corcoran}, M.~F., \& {Ishibashi}, K. 1998,
  \mnras, 299, L5+

\bibitem[\protect\citeauthoryear{{Price}}{{Price}}{2008}]{Price:2008}
{Price}, D.~J. 2008, Journal of Computational Physics, 227, 10040

\bibitem[\protect\citeauthoryear{{Quirk}}{{Quirk}}{1994}]{Quirk:1994}
{Quirk}, J.~J. 1994, \ijnmf, 18, 555

\bibitem[\protect\citeauthoryear{{Read}, {Hayfield} \& {Agertz}}{{Read}
  et~al.}{2009}]{Read:2009}
{Read}, J.~I., {Hayfield}, T., \& {Agertz}, O. 2009, ArXiv e-prints

\bibitem[\protect\citeauthoryear{{Reyes-Iturbide}, {Vel{\'a}zquez}, {Rosado},
  {Rodr{\'{\i}}guez-Gonz{\'a}lez}, {Gonz{\'a}lez} \&
  {Esquivel}}{{Reyes-Iturbide} et~al.}{2009}]{Reyes-Iturbide:2009}
{Reyes-Iturbide}, J., {Vel{\'a}zquez}, P.~F., {Rosado}, M.,
  {Rodr{\'{\i}}guez-Gonz{\'a}lez}, A., {Gonz{\'a}lez}, R.~F., \& {Esquivel}, A.
  2009, \mnras, 394, 1009

\bibitem[\protect\citeauthoryear{{Robertson}, {Kravtsov}, {Gnedin},
    {Abel}, {Rudd}}{{Robertson} et~al.}{2005}]{Robertson:2010}
{Robertson}, B.~E. and {Kravtsov}, A.~V. and {Gnedin}, N.~Y. and 
	{Abel}, T. and {Rudd}, D.~H. 2010, \mnras, 401, 2463 

\bibitem[\protect\citeauthoryear{{Rockefeller}, {Fryer}, {Melia} \&
  {Wang}}{{Rockefeller} et~al.}{2005}]{Rockefeller:2005}
{Rockefeller}, G., {Fryer}, C.~L., {Melia}, F., \& {Wang}, Q.~D. 2005, \apj,
  623, 171

\bibitem[\protect\citeauthoryear{{Rodr{\'{\i}}guez-Gonz{\'a}lez}, {Esquivel},
  {Raga} \& {Cant{\'o}}}{{Rodr{\'{\i}}guez-Gonz{\'a}lez}
  et~al.}{2008}]{Rodriquez-Gonzalez:2008}
{Rodr{\'{\i}}guez-Gonz{\'a}lez}, A., {Esquivel}, A., {Raga}, A.~C., \&
  {Cant{\'o}}, J. 2008, \apj, 684, 1384

\bibitem[\protect\citeauthoryear{{Rosswog} \& {Price}}{{Rosswog} \&
  {Price}}{2007}]{Rosswog:2007}
{Rosswog}, S. \& {Price}, D. 2007, \mnras, 379, 915

\bibitem[\protect\citeauthoryear{{Saitoh} \& {Makino}}{{Saitoh} \&
  {Makino}}{2009}]{Saitoh:2009}
{Saitoh}, T.~R. \& {Makino}, J. 2009, \apjl, 697, L99

\bibitem[\protect\citeauthoryear{{Shang}, {Allen}, {Li}, {Liu}, {Chou} \&
  {Anderson}}{{Shang} et~al.}{2006}]{Shang:2006}
{Shang}, H., {Allen}, A., {Li}, Z., {Liu}, C., {Chou}, M., \& {Anderson}, J.
  2006, \apj, 649, 845

\bibitem[\protect\citeauthoryear{{Shin}, {Stone} \& {Snyder}}{{Shin}
  et~al.}{2008}]{Shin:2008}
{Shin}, M., {Stone}, J.~M., \& {Snyder}, G.~F. 2008, \apj, 680, 336

\bibitem[\protect\citeauthoryear{{Stevens}, {Blondin} \& {Pollock}}{{Stevens}
  et~al.}{1992}]{Stevens:1992}
{Stevens}, I.~R., {Blondin}, J.~M., \& {Pollock}, A.~M.~T. 1992, \apj, 386, 265

\bibitem[\protect\citeauthoryear{{Stone}, {Gardiner}, {Teuben}, {Hawley} \&
  {Simon}}{{Stone} et~al.}{2008}]{Stone:2008}
{Stone}, J.~M., {Gardiner}, T.~A., {Teuben}, P., {Hawley}, J.~F., \& {Simon},
  J.~B. 2008, \apjs, 178, 137

\bibitem[\protect\citeauthoryear{{Strickland} \& {Heckman}}{{Strickland} \&
  {Heckman}}{2009}]{Strickland:2009}
{Strickland}, D.~K. \& {Heckman}, T.~M. 2009, \apj, 697, 2030

\bibitem[\protect\citeauthoryear{{Strickland}, {Heckman}, {Weaver} \&
  {Dahlem}}{{Strickland} et~al.}{2000}]{Strickland:2000b}
{Strickland}, D.~K., {Heckman}, T.~M., {Weaver}, K.~A., \& {Dahlem}, M. 2000,
  \aj, 120, 2965

\bibitem[\protect\citeauthoryear{{Strickland} \& {Stevens}}{{Strickland} \&
  {Stevens}}{2000}]{Strickland:2000}
{Strickland}, D.~K. \& {Stevens}, I.~R. 2000, \mnras, 314, 511

\bibitem[\protect\citeauthoryear{{Strickland} \& {Blondin}}{{Strickland} \&
  {Blondin}}{1995}]{Strickland:1995}
{Strickland}, R. \& {Blondin}, J.~M. 1995, \apj, 449, 727

\bibitem[\protect\citeauthoryear{{Sutherland} \& {Bicknell}}{{Sutherland} \&
  {Bicknell}}{2007}]{Sutherland:2007}
{Sutherland}, R.~S. \& {Bicknell}, G.~V. 2007, \apjs, 173, 37

\bibitem[\protect\citeauthoryear{{Tang}, {Wang}, {Mac Low} \& {Joung}}{{Tang}
  et~al.}{2009}]{Tang:2009}
{Tang}, S., {Wang}, Q.~D., {Mac Low}, M., \& {Joung}, M.~R. 2009, \mnras, 398,
  1468

\bibitem[\protect\citeauthoryear{{Tasker}, {Brunino}, {Mitchell}, {Michielsen},
  {Hopton}, {Pearce}, {Bryan} \& {Theuns}}{{Tasker} et~al.}{2008}]{Tasker:2008}
{Tasker}, E.~J., {Brunino}, R., {Mitchell}, N.~L., {Michielsen}, D., {Hopton},
  S., {Pearce}, F.~R., {Bryan}, G.~L., \& {Theuns}, T. 2008, \mnras, 390, 1267

\bibitem[\protect\citeauthoryear{{Tenorio-Tagle}, {Mu{\~n}oz-Tu{\~n}{\'o}n},
  {P{\'e}rez}, {Silich} \& {Telles}}{{Tenorio-Tagle}
  et~al.}{2006}]{Tenorio-Tagle:2006}
{Tenorio-Tagle}, G., {Mu{\~n}oz-Tu{\~n}{\'o}n}, C., {P{\'e}rez}, E., {Silich},
  S., \& {Telles}, E. 2006, \apj, 643, 186

\bibitem[\protect\citeauthoryear{{Tenorio-Tagle}, {Rozyczka}, {Franco} \&
  {Bodenheimer}}{{Tenorio-Tagle} et~al.}{1991}]{Tenorio-Tagle:1991}
{Tenorio-Tagle}, G., {Rozyczka}, M., {Franco}, J., \& {Bodenheimer}, P. 1991,
  \mnras, 251, 318

\bibitem[\protect\citeauthoryear{{Tenorio-Tagle}, {Silich} \&
  {Mu{\~n}oz-Tu{\~n}{\'o}n}}{{Tenorio-Tagle} et~al.}{2003}]{Tenorio-Tagle:2003}
{Tenorio-Tagle}, G., {Silich}, S., \& {Mu{\~n}oz-Tu{\~n}{\'o}n}, C. 2003, \apj,
  597, 279

\bibitem[\protect\citeauthoryear{{Teyssier}}{{Teyssier}}{2002}]{Teyssier:2002}
{Teyssier}, R. 2002, \aap, 385, 337

\bibitem[\protect\citeauthoryear{{van Loo}, {Falle}, {Hartquist} \&
  {Moore}}{{van Loo} et~al.}{2007}]{vanLoo:2007}
{van Loo}, S., {Falle}, S.~A.~E.~G., {Hartquist}, T.~W., \& {Moore}, T.~J.~T.
  2007, \aap, 471, 213

\bibitem[\protect\citeauthoryear{{van Loo}, {Falle}, \&
    {Hartquist}}{{van Loo} et~al.}{2010}]{vanLoo:2010} {van Loo}, S.,
  {Falle}, S.~A.~E.~G., \& {Hartquist}, T.~W. 2010, arXiv1003.5843V

\bibitem[\protect\citeauthoryear{{Vishniac}}{{Vishniac}}{1983}]{Vishniac:1983}
{Vishniac}, E.~T. 1983, \apj, 274, 152

\bibitem[\protect\citeauthoryear{{Vishniac}}{{Vishniac}}{1994}]{Vishniac:1994}
{Vishniac}, E.~T. 1994, \apj, 428, 186

\bibitem[\protect\citeauthoryear{{Wadsley}, {Veeravalli} \&
  {Couchman}}{{Wadsley} et~al.}{2008}]{Wadsley:2008}
{Wadsley}, J.~W., {Veeravalli}, G., \& {Couchman}, H.~M.~P. 2008, \mnras, 387,
  427

\bibitem[\protect\citeauthoryear{{Walder} \& {Folini}}{{Walder} \&
  {Folini}}{1996}]{Walder:1996}
{Walder}, R. \& {Folini}, D. 1996, \aap, 315, 265

\bibitem[\protect\citeauthoryear{{Walder} \& {Folini}}{{Walder} \&
  {Folini}}{1998}]{Walder:1998}
{Walder}, R. \& {Folini}, D. 1998, \aap, 330, 21

\bibitem[\protect\citeauthoryear{{Walder} \& {Folini}}{{Walder} \&
  {Folini}}{2000}]{Walder:2000}
{Walder}, R. \& {Folini}, D. 2000, \apss, 274, 343

\bibitem[\protect\citeauthoryear{{Walder} \& {Folini}}{{Walder} \&
  {Folini}}{2002}]{Walder:2002}
{Walder}, R. \& {Folini}, D. 2002, in {A.~F.~J.~Moffat \& N.~St-Louis} ed.,
  Interacting Winds from Massive Stars Vol.~260 of Astronomical Society of the
  Pacific Conference Series, {Theoretical Considerations on Colliding Clumped
  Winds}.
pp 595--+

\bibitem[\protect\citeauthoryear{{Westmoquette}, {Smith}, {Gallagher} III,
  {O'Connell}, {Rosario} \& {de Grijs}}{{Westmoquette}
  et~al.}{2007}]{Westmoquette:2007}
{Westmoquette}, M.~S., {Smith}, L.~J., {Gallagher}, III, J.~S., {O'Connell},
  R.~W., {Rosario}, D.~J., \& {de Grijs}, R. 2007, \apj, 671, 358

\bibitem[\protect\citeauthoryear{{W{\"u}nsch}, {Tenorio-Tagle}, {Palou{\v s}}
  \& {Silich}}{{W{\"u}nsch} et~al.}{2008}]{Wunsch:2008}
{W{\"u}nsch}, R., {Tenorio-Tagle}, G., {Palou{\v s}}, J., \& {Silich}, S. 2008,
  \apj, 683, 683

\bibitem[\protect\citeauthoryear{{Yirak}, {Frank} \& {Cunningham}}{{Yirak}
  et~al.}{2009}]{Yirak:2009}
{Yirak}, K., {Frank}, A., \& {Cunningham}, A.~J. 2009, arXiv:0912.4777

\bibitem[\protect\citeauthoryear{{Zhekov} \& {Myasnikov}}{{Zhekov} \&
    {Myasnikov}}{1998}]{Zhekov:1998} {Zhekov}, S.~A.\& {Myasnikov},
  A.~V. 1998, New Astronomy, 3, 57

\end{thebibliography}

\label{lastpage}


\end{document}